\documentclass[11pt]{article}
\usepackage[a4paper,margin=1in]{geometry}
\usepackage{amsmath,amsthm,amssymb,mathtools,bm}
\usepackage{microtype}
\usepackage{natbib}
\usepackage{hyperref}
\hypersetup{colorlinks=true,linkcolor=blue,citecolor=blue,urlcolor=blue}
\usepackage[nofiglist,notablist,nomarkers]{endfloat}
\usepackage{tikz}
\usetikzlibrary{
  arrows.meta,        
  positioning,        
  shapes.geometric,   
  calc,
  decorations.pathreplacing 
}
\usepackage{subcaption}

\usepackage[utf8]{inputenc}
\usepackage[T1]{fontenc}

\title{Linear Risk Sharing on Networks}
\author{%
  Arthur Charpentier\textsuperscript{1,2,*} \and
  Philipp Ratz\textsuperscript{1}
}
\date{\begin{flushleft}
    \footnotesize
    \textsuperscript{*}~Corresponding author, \texttt{charpentier.arthur@quam.ca}\\
    \textsuperscript{1}~Université du Québec à Montréal, Canada\\
    \textsuperscript{2}~Kyoto University, Japan
  \end{flushleft}}

\newtheorem{definition}{Definition}
\newtheorem{proposition}{Proposition}
\newtheorem{lemma}{Lemma}
\newtheorem{remark}{Remark}
\newtheorem{corollary}{Corollary}
\newtheorem{example}{Example}
\newtheorem{theorem}{Theorem}

\begin{document}
\maketitle

\begin{abstract}
Over the past decade alternatives to traditional insurance and banking have grown in popularity. The desire to encourage local participation has lead products such as peer-to-peer insurance, reciprocal contracts, and decentralized finance platforms to increasingly rely on network structures to redistribute risk among participants. In this paper, we develop a comprehensive framework for \emph{linear risk sharing} (LRS), where random losses are reallocated through nonnegative linear operators which can accommodate a wide range of networks. Building on the theory of stochastic and doubly stochastic matrices, we establish conditions under which constraints such as budget balance, fairness, and diversification are guaranteed. The convex order framework allows us to compare different allocations rigorously, highlighting variance reduction and majorization as natural consequences of doubly
stochastic mixing. We then extend the analysis to network-based sharing, showing how their topology shapes risk outcomes in complete, star, ring, random, and scale-free graphs. A second layer of randomness, where the sharing matrix itself is random, is introduced via Erdős--Rényi and preferential-attachment networks, connecting
risk-sharing properties to degree distributions. Finally, we study convex combinations of identity and network-induced operators, capturing the trade-off between self-retention and diversification. Our results provide design principles for fair and efficient peer-to-peer insurance and network-based risk pooling, combining mathematical soundness with economic interpretability.
\end{abstract}


\ 



\section{Introduction}

\subsection{Motivation and Related Literature}

The redistribution of risk lies at the heart of insurance and financial stability. Classical actuarial theory models a homogeneous pool of individuals who agree to share random losses symmetrically, ensuring that each participant faces the same distribution of residual risk. Such models rely heavily on anonymity and exchangeability: agents are indistinguishable, large in numbers and sharing rules are symmetric. This paradigm has been formalized using convex order and majorization, which provide natural criteria to compare risk allocations under risk-averse preferences \citep{Ohlin1969,hardy1934inequalities,ShakedShanthikumar2007,denuit2006actuarial}. The mathematical link between convex order and doubly stochastic matrices has long been recognized as a cornerstone for understanding diversification in insurance portfolios \citep{Dahl1999,Carlier2002}.

Whereas much of the theory around risk-sharing developed around these paradigms, risk-sharing in practice rarely occurs in anonymous pools. On the contrary, modern risk-sharing arrangements often involve heterogeneous structures such as peer-to-peer insurance, reciprocal contracts, mutual aid platforms, or reinsurance treaties, where each participant has a distinct position defined by local and known connections. Examples include online platforms such as {Friendsurance} or {Lemonade} where risk-sharing is explicitly performed with known connections, as well as blockchain-based decentralized finance (DeFi) protocols, where connections are verified through a communal approach. 

As we will discuss in this article, a crucial insight in such contexts is that \emph{network structure matters}: diversification depends crucially on local degrees and connectivity, with scale-free structures, which often appear in the real world, leading to persistent inequities between hubs and peripheral agents \citep{BarabasiAlbert1999}. A second key insight is that \emph{matrix representations provide clarity} over the underlying mechanisms. As we will show, many decentralized schemes can be expressed as linear operators acting on the loss vector, with nonnegativity and stochasticity constraints encoding budget balance, fairness, and convexity. Roughly speaking: row-stochasticity enforces per-agent fairness (convex combinations), column-stochasticity enforces budget balance, and doubly-stochasticity guarantees both. This guide will serve as a roadmap for the technical sections that follow.


\subsection{Agenda}

The present paper builds on these insights by developing a unified framework for \emph{linear risk sharing} (LRS). Here post-sharing allocations are obtained by applying a nonnegative matrix to the vector of pre-sharing losses. By encoding allocation rules as the nonnegative matrices, we connect classical stochastic-order arguments with modern network models, and demonstrate how algebraic constraints on the sharing operator translate directly into economic properties such as variance reduction or incentive compatibility. Our main contributions can be summarized as follows:

\begin{enumerate}
    \item \textbf{General framework.} We characterize the algebraic properties of sharing matrices (namely nonnegativity, row-stochasticity, column-stochasticity, and doubly-stochasticity)
    and establish precise conditions under which allocations preserve budget balance, fairness, 
    and mean-preservation. These properties connect directly to convex order dominance, variance 
    inequalities, and classical results on majorization.

    \item \textbf{Convex ordering of allocations.} Using convex order and componentwise convex 
    order, we provide a hierarchy of comparisons between different schemes. Doubly-stochastic 
    matrices emerge as the ``safe class,'' ensuring universal variance reduction and fairness, 
    while weaker stochasticity conditions lead to partial or representative-agent guarantees.

    \item \textbf{Networks and random graphs.} We specialize the framework to network-induced 
    sharing rules, showing how topology shapes diversification. Complete, star, ring, 
    Erdős–Rényi, and scale-free graphs are analyzed, highlighting the role of degree distributions 
    in determining fairness and variance outcomes. A second layer of randomness, where the 
    sharing matrix itself is random, provides a tractable model for peer-to-peer systems with 
    stochastic participation.

    \item \textbf{Hybrid mechanisms and incentives.} We study convex combinations of 
    self-retention (identity) and network-induced sharing. This ``$\lambda$-mix'' balances 
    autarky and diversification, delivers monotone variance reduction under doubly-stochastic 
    mixers, and aligns individual incentives with collective welfare. Closed-form variance 
    formulas and spectral characterizations yield practical design principles for decentralized 
    insurance.
\end{enumerate}

Together, these results connect classical stochastic-order methods with modern network-based 
insurance models, offering both mathematical soundness and economic interpretability. The 
framework applies broadly to peer-to-peer insurance, mutual aid, and decentralized financial 
risk-sharing systems.

The remainder of this article is structured as follows: Section~\ref{sec:properties} develops the general mathematical properties of LRS and identifies desirable conditions for well-behaved mechanisms. Section~\ref{sec:convex order} introduces convex ordering and its implications for stochastic and doubly stochastic operators. Section~\ref{sec:networks} studies specific network topologies and their diversification properties. Section~\ref{sec:random} extends the framework to random graphs, including Erdős–Rényi and preferential-attachment models. Section~\ref{sec:selfretention} analyzes convex combinations of identity and network-induced sharing, connecting diversification to incentives. Finally, we concludes with key findings and avenues for future research.

\section{Theoretical Framework for Linear Risk Sharing}\label{sec:properties}

This section develops the necessary definitions and properties that any well-behaved linear risk-sharing (LRS) mechanism should satisfy. We prove each property in detail and provide remarks on their implications. Throughout, let $\boldsymbol{X}=(X_1,\ldots,X_n)^\top\in\mathbb{R}^n_+$ be the vector of pre-sharing
losses (on some probabilistic space $(\Omega_X,\mathcal{F}_X,\mathbb{P}_X)$) and let $\mathrm{M}\in\mathbb{R}^{n\times n}_+$ be a nonnegative matrix that generates
post-sharing allocations $\boldsymbol{\xi}= \mathrm{M}\boldsymbol{X}$. Until Section~\ref{sec:prob:second:layer}, consider $\mathrm M$ as a fixed (deterministic) nonnegative matrix.

\subsection{Basic Definitions and Model Setup}

\begin{definition}[Exchangeability]
A random vector $\boldsymbol X=(X_1,\ldots,X_n)$ is  \emph{exchangeable} if 
\[
 (X_{\pi(1)},\ldots,X_{\pi(n)}) \ \stackrel{d}{=} \ (X_1,\ldots,X_n)
\]
for every permutation $\pi$ of $\{1,\ldots,n\}$, where $\stackrel{d}{=}$ denotes equality in distribution.
In words, the joint distribution of $\boldsymbol X$ is invariant under permutations of its components.
\end{definition}

A stronger assumption is actually independence (see \cite{chow2012probability} for more details and characterizations).

\begin{definition}[Risk-sharing scheme (RSS)]
A random vector $\boldsymbol{\xi}=(\xi_1,\ldots,\xi_n)^\top$ is a
\emph{risk-sharing scheme} (RSS) for $\boldsymbol{X}$ if
\[
\mathbf{1}^\top\boldsymbol{\xi}=\mathbf{1}^\top\boldsymbol{X}
\text{, or }
\sum_{i=1}^n \xi_i \;=\; \sum_{i=1}^n X_i ,~\text{almost surely.}
\]
\end{definition}

\begin{definition}[Linear risk-sharing (LRS)]
A RSS is \emph{linear} if there exists a nonnegative matrix $\mathrm{M}\in\mathbb{R}_+^{n\times n}$ (entrywise) with
\[
\boldsymbol{\xi} \;=\; \mathrm{M}\boldsymbol{X},~\text{almost surely.}
\]
\end{definition}

We will use the following terminology for matrices $\mathrm{M}$: 

\begin{definition}[Risk Sharing Matrices]
    Let $n\in\mathbb{N}$ and $\mathbf{1}=(1,\dots,1)^\top\in\mathbb{R}^n$,
\begin{align*}
\mathcal{M}_n^+ 
&:= \{ \mathrm{M}\in\mathbb{R}_+^{n\times n} : \text{i.e., }\mathrm{M}_{ij}\ge 0\ \ \forall i,j\} 
&& \text{(entrywise nonnegative matrices)}, \\[4pt]
\mathcal{RS}_n 
&:= \{ \mathrm{M}\in \mathcal{M}_n^+ : \mathrm{M} \mathbf{1} = \mathbf{1}\} 
&& \text{(row-stochastic: each row sums to $1$)}, \\[4pt]
\mathcal{CS}_n 
&:= \{ \mathrm{M}\in \mathcal{M}_n^+ : \mathrm{M}^\top \mathbf{1} = \mathbf{1}\} 
&& \text{(column-stochastic: each column sums to $1$)}, \\[4pt]
\mathcal{DS}_n 
&:= \mathcal{RS}_n \cap \mathcal{CS}_n 
&& \text{(doubly-stochastic: both row and column sums $=1$)}.
\end{align*}
\end{definition}

The use of majorization and doubly stochastic operators connects directly with classical results in inequalities \citep{hardy1934inequalities}.

\begin{definition}[Permutation matrices]
The set of $n\times n$ permutation matrices is
\[
\mathcal{P}_n := \{ \mathrm{P}\in\{0,1\}^{n\times n} : \mathrm{P}\mathbf{1}=\mathbf{1},\ \mathrm{P}^\top\mathbf{1}=\mathbf{1}\}.
\]
Each $\mathrm{P}\in \mathcal{P}_n$ has exactly one entry equal to $1$ in each row and column,
and $0$ elsewhere. Thus $\mathcal{P}_n\subset \mathcal{DS}_n$.
\end{definition}

\begin{remark}
$\mathcal{DS}_n$ is the convex hull of $\mathcal{P}_n$ (Birkhoff–von Neumann theorem).
$\mathcal{RS}_n$ and $\mathcal{CS}_n$ are convex polytopes as well, but larger: $\mathcal{RS}_n$ allows 
  arbitrary row-probability vectors, $\mathcal{CS}_n$ arbitrary column-probability vectors.
In our context, $\mathcal{M}_n^+$ encodes feasibility (nonnegative weights); 
  stochasticity restrictions enforce budget balance (CS) and fairness/convexity (RS).
\end{remark}


\subsubsection{Budget balance and conservation}

Roughly speaking, budget balance ensures that all losses can be distributed throughout the scheme. Formally, consider the following proposition:

\begin{proposition}[Budget balance $\Longleftrightarrow$ column-stochasticity]
\label{prop:budget-CS}
For a linear scheme $\boldsymbol{\xi}=\mathrm{M}\boldsymbol{X}$, the following are equivalent:
\begin{enumerate}
\item[(i)] $\boldsymbol{\xi}$ is a risk-sharing scheme, i.e., $\displaystyle\sum_i \xi_i = \sum_i X_i$ almost surely (for all realizations of $X$).
\item[(ii)] $\mathrm{M}$ is column-stochastic, i.e., $\mathrm{M}^\top\mathbf{1}=\mathbf{1}$.
\end{enumerate}
\end{proposition}

\begin{proof}
Since $\boldsymbol{\xi}= \mathrm{M}\boldsymbol{X}$, we have $\mathbf{1}^\top \boldsymbol{\xi}=\mathbf{1}^\top \mathrm{M}\boldsymbol{X} = (\mathrm{M}^\top \mathbf{1})^\top \boldsymbol{X}$.
Thus $\mathbf{1}^\top \boldsymbol{\xi}=\mathbf{1}^\top \boldsymbol{X}$ for all $\boldsymbol{X}$ if and only if $\mathrm{M}^\top \mathbf{1} = \mathbf{1}$.
\end{proof}

\begin{remark}
If $\mathrm{M}^\top\mathbf{1}\neq \mathbf{1}$, then $\displaystyle\sum_i \xi_i = (\mathrm{M}^\top\mathbf{1})^\top \boldsymbol{X}$ can systematically
\emph{create or destroy} total loss depending on $\boldsymbol{X}$; this violates budget balance and is
typically unacceptable in insurance.
\end{remark}

\subsubsection{Identifiability of the sharing operator.}

Suppose $\boldsymbol\xi^{(1)}=\mathrm{M}_1 \boldsymbol{X}$ and $\boldsymbol\xi^{(2)}=\mathrm{M}_2 \boldsymbol{X}$ are two linear risk-sharing (LRS) schemes 
built from the same underlying random vector $\boldsymbol{X}=(X_1,\dots,X_n)^\top$.

\begin{proposition}[Almost-sure equality]
If $\boldsymbol\xi^{(1)}=\boldsymbol\xi^{(2)}$ almost surely for all realizations of $\boldsymbol{X}$ (not just in law),
then $\mathrm{M}_1=\mathrm{M}_2$.
\end{proposition}

\begin{proof}
The condition means $(\mathrm{M}_1-\mathrm{M}_2)\boldsymbol{X}=\boldsymbol{0}$ a.s.\ for all integrable $\boldsymbol{X}$.  
Since this must hold for every possible $\boldsymbol{X}$, the only possibility is $\mathrm{M}_1-\mathrm{M}_2=\mathrm{0}$, i.e. $\mathrm{M}_1-\mathrm{M}_2$. It is the extension on the set of random vectors of the standard property in linear algebra, that the only linear map that vanishes on all vectors is the zero map, see Theorem 2.1.4. in \cite{horn2012matrix}.
\end{proof}

\begin{remark}[Equality in distribution]
If instead $\boldsymbol\xi^{(1)}\stackrel{d}{=}\boldsymbol\xi^{(2)}$, then $\mathrm{M}_1$ and $\mathrm{M}_2$ need not coincide.  
They can differ while inducing the same law for $\boldsymbol\xi$. For example, if $\boldsymbol{X}$ has a spherically symmetric distribution 
(e.g.\ i.i.d.\ Gaussian components), any orthogonal transformation of $\boldsymbol{X}$ yields the same law (for spherical laws such as Gaussian, see \cite{fang1990symmetric}).  
Thus distinct $\mathrm{M}_1,\mathrm{M}_2$ may produce identical distributions of outcomes.
\end{remark}

\begin{remark}[Exchangeable $\boldsymbol{X}$]
If $X$ is exchangeable (all finite-dimensional distributions invariant under permutations of coordinates),
then $\xi^{(1)}\stackrel{d}{=}\xi^{(2)}$ can occur whenever $\mathrm{M}_1$ and $\mathrm{M}_2$ are related by permutation of rows.  
Indeed, if $\mathrm P$ is a permutation matrix, then $\mathrm{M}_2=\mathrm{P}\mathrm{M}_1$ yields $\xi^{(2)}=\mathrm{M}_2\boldsymbol X=\mathrm{P}\mathrm{M}_1\boldsymbol X\stackrel{d}{=}\mathrm{M}_1\boldsymbol X=\boldsymbol\xi^{(1)}$
because exchangeability makes $\mathrm{P}\boldsymbol X\stackrel{d}{=}\boldsymbol X$. 
More generally, equality in distribution of $\mathrm{M}_1\boldsymbol X$ and $\mathrm{M}_2\boldsymbol X$ characterizes the invariance group of the law of $\boldsymbol X$:
two operators $\mathrm{M}_1,\mathrm{M}_2$ are indistinguishable in distribution if they differ only by a transformation under which $\boldsymbol X$ is invariant.
\end{remark}


\subsubsection{Convex-combination structure and Jensen}

\begin{proposition}[Row-stochasticity $\Rightarrow$ per-agent convex combinations]
\label{prop:RS-convex-combo}
If $\mathrm{M}$ is row-stochastic (and nonnegative), then for each $i$ and every realization of $\boldsymbol X$,
\[
\xi_i \;=\; \sum_{j=1}^n \mathrm{M}_{ij} X_j
~ \text{with}~  \sum_{j=1}^n \mathrm{M}_{ij}=1, \; \mathrm{M}_{ij}\ge0,
\]
i.e., $\xi_i$ is a convex combination of $\boldsymbol X=(X_1,\ldots,X_n)$.
Consequently, for every convex $f:\mathbb{R}\to\mathbb{R}$,
\[
f(\xi_i)\;=\; f\Big(\sum_{j=1}^n \mathrm{M}_{ij} X_j\Big) \;\le\; \sum_{j=1}^n \mathrm{M}_{ij} f(X_j)
~ \text{(Jensen).}
\]
\end{proposition}

\begin{proof}
Immediate from the RS property and nonnegativity of the row entries; the Jensen inequality
is applied pathwise for fixed $\boldsymbol X$.
\end{proof}

\begin{remark}
If a row does not sum to $1$, then $\xi_i$ is an \emph{affine} (not convex) combination of $X$;
Jensen’s inequality does not apply in the usual way. If the row sum exceeds $1$, the mechanism
amplifies the realization $X$ for agent $i$; if it is below $1$, it attenuates it. In either case,
the convex order and variance-reduction guarantees below can fail.
\end{remark}

\subsubsection{Mean-preserving properties}

\begin{proposition}[Individual mean preservation under identical means]
\label{prop:mean-preservation}
Suppose $\mathbb{E}[X_i]=\mu$ for all $i$ (not necessarily independent).
If $\mathrm{M}$ is row-stochastic, then $\mathbb{E}[\xi_i]=\mu$ for every $i$.
\end{proposition}

\begin{proof}
$\mathbb{E}[\xi_i]=\mathbb{E}[\sum_j \mathrm{M}_{ij} X_j]=\sum_j \mathrm{M}_{ij} \mathbb{E}[X_j]
= \sum_j \mathrm{M}_{ij}\mu = \mu$ because the $i$-th row sums to $1$.
\end{proof}

\begin{proposition}[Total-mean preservation]
\label{prop:total-mean}
For any integrable $X$, $\mathbb{E}\big[\displaystyle\sum_i \xi_i\big]=\mathbb{E}\big[\sum_i X_i\big]$ for all $X$
if and only if $\mathrm{M}$ is column-stochastic.
\end{proposition}

\begin{proof}
$\mathbb{E}[\displaystyle\sum_i \xi_i] = \mathbb{E}[\mathbf{1}^\top \mathrm{M}\boldsymbol{X}]= \mathbb{E}[ (\mathrm{M}^\top\mathbf{1})^\top \boldsymbol{X}]$.
Equality to $\mathbb{E}[\mathbf{1}^\top \boldsymbol{X}]$ for all integrable $X$ is equivalent to $M^\top\mathbf{1}=\mathbf{1}$.
\end{proof}

\subsubsection{Variance and covariance structure}

\begin{proposition}[Variance-covariance transformation]
\label{prop:var-structure}
If $X$ has covariance matrix $\Sigma=\mathrm{Var}[\boldsymbol{X}]$ (finite), then
\[
\mathrm{Var}[\boldsymbol{\xi}] \;=\; \mathrm{M}\,\Sigma\,\mathrm{M}^\top.
\]
In particular, $\mathrm{tr}\big(\mathrm{Var}[\boldsymbol{\xi}]\big)=\mathrm{tr}\big(\mathrm{M} \Sigma \mathrm{M}^\top\big)$.
\end{proposition}

\begin{proof}
Since $\boldsymbol{\xi}=\mathrm{M}\boldsymbol{X}$, $\mathrm{Cov}(\boldsymbol{\xi})=\mathrm{M}\,\mathrm{Cov}(\boldsymbol{X})\,\mathrm{M}^\top$ by linearity of covariance.
\end{proof}

\begin{proposition}[Trace contraction under DS]\label{prop:trace-DS}
Let $\Sigma\succeq 0$ and let $\mathrm D$ be a doubly-stochastic matrix. Then
\[
\operatorname{tr}(\mathrm D\,\Sigma\,\mathrm D^\top)\ \le\ \operatorname{tr}(\Sigma).
\]
\end{proposition}

\begin{proof}
See Appendix~\ref{app:prop:trace-DS}.
\end{proof}

\begin{remark}
If only CS holds, budget balance is achieved, but $h(\mathrm{M})=\mathrm{tr}(\mathrm{M}\Sigma \mathrm{M}^\top)$ need not be
$\le \mathrm{tr}(\Sigma)$; similarly with only RS. The convex order and per-agent variance
reductions (next section) typically \emph{require} RS (for Jensen inequality) and \emph{use} i.i.d./exchangeability
for comparisons to the no-sharing baseline. DS combines CS (budget balance) with RS (convex-combo rows),
hence it is the natural ``safe'' class for strong dominance results.
\end{remark}

\subsection{Network Representation of Risk-Sharing Mechanisms}

To represent arbitrary networks, we rely on the definitions of a graph $G$ and its representation via the adjacency matrix $A$, formally defined below. The matrix representation of a network allows us to formalize diverse mechanisms of risk sharing. In its simplest form, the adjacency matrix encodes reciprocal commitments: two policyholders connected by an edge agree ex ante to share part of their risks. Such \emph{reciprocal contracts} are the network analogue of mutual aid agreements: if $i$ and $j$ are connected, then $i$ commits 
to covering a fixed share of $j$'s loss whenever it materializes, and symmetrically $j$ commits to cover the same share of $i$'s loss. 

This formalization captures peer-to-peer insurance mechanisms that have emerged in practice (e.g., Friendsurance, Lemonade), and also connects to the actuarial literature on P2P ({\em peer--to--peer}) insurance 
\citep{DenuitRobert2020,AbdikerimovaFeng2022,FengLiuTaylor2022}. The novelty here is that 
allocations are shaped not by an anonymous pool but by the local degree structure: the more 
neighbours an agent has, the more finely her risk can be spread, but also the more reciprocal 
obligations she must assume.

Reciprocal contracts have two important features:
\begin{itemize}
  \item They are \emph{symmetric by design}, unlike traditional insurance contracts. This symmetry 
  naturally enforces trust and mitigates moral hazard through \emph{social collateral}: reneging on a 
  commitment damages not only the financial position of the counterpart but also the social link 
  itself \citep{Granovetter2005,Karlan2009,Liu2020}.
  \item They are \emph{local rather than global}, emphazising the peer-to-peer nature: agents only exchange risk with their neighbours. 
  This locality explains why the variance reduction is strongest in regular networks but can fail in 
  highly heterogeneous, scale-free topologies \citep{BarabasiAlbert1999}.
\end{itemize}

Although this seems simplistic at first, the framework can be extended in several ways. For example, one can allow for heterogeneous contracts on each edge, optimized by a planner subject to budget and fairness constraints (which can conveniently be expressed as a linear program). Another extension is to include \emph{friends-of-friends} contributions: if $i$ and $j$ share a common neighbour $k$, 
they may enter a second-order reciprocal contract with lower weight. This generalization expands the effective risk-sharing network, while recognizing that trust decays with distance. 

In sum, risk sharing on networks formalizes how P2P insurance reallocates risk not globally but 
through a web of reciprocal obligations. This approach links actuarial models to sociological ideas 
of trust, reciprocity, and social collateral, and helps to explain both the promise and the fragility 
of network-based insurance.

 
\begin{definition}[Network-induced sharing]
Let $G=(V,E)$ be a graph on $\{1,\ldots,n\}$ with adjacency matrix $A$.
We say $\mathrm{M}$ is \emph{compatible} with $G$ if $\mathrm{M}_{ij}=0$ whenever $\{i,j\}\notin E$ (i.e., $\mathrm{A}_{ij}=0$). Sharing is symmetrical if $\mathrm M$ is symmetric. 
\end{definition}

\begin{proposition}[Budget/convexity on networks]
\label{prop:network-stochastic}
If $\mathrm{M}$ is compatible with $G$ and is column-stochastic, the LRS is budget balanced.
If, in addition, $\mathrm{M}$ is row-stochastic, then each $\xi_i$ is a convex combination
of $\{X_j:\{i,j\}\in E\}$, i.e., only neighbors of $i$ (including $i$ if allowed).
\end{proposition}

\begin{proof}
Budget balance follows from Proposition~\ref{prop:budget-CS}.
Row-stochasticity plus nonnegativity imply each $i$-th row is a probability vector
supported on $\{j:\{i,j\}\in E\}$, yielding a convex combination over neighbors.
\end{proof}


\begin{remark}
For undirected $G$, one may impose $\mathrm{M}_{ij}=\mathrm{M}_{ji}$ to reflect reciprocal 
agreements; this symmetry is \emph{not} required by our general results. In particular, 
intergenerational risk sharing schemes (such as pay-as-you-go pension systems) are 
typically non-reciprocal: young cohorts contribute to finance the benefits of retirees, 
without receiving symmetric transfers in return \citep{Diamond2004}. Our framework 
allows for such directed structures through non-symmetric matrices $\mathrm M$.
\end{remark}


\subsection{Properties of Sharing Matrices: Fairness, Budget Balance, and Diversification}

Optimal design of insurance contracts under convex order has already been studied in the context of general utility‐based allocation problems \citep{Carlier2002}. To show its utility in our context, we summarize the roles of the three stochasticity notions:

\paragraph{Column-stochastic (CS):} Ensures \emph{budget balance} and \emph{total-mean preservation}
(Propositions~\ref{prop:budget-CS} and \ref{prop:total-mean}). Without CS, the mechanism can
create/destroy aggregate loss.

\paragraph{Row-stochastic (RS):} Ensures each $\xi_i$ is a \emph{convex combination} of the inputs
(Proposition~\ref{prop:RS-convex-combo}), enabling pathwise Jensen and, under symmetry assumptions
(i.i.d.\ or exchangeability), convex order comparisons and individual mean preservation
(Proposition~\ref{prop:mean-preservation}). Without RS, per-agent convexity and the associated
risk-order guarantees fail.

\paragraph{Doubly-stochastic (DS):} Combines budget balance and convex-combo rows, enabling
\emph{strong} orderings (Section~\ref{sec:convex order}) such as componentwise convex order
improvement versus no-sharing and trace-of-variance reduction (Proposition~\ref{prop:trace-DS}).
If DS is weakened to CS only, one can still obtain \emph{representative-agent} variance reductions
under independence (via trace comparisons with additional arguments), but not universal
per-agent improvements.

\begin{lemma}[Row-stochasticity and mean preservation under dependence]
\label{lem:RS-mean}
Assume $\mathbb{E}[X_i]=\mu$ for all $i$ (no independence required).
If $\mathrm{M}$ is row-stochastic, then for each agent $i$,
\[
\mathbb{E}[\xi_i] \;=\; \mathbb{E}\!\Big[\sum_{j=1}^n \mathrm{M}_{ij} X_j\Big]
\;=\; \sum_{j=1}^n \mathrm{M}_{ij}\,\mathbb{E}[X_j]
\;=\; \mu \sum_{j=1}^n \mathrm{M}_{ij}
\;=\; \mu.
\]
Hence each agent’s expected post-sharing allocation equals its pre-sharing mean.
\end{lemma}

\begin{proof}
The first equality is based on the linearity of the expectation operator. The second holds because each $\mathbb{E}[X_j]=\mu$. The third uses row-stochasticity ($\displaystyle\sum_j \mathrm{M}_{ij}=1$). No independence assumptions are needed.
\end{proof}
 
\begin{example}\label{IJ:ex}
Let $\mathbb{I}$ denote the $n\times n$ identity matrix, $\mathbb{I}:=\mathrm{diag}(\mathbf{1})$, and $\mathrm J:=\mathbf{1}\mathbf{1}^\top\in\mathbb{R}^{n\times n}$ denote the
all-ones matrix, where $\mathbf{1}=(1,\dots,1)^\top\in\mathbb{R}^n$.
The matrix $\mathrm J/n$ is the \emph{averaging operator}: for any $\boldsymbol x\in\mathbb{R}^n$,
\[
\frac{\mathrm J}{n}\,x=\Big(\frac{1}{n}\sum_{i=1}^n x_i\Big)\mathbf{1}=\overline{\boldsymbol{x}}\mathbf{1},
\]
so it projects $\boldsymbol{x}$ onto the span of $\mathbf{1}$. It is symmetric, rank-one,
idempotent $\displaystyle\big(\frac{\mathrm J}{n}\big)^2=\frac{\mathrm J}{n}$, and \emph{doubly stochastic}
(all rows and columns sum to $1$). 
\end{example}

\begin{example}[Failure without row-stochasticity]
Let $n=2$, $X_1,X_2$ be i.i.d.\ with mean $\mu$.
Consider
\[
M = \begin{pmatrix}
2 & 0 \\
0 & 0
\end{pmatrix},
~  \boldsymbol{\xi} = \mathrm{M}\boldsymbol{X} = (2X_1,0)^\top.
\]
Here $\mathrm{M}$ is column-stochastic (each column sums to $1$) but not row-stochastic.
Then $\mathbb{E}[\xi_1]=2\mu \neq \mu$ and $\mathbb{E}[\xi_2]=0 \neq \mu$.
Thus, even though total budget balance holds ($\xi_1+\xi_2=X_1+X_2$ in expectation),
individual means are distorted and one agent systematically bears more than its fair share.
\end{example}

\begin{remark}
This illustrates why \emph{row-stochasticity} is critical: it enforces per-agent
mean preservation (Lemma~\ref{lem:RS-mean}) and ensures that each allocation is a convex combination of the original losses, which in turn is essential for convex order comparisons and fairness guarantees.
\end{remark}

\subsection{Birkhoff--von Neumann decomposition and bipartite expansion}\label{sec:sec:BvN}

The celebrated Birkhoff--von Neumann theorem states that the set of $n\times n$ 
doubly stochastic matrices is exactly the convex hull of permutation matrices
\citep{Birkhoff1946,horn2012matrix}:
\[
\mathcal{D}\mathcal{S}_n \ = \ \Big\{ \mathrm D \in \mathbb{R}_+^{n\times n}: \mathrm D\mathbf{1}=\mathbf{1}, 
\mathrm D^\top \mathbf{1}=\mathbf{1}\Big\}
\ = \ \mathrm{conv}\{\mathcal{P}_n\}.
\]
Hence any $D\in\mathcal{D}\mathcal{S}_n$ admits a representation
\[
\mathrm D \ = \ \sum_{r=1}^R \omega_r \mathrm P_r,
\text{ with } \omega_r \ge 0,\ \sum_r \omega_r=1,
\]
where $\mathrm P_r$'s are a collection of permutation matrices, for some finite $R$. This decomposition has a useful graph-theoretic 
interpretation via the \emph{bipartite expansion} of a network $G$.

Given a network adjacency matrix $\mathrm A$, form the bipartite graph $B(G)$ with left 
vertices corresponding to rows and right vertices to columns. Place an edge $(i,j)$ 
with weight $\mathrm A_{ij}$ whenever $i$ is connected to $j$ in $G$. If $\mathrm D$ is a DS scaling 
of $\mathrm A$, then each row-node and column-node has weighted degree $1$, and a permutation 
matrix corresponds to a perfect matching in $B(G)$ 
\citep{Schrijver2003,Brualdi2006,LovaszPlummer1986}. Thus the Birkhoff--von Neumann 
theorem shows that $\mathrm D$ can be interpreted as a convex combination of perfect matchings 
of the bipartite expansion $B(G)$.

In the context of risk sharing, this means that a DS operator reallocates risk as if 
by randomizing over one-to-one matchings of agents, with probabilities given by 
$(\omega_r)$. This combinatorial perspective explains why convex order properties and 
variance reductions hold for DS matrices: they follow by convexity from the extreme 
cases given by permutations.

\subsection{Matrix scaling and fairness adjustment}\label{sec:sec:scaling:sin}

In practice, network-based allocation matrices may fail to be doubly stochastic:
some matrices are only row-stochastic (RS) or column-stochastic (CS). A natural
question is whether such a matrix can be ``adjusted'' into a doubly stochastic (DS)
form without changing its sparsity pattern. The classical result of 
\cite{SinkhornKnopp1967} provides a positive answer under mild conditions.

\begin{theorem}[Sinkhorn--Knopp scaling]\label{thm:sink}
Let $\mathrm A$ be a square matrix with nonnegative entries. If $\mathrm A$ has \emph{total support},
then there exist positive diagonal matrices $\mathrm D_1,\mathrm D_2$ such that
\[
\mathrm B = \mathrm D_1 \mathrm A \mathrm D_2
\]
is doubly stochastic. Moreover, if $\mathrm A$ is positive, the resulting $\mathrm B$ is unique.
\end{theorem}

\begin{proof}
    See Appendix~\ref{app:sin} for a technical discussion of the implications of that result. See also \cite{BapatRaghavan1997,Knight2008,FranklinLorenz1989,Idel2016}
for more details.
\end{proof}

The scaling $(\mathrm D_1,\mathrm D_2)$ can be obtained by iteratively normalizing rows and columns,
a process that converges to a DS matrix. In the context of risk sharing, this theorem
shows that any feasible network operator $\mathrm M$ with sufficient connectivity can be
rescaled into a ``fair'' DS mechanism, ensuring budget balance and symmetry of
allocation while preserving the structure of connections.

\medskip
\noindent
In Section~\ref{sec:convex order}, we exploit these properties—together with i.i.d.\ or exchangeability
assumptions—to derive convex order dominance results and variance inequalities for both individuals
and the portfolio.

\section{Risk Comparison via Convex Order}
\label{sec:convex order}

Convex ordering provides a natural framework to compare random allocations
induced by different linear risk sharing (LRS) mechanisms. It allows us to formalize
when one scheme is ``less risky'' than another, either from an individual
perspective or from a representative-agent perspective \citep{Ohlin1969,ShakedShanthikumar2007,denuit2006actuarial}.

Throughout this section we work under the following standing assumption unless
explicitly stated otherwise.


\subsection{Assumptions at a glance}

For clarity, we recall the standing assumptions needed for the results in this section:
\begin{itemize}
    \item[(A1)] Losses $X_1,\dots,X_n$ are i.i.d., integrable random variables (used for convex order comparisons).
    \item[(A2)] Variances are finite, $\mathrm{Var}(X_i)<\infty$.
    \item[(A3)] Exchangeability of $\boldsymbol\xi$ 
\end{itemize}

As we will see, (A1) is essential for the strong convex order results: without i.i.d. (or at least exchangeability of $X$), one will lose the symmetry needed to compare each row’s convex combination to a single $X_i$. (A2) is a technical necessity if we want variance-based statements (trace inequalities, monotonicity of variance with $\lambda$). For convex order definitions alone, integrability is enough. (A3) is only invoked in Proposition \ref{prop:postmix}. 

\medskip

\begin{corollary}[Variance reductions under LRS]\label{cor:variance-summary}
Under Assumptions~(A1)--(A2), let $\boldsymbol{\xi}=\mathrm{M}\boldsymbol{X}$ be an LRS allocation.
\begin{enumerate}
    \item \textbf{Individual variance reduction (doubly stochastic case).}  
    If $\mathrm{M}$ is doubly stochastic, then
    \[
    \mathrm{Var}(\xi_i) \;\le\; \mathrm{Var}(X_i), ~  \forall i.
    \]
    \item \textbf{Representative-agent variance reduction (column-stochastic case).}  
    If $\mathrm{M}$ is column-stochastic and the $X_i$ are independent with common variance $\sigma^2$, then
    \[
    \mathrm{Var}({\xi}') \;=\; \frac{1}{n}\,\mathrm{tr}\!\big(\mathrm{Var}[\boldsymbol{\xi}]\big) \;\le\; \sigma^2,
    \]
    where ${\xi}'={\xi}_I$ for $I\sim\mathrm{Unif}\{1,\dots,n\}$.
    \item \textbf{Trace-of-variance reduction (doubly-stochastic post-mixing).}  
    If $\boldsymbol\xi_2=\mathrm D\boldsymbol\xi_1$ with a doubly-stochastic $\mathrm D$, then
    \[
    \mathrm{tr}\!\big(\mathrm{Var}[\boldsymbol\xi_2]\big) \;\le\; \mathrm{tr}\!\big(\mathrm{Var}[\boldsymbol\xi_1]\big).
    \]
\end{enumerate}
\end{corollary}

\begin{proof}
Part (1) is Proposition~\ref{prop:var-ind};  
Part (2) is Proposition~\ref{prop:rep-agent-variance};  
Part (3) is Proposition~\ref{prop:trace}.
\end{proof}

\subsection{Properties of Convex and Componentwise Convex Orders}

\subsubsection{Convex Order and Componentwise Convex Order}

The convex order is the canonical partial order for comparing insurance risks.
It reflects the idea that, under risk aversion, one allocation is preferred to
another if it yields a smaller expectation for every convex utility function.
Formally, if $X \leq_{{CX}} Y$, then $\mathbb{E}[u(X)] \leq \mathbb{E}[u(Y)]$
for every convex $u$, meaning that all risk-averse agents (with concave utility)
weakly prefer $X$ to $Y$. This connection explains why convex order is widely
regarded as the \emph{natural order for risk comparisons in insurance and
actuarial science}; see \cite{Ohlin1969,DenuitRobert2020,ShakedShanthikumar2007} for
classic treatments.
Following \cite{denuit2006actuarial}, define several concepts related to the convex order. 

\begin{definition}[Convex order]
Given two integrable random variables $U$ and $V$, we say that
$U$ is smaller than $V$ in the \emph{convex order}, written $U \preceq_{CX} V$,
if
\[
\mathbb{E}[f(U)] \le \mathbb{E}[f(V)]
~ \text{for all convex functions } f:\mathbb{R}\to\mathbb{R}
\text{ for which the expectations exist.}
\]
\end{definition}

The componentwise convex order was introduced in the study of multivariate 
risk comparisons by \cite{Tchen1980} and is systematically treated in 
\cite{ShakedShanthikumar2007}. See also \cite{DENUIT2012265} for discussions and motivations with some actuarial perspectives,

\begin{definition}[Componentwise convex order]
For two $\mathbb{R}^n$-valued random vectors $\boldsymbol{\xi}=(\xi_1,\dots,\xi_n)$
and $\boldsymbol\zeta=(\zeta_1,\dots,\zeta_n)$, we write
$\boldsymbol{\xi}\preceq_{CCX} \boldsymbol\zeta$ if $\xi_i \preceq_{CX} \zeta_i$ for all $i=1,\dots,n$.
\end{definition}

\begin{proposition}[Basic properties of $\preceq_{CX}$]\label{prop:cx-basic}
If $U \preceq_{CX} V$, then $\mathbb{E}[U]=\mathbb{E}[V]$ and
$\mathrm{Var}(U)\le \mathrm{Var}(V)$ (whenever the variances are finite).
\end{proposition}

\begin{proof}
Take $f(x)=x$ and $f(x)=-x$, which are convex; the two inequalities imply
$\mathbb{E}[U]=\mathbb{E}[V]$. For the variance, use $f(x)=x^2$ (convex) to get
$\mathbb{E}[U^2]\le \mathbb{E}[V^2]$ and combine with equality of means.
\end{proof}

\subsubsection{Majorization and stochastic matrices}

\begin{definition}[Majorization]
For $\boldsymbol{a},\boldsymbol{b}\in\mathbb{R}^n$, write $a_{[1]}\ge \cdots \ge a_{[n]}$ for the decreasing rearrangement.
We say that $\boldsymbol{a}$ is majorized by $\boldsymbol{b}$ (and write $\boldsymbol a\preceq \boldsymbol b$) if
\[
\sum_{i=1}^k a_{[i]}\le \sum_{i=1}^k b_{[i]}~  (k=1,\dots,n-1),
~\text{and}~ \sum_{i=1}^n a_i=\sum_{i=1}^n b_i.
\]
\end{definition}

\begin{theorem}[Hardy--Littlewood--P\'olya]
$\boldsymbol{a}\preceq \boldsymbol{b}$ if and only if there exists a doubly-stochastic matrix $\mathrm D$ such that $\boldsymbol a=\mathrm D\boldsymbol b$.
\end{theorem}
\begin{proof}
Chapter II, Theorem 1.B.2, in~\cite{hardy1934inequalities}.
\end{proof}

\subsection{Case of column-stochastic matrices}

The connection between linear risk sharing and matrix majorization has been emphasized in earlier works on stochastic matrices \citep{Dahl1999}.


\begin{definition}[Weak ordering of LRS via column-stochastic post-composition]
Let $\boldsymbol\xi_1,\boldsymbol\xi_2$ be two LRS allocations (of the same risk vector $\boldsymbol X$). We write $\boldsymbol\xi_2 \preceq_{wCX} \boldsymbol\xi_1$ if there exists
a column-stochastic matrix $\mathrm{C}$ (i.e., $\mathrm{C}^\top\mathbf{1}=\mathbf{1}$, $\mathrm{C}\ge 0$) such that
$\boldsymbol\xi_2=\mathrm{C}\boldsymbol\xi_1$.
\end{definition}

This ordering preserves global budget balance but does not necessarily give individual
convex order improvement relative to \emph{each} component of $\boldsymbol\xi_1$ without further symmetry.
Nevertheless, variance comparison is available at the ``representative agent'' level.

\begin{proposition}[Variance of a randomly selected agent]\label{prop:rep-agent-variance}
Let $I$ be independent of everything else and uniform on $\{1,\dots,n\}$.
For any $\mathbb{R}^n$-valued random vector $\boldsymbol\eta$, define $\eta'=\eta_I$.
Then
\[
\mathrm{Var}[\eta'] \;=\; \frac{1}{n}\,\mathrm{tr}\!\big(\mathrm{Var}[\eta]\big).
\]
In particular, if $\boldsymbol\xi_2=\mathrm C\boldsymbol\xi_1$ with a column-stochastic $\mathrm C$ and the $X_i$ are independent with
$\mathrm{Var}(X_i)=\sigma^2<\infty$, then
\[
\mathrm{Var}[\xi_2'] \;\le\; \mathrm{Var}[\xi_1'].
\]
\end{proposition}

\begin{proof}
See Appendix~\ref{app:prop:rep-agent-variance}.
\end{proof}

\subsection{Strong Guarantees under Doubly Stochastic Operators}\label{sec:sec:dsm}

\subsubsection{Strong ordering results}

We begin with a key lemma that will be the workhorse for convex order statements.

\begin{lemma}\label{lem:prob-weight-cx}
Let $\boldsymbol{p}=(p_1,\dots,p_n)$ be a probability vector ($p_j\ge 0$, $\displaystyle\sum_j p_j=1$). Under Assumption~(A1),
\[
\boldsymbol{p}^\top \boldsymbol{X} \;\preceq_{CX}\; X_1.
\]
\end{lemma}

\begin{proof}
Let $J$ be an index, independent of $X$, with $\mathbb{P}(J=j)=p_j$.
Then for any convex $f$ and for each realization of $X$,
\[
f\big(p^\top X\big) \;=\; f\!\Big(\,\mathbb{E}[\,X_J\mid X\,]\,\Big)
\;\le\; \mathbb{E}\big[\,f(X_J)\mid X\,\big],
\]
by Jensen's inequality applied conditionally on $X$. Taking expectations gives
$\mathbb{E}[f(\boldsymbol{p}^\top \boldsymbol{X})] \le \mathbb{E}[f(X_J)]$.
By i.i.d. of the $X_i$, $\mathbb{E}[f(X_J)]=\sum_j p_j \mathbb{E}[f(X_j)]=\mathbb{E}[f(X_1)]$,
hence $p^\top X \preceq_{CX} X_1$.
\end{proof}

\begin{proposition}[Desirability under doubly-stochastic mixing]\label{prop:ds-desirable}
Let $\boldsymbol{\xi}=\mathrm{M}\boldsymbol{X}$ with $\mathrm{M}$ doubly-stochastic (nonnegative, $M\mathbf{1}=\mathbf{1}$ and $M^\top\mathbf{1}=\mathbf{1}$).
Under Assumption~(A1),
\[
\boldsymbol\xi \;\preceq_{CCX}\; \boldsymbol X.
\]
\end{proposition}

\begin{proof}
Write the $i$-th row of $\mathrm{M}$ as a probability vector $p^{(i)}$ (nonnegative and sums to~1).
Then $\xi_i = p^{(i)\top} \boldsymbol X$.
By Lemma~\ref{lem:prob-weight-cx}, $\xi_i \preceq_{CX} X_1$, and by i.i.d.\ of $\boldsymbol X$,
$X_1 \stackrel{d}{=} X_i$ for each $i$. Hence $\xi_i \preceq_{CX} X_i$ for all $i$, i.e.,
$\boldsymbol{\xi}\preceq_{CCX} \boldsymbol X$.
\end{proof}

\begin{proposition}[Monotonicity under post-mixing with exchangeability]\label{prop:postmix}
Suppose $\boldsymbol\xi_1=\mathrm{M}_1\boldsymbol X$ with Assumption~(A1), and assume the components of $\boldsymbol\xi_1$ are exchangeable
(Assumption~(A3), e.g., each row of $\mathrm{M}_1$ is a permutation of a fixed probability vector).
If $\boldsymbol\xi_2 =\mathrm  D\boldsymbol\xi_1$ with $\mathrm D$ doubly-stochastic, then
\[
\boldsymbol\xi_2 \;\preceq_{CCX}\; \boldsymbol\xi_1.
\]
\end{proposition}

\begin{proof}
Fix $i$. Let $\boldsymbol d^{(i)}$ be the $i$-th row of $\mathrm D$; it is a probability vector.
Then $\xi_{2,i} = \displaystyle\sum_{k=1}^n d^{(i)}_k \,\xi_{1,k}$.
Let $K$ be independent of $\boldsymbol\xi_1$ with $\mathbb{P}(K=k)=d^{(i)}_k$.
For convex $f$,
\[
f(\xi_{2,i}) \;=\; f\!\Big(\,\mathbb{E}[\xi_{1,K}\mid \xi_1]\,\Big) \;\le\; \mathbb{E}[\,f(\xi_{1,K})\mid \xi_1],
\]
hence $\mathbb{E}[f(\xi_{2,i})] \le \mathbb{E}[f(\xi_{1,K})]$.
By exchangeability of $\boldsymbol \xi_1$, $\xi_{1,K}\stackrel{d}{=}\xi_{1,i}$, so
$\mathbb{E}[f(\xi_{1,K})]=\mathbb{E}[f(\xi_{1,i})]$. Therefore $\xi_{2,i}\preceq_{CX}\xi_{1,i}$.
\end{proof}

\begin{remark}
Without exchangeability of $(\xi_{1,1},\dots,\xi_{1,n})$, post-mixing by $\mathrm D$
gives $\boldsymbol\xi_{2,i} \preceq_{CX} \boldsymbol\xi_{1,K}$ (mixture across $k$) but not necessarily
$\xi_{2,i} \preceq_{CX} \xi_{1,i}$ for each fixed $i$.
\end{remark}

\subsubsection{Variance reduction for individuals}

\begin{proposition}\label{prop:var-ind}
Under Assumptions~(A1) and (A2), if $\boldsymbol{\xi}=\mathrm{M}\boldsymbol{X}$ with $\mathrm{M}$ doubly-stochastic, then for each $i$,
\[
\mathbb{E}[\xi_i] = \mathbb{E}[X_i]~\text{and}~ \mathrm{Var}(\xi_i)\le \mathrm{Var}(X_i).
\]
\end{proposition}

\begin{proof}
Mean preservation follows from $\mathrm M^\top\mathbf{1}=\mathbf{1}$ (column sums equal 1).
For the variance, combine Proposition~\ref{prop:ds-desirable} with
Proposition~\ref{prop:cx-basic}: $\xi_i \preceq_{CX} X_i$ implies equal means and
variance inequality.
\end{proof}

\subsubsection{Trace-of-variance inequalities}

\begin{proposition}[Covariance trace reduction under doubly-stochastic post-mixing]\label{prop:trace}
Let $\boldsymbol\xi_1$ be any $\mathbb{R}^n$-valued random vector with $\mathrm{Var}[\boldsymbol\xi_1]$ finite and
positive semidefinite, and let $\boldsymbol\xi_2 = \mathrm D \boldsymbol\xi_1$ for a doubly-stochastic matrix $\mathrm D$.
Then
\[
\mathrm{tr}\big(\mathrm{Var}[\boldsymbol\xi_2]\big) \;\le\; \mathrm{tr}\big(\mathrm{Var}[\boldsymbol\xi_1]\big).
\]
\end{proposition}

\begin{proof}
See Appendix~\ref{app:prop:trace}.
\end{proof}

\subsection{Consequences for weights and majorization}

The previous results can be reframed as Schur-convexity statements in the
weight vectors. Under Assumption~(A1), if $\boldsymbol a,\boldsymbol b\in\mathbb{R}^n_+$ are probability
vectors with $\boldsymbol a\preceq \boldsymbol b$ (i.e., $\boldsymbol a=\mathrm D\boldsymbol b$ for some doubly-stochastic $\mathrm D$), then
$\boldsymbol a^\top \boldsymbol X \preceq_{CX} \boldsymbol b^\top \boldsymbol X$ (apply Lemma~\ref{lem:prob-weight-cx} iteratively
via Hardy--Littlewood--P\'olya).
This provides a convenient way to compare different rows of sharing matrices
through the majorization order on their weights.

\section{Network Topologies and Diversification Outcomes}\label{sec:networks}

We now specialize the general linear risk-sharing (LRS) framework to particular
network topologies. These examples illustrate how the algebraic properties of
the sharing matrix $\mathrm{M}$ translate into economic and risk-theoretic interpretations. Throughout, let $\boldsymbol{X}=(X_1,\ldots,X_n)^\top$ be i.i.d.\ integrable random losses with
mean $\mu$ and variance $\sigma^2$, and $\boldsymbol{\xi}=\mathrm{M}\boldsymbol{X}$ be the allocation.



\subsection{Complete Network (Equal Sharing)}

The simplest form of networks we study is the complete network. Here, every agent is connected to every other agent, akin to risk sharing within a pool. This allows us to consider classical risk sharing mechanisms as a special case of risk sharing on networks. 

\begin{definition}[Complete pooling]
In a complete network, each agent shares equally with every other.
The sharing matrix is
\[
\mathrm M = \mathrm J = \frac{1}{n}\mathbf{1}\mathbf{1}^\top ,
\]
(from Example~\ref{IJ:ex}) so that $\xi_i = \displaystyle\frac{1}{n}\sum_{j=1}^n X_j$ for all $i$.
\end{definition}

\begin{proposition}[Properties of complete pooling]
\label{prop:complete}
Under complete pooling:
\begin{enumerate}
    \item $\mathrm{M}$ is doubly-stochastic, symmetric, and idempotent.
    \item $\mathbb{E}[\xi_i]=\mu$ for all $i$.
    \item $\mathrm{Var}(\xi_i)=\displaystyle\frac{1}{n}\sigma^2$ for all $i$.
\end{enumerate}
\end{proposition}

\begin{proof}
(1) Each row and column sums to $1$; symmetry and idempotence are immediate from the rank-one structure.  
(2) Mean preservation holds by row-stochasticity.  
(3) Since $X_j$ are i.i.d., $$\mathrm{Var}(\displaystyle\frac{1}{n}\sum_j X_j)=\frac{1}{n^2}\sum_j \mathrm{Var}(X_j)=\frac{1}{n}\sigma^2.$$
\end{proof}

Heuristically, complete pooling maximizes diversification: every agent bears the same average.
Variance shrinks at rate $1/n$, so large pools nearly eliminate idiosyncratic risk. The asymptotic properties of peer‐to‐peer insurance pools have been studied in \citep{DenuitRobert2020}, emphasizing sustainability in large markets.

\subsection{Star Network}

\begin{definition}[Star network]
A star has one central node (say node $1$) connected to all others, but no peripheral-to-peripheral links.
A simple LRS choice is:
\[
\mathrm{M}_{11}=1, ~  \mathrm{M}_{i1}=1 \text{ for } i>1, ~ 
\mathrm{M}_{ii}=0, \; \mathrm{M}_{ij}=0 \text{ otherwise}.
\]
Thus $\xi_1=X_1+\displaystyle\sum_{j=2}^n X_j$, while $\xi_i=0$ for $i>1$.
\end{definition}

This raw star allocation is budget balanced (CS) but not row-stochastic: the center absorbs all losses.

\begin{proposition}[Variance distortion in the naive star]\label{prop:star-variance}
Assume $X_1,\dots,X_n$ are independent with $\operatorname{Var}(X_i)=\sigma^2<\infty$.
Consider the star mechanism with center $1$ given by
$M_{11}=1$, $M_{i1}=1$ for $i>1$, and $M_{ij}=0$ otherwise, so $\xi_1=\sum_{j=1}^n X_j$
and $\xi_i\equiv 0$ for $i>1$. Then
\[
\operatorname{Var}(\xi_1)=n\,\sigma^2,\qquad \operatorname{Var}(\xi_i)=0\quad (i>1).
\]
\end{proposition}

\begin{proof}
By independence and finite variance,
$\operatorname{Var}\!\left(\sum_{j=1}^n X_j\right)=\sum_{j=1}^n \operatorname{Var}(X_j)=n\sigma^2$.
The remaining components are constants, hence have variance $0$.
\end{proof}

Heuristically, The naive star is \emph{unfair}: the center is overburdened.
A more symmetric design assigns weights $\frac{1}{2}$ along each spoke, making each edge a reciprocal contract. This yields a DS $\mathrm{M}$ where the center diversifies across all spokes and peripherals contribute equally. Such refinements restore fairness and convex order desirability.

\subsection{Ring or Chain Network}

\begin{definition}[Ring network]
Nodes are arranged on a cycle, each connected to two neighbors.
A natural LRS uses equal weights to neighbors and self:
\[
\mathrm{M}_{ii}= \frac{1}{3}, ~  \mathrm{M}_{i,i-1}=\mathrm{M}_{i,i+1}=\frac{1}{3}, 
\]
with indices modulo $n$.
\end{definition}

\begin{proposition}[Variance in ring]
If $X_i$ are independent with variance $\sigma^2$, then
\[
\mathrm{Var}(\xi_i) = \frac{1}{3^2}( \sigma^2+\sigma^2+\sigma^2 ) = \frac{1}{3}\sigma^2.
\]
\end{proposition}

\begin{proof}
$\xi_i=\frac{1}{3}(X_{i-1}+X_i+X_{i+1})$.
By independence, variances add, giving $\frac{1}{9}(3\sigma^2)=\frac{1}{3}\sigma^2$.
\end{proof}

Heuristically, the ring diversifies partially: each agent shares with only two neighbors.
Variance shrinks to $\frac{1}{3}\sigma^2$, better than autarky but worse than complete pooling.
Local structure limits diversification.

\subsection{Erd\H{o}s–R\'enyi and Scale-Free Networks}

\begin{definition}[Erd\H{o}s–R\'enyi]
An Erd\H{o}s–R\'enyi $G(n,p)$ network has each edge $\{i,j\}$ independently present with probability $p$.
\end{definition}

\begin{definition}[Scale-free networks]
A scale-free (Barab\'asi–Albert) network grows by preferential attachment, producing hubs with high degree.
\end{definition}

\begin{proposition}[Expected degree-variance link]
For a random network with adjacency matrix $\mathrm A$, let $d_i=\sum_j \mathrm  A_{ij}$.
If $\xi_i = \displaystyle\frac{1}{d_i+1}\big(X_i+\sum_{j\in N(i)} X_j\big)$ (equal-share with neighbors, where $N(i)$ are neighbors of node $i$),
then
\[
\mathbb{E}[\mathrm{Var}(\xi_i)\mid d_i] = \frac{1}{(d_i+1)^2}(d_i+1)\sigma^2 = \frac{\sigma^2}{d_i+1}.
\]
\end{proposition}

\begin{proof}
Conditional on degree $d_i$, $\xi_i$ averages $d_i+1$ independent risks.
Variance of the average is $\sigma^2/(d_i+1)$.
\end{proof}

Heuristically, variance reduction depends inversely on degree.
In Erd\H{o}s–R\'enyi graphs, degrees are roughly homogeneous, so risk-sharing gains are balanced.
In scale-free networks, hubs achieve near-complete diversification (low variance), while peripheral nodes with small $d_i$ remain poorly diversified — leading to inequity.

\subsection{Numerical Illustration}

From the previous parts, we have seen that these mechanism illustrate a spectrum:
\begin{itemize}
    \item Complete networks maximize diversification and fairness (all agents identical).
    \item Stars highlight the danger of unbalanced structures (unless weights are adjusted).
    \item Rings provide local sharing, with variance reduction limited by degree.
    \item Random/scale-free networks show heterogeneity: diversification depends strongly on degree, producing hubs vs.\ periphery dynamics.
\end{itemize}

\begin{figure}[!htbp]
    \centering
    \begin{minipage}{0.45\textwidth}
        \centering
        \includegraphics[width=0.48\linewidth]{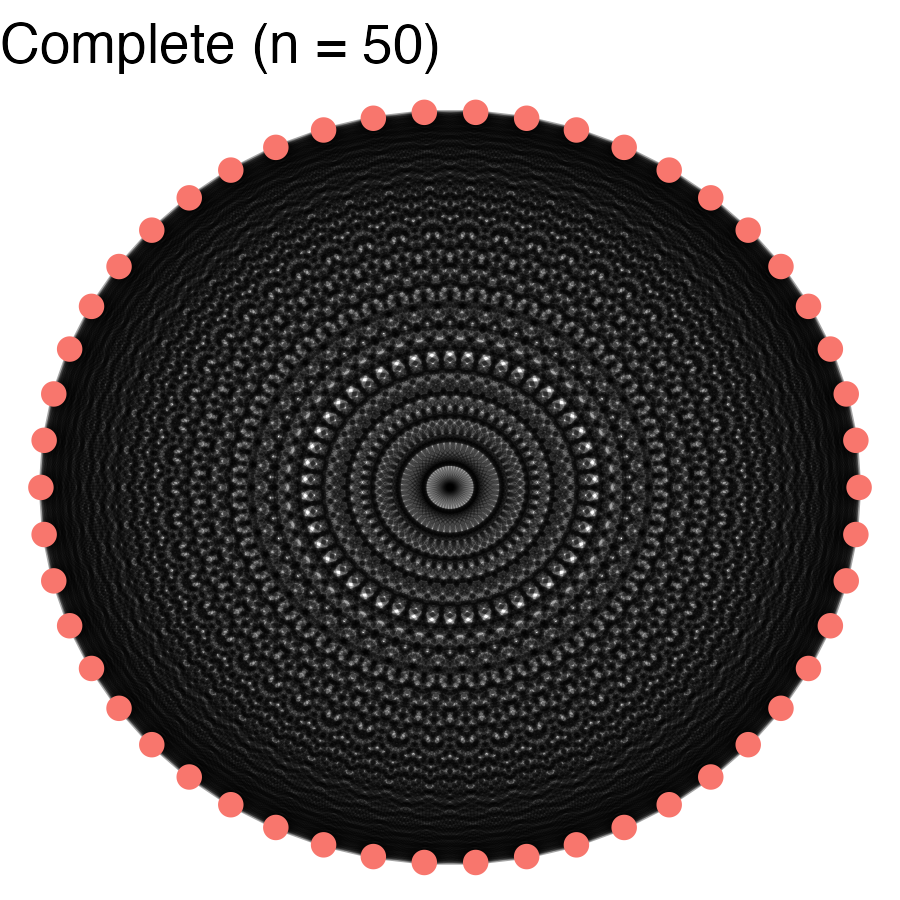}
        \includegraphics[width=0.48\linewidth]{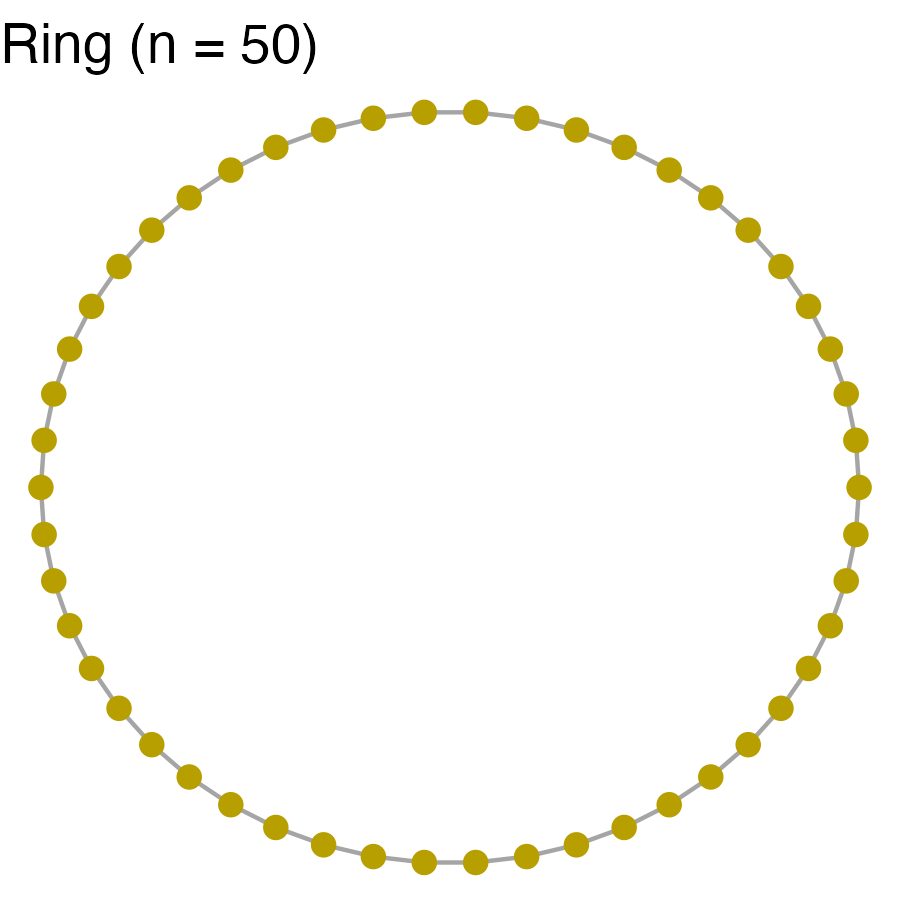} \\
        \includegraphics[width=0.48\linewidth]{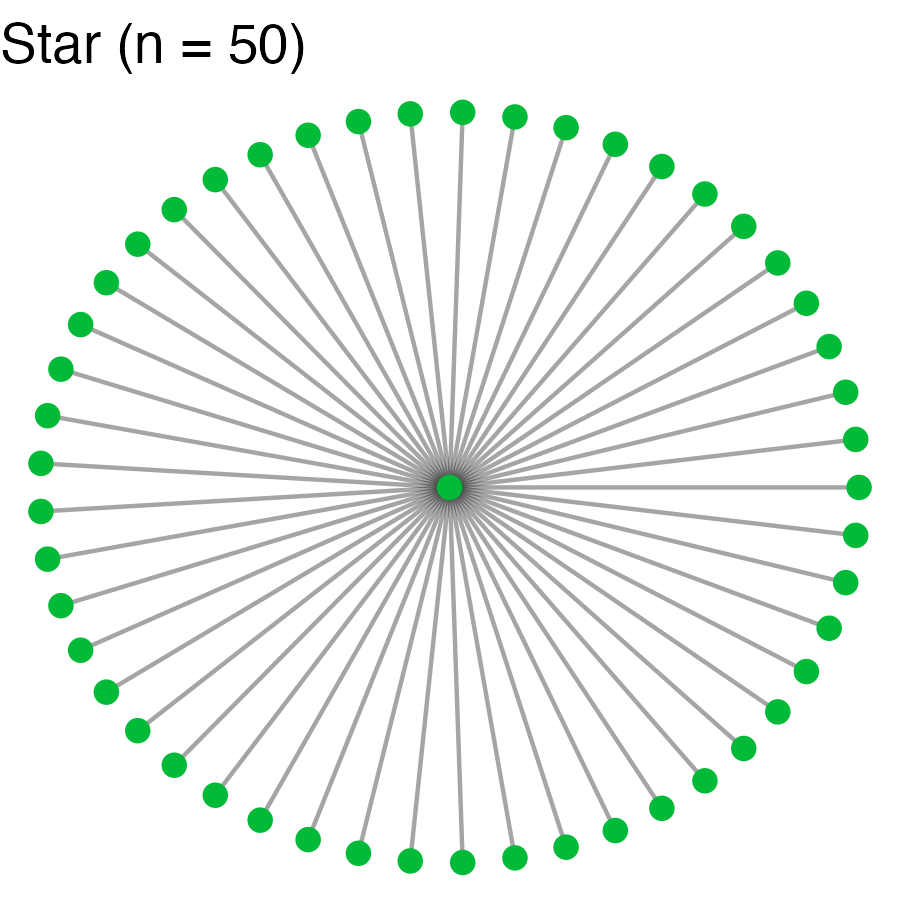}
        \includegraphics[width=0.48\linewidth]{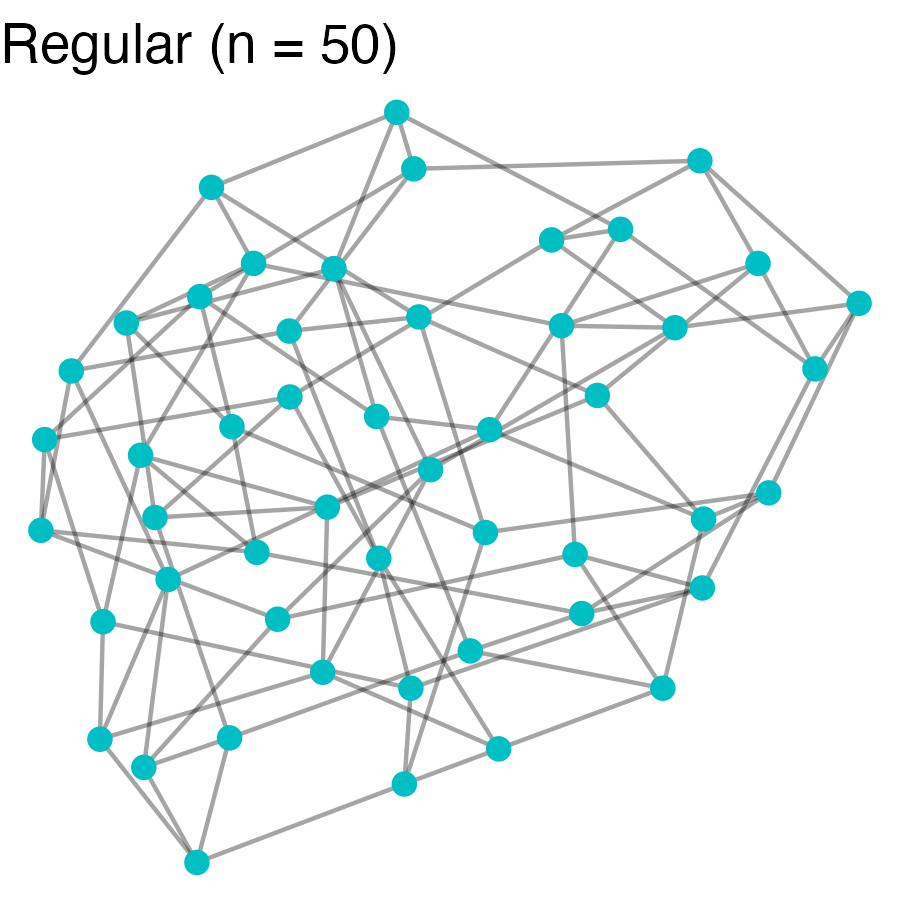} \\
        \includegraphics[width=0.48\linewidth]{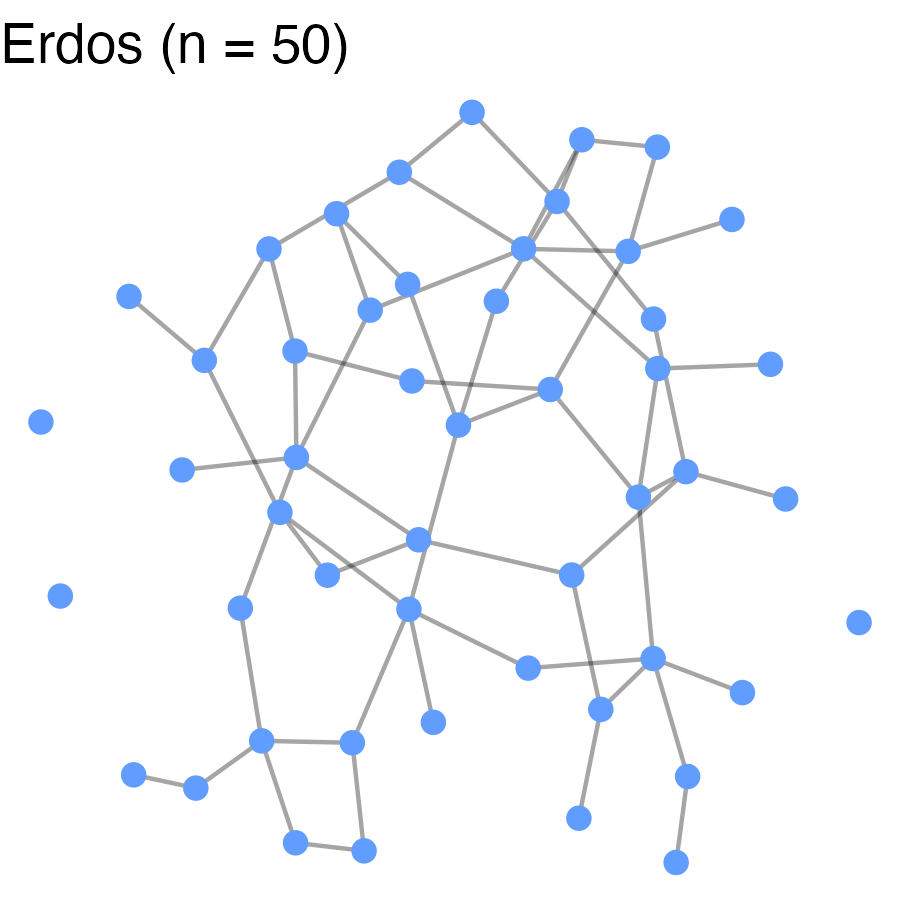}
        \includegraphics[width=0.48\linewidth]{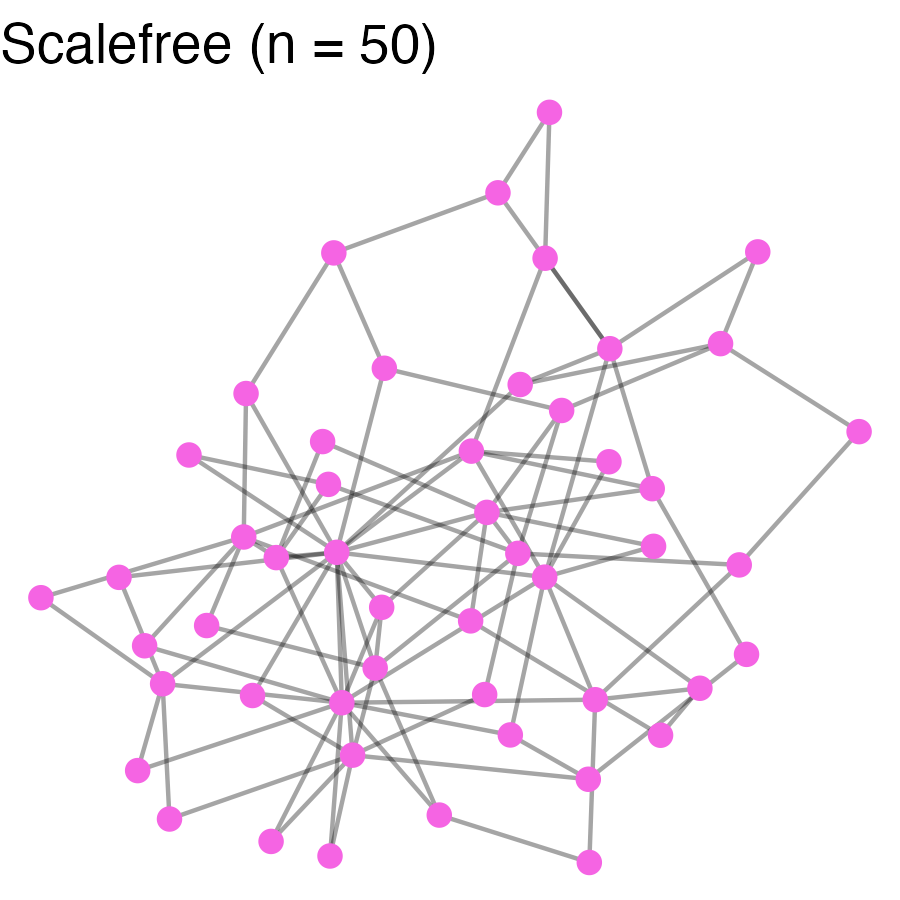}
    \end{minipage}
    \begin{minipage}{0.5\textwidth}
        \centering
        \includegraphics[width=\linewidth]{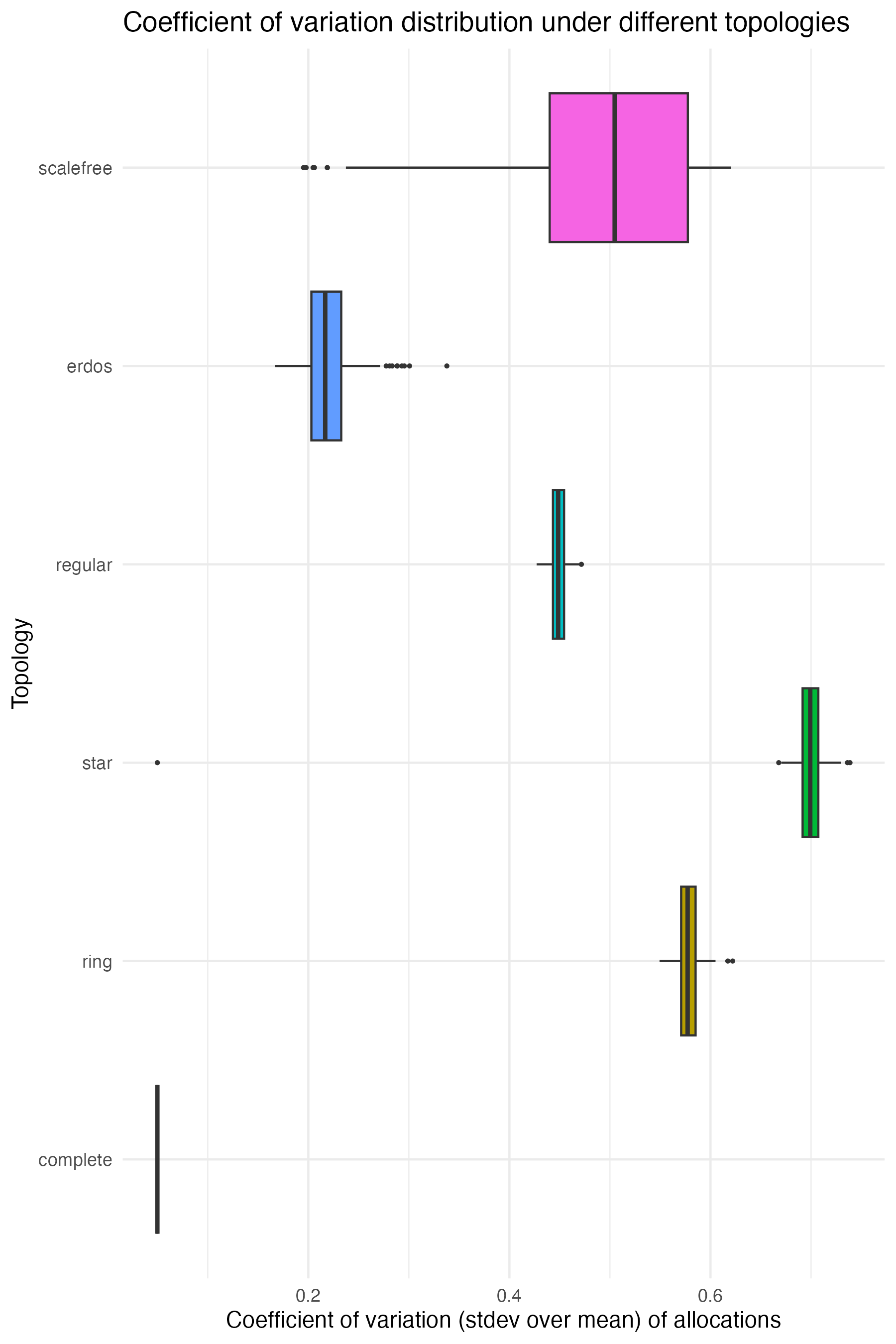}
    \end{minipage}
    \caption{Graph topologies with 
$n=50$ nodes (left) and the distribution of risk-sharing outcomes (right). The left column illustrates six canonical networks: complete, ring, star, random 4-regular, Erdős–Rényi with $p=0.05$, and Barabási–Albert scale-free with $m=2$. The right panel shows the distribution of the coefficient of variation (standard deviation divided by mean) of allocations under each topology, with node colors matching between the two panels.}
    \label{fig:topologies_variance}
\end{figure}

Figure~\ref{fig:topologies_variance} contrasts structural properties of canonical networks with their
implications for linear risk sharing. All networks are generated with $n=50$ nodes. The complete
graph serves as the benchmark of full pooling. The ring connects each node to two neighbors, while
the star concentrates all connections on a single hub. The random regular network fixes degree at
$k=4$, ensuring homogeneous diversification. The Erd\H{o}s--R\'enyi graph is drawn with edge
probability $p=0.05$, yielding an average degree close to $2.5$. The scale-free network follows the
Barab\'asi--Albert construction with attachment parameter $m=2$, producing a heavy-tailed degree
distribution with visible hubs. For each topology, we simulate $B=2000$ independent exponential
losses and compute the allocation vector $\boldsymbol\xi =\mathrm M\boldsymbol X$ under equal-weight reciprocal sharing rules.
The right panel displays the coefficient of variation (standard deviation divided by mean) across agents. Complete
pooling yields minimal dispersion, while the star highlights extreme inequity between the center
and periphery. The ring and Erd\H{o}s--R\'enyi graphs produce intermediate outcomes, and the
scale-free graph exhibits the widest spread, reflecting persistent inequality between hubs and
peripheral nodes.

Such heuristics reinforce the importance of matrix stochasticity conditions and network design
in ensuring both fairness and efficiency in peer-to-peer risk-sharing mechanisms.

\subsection{Design Principles for Network-Based Risk Sharing}

The case studies above suggest several general principles for designing
peer-to-peer risk-sharing mechanisms:

\begin{enumerate}
    \item \textbf{Doubly-stochasticity ensures fairness and efficiency.}  
    When $\mathrm{M}$ is doubly-stochastic (DS), each agent’s allocation is a convex combination
    of all losses (row-stochasticity) and aggregate budget is preserved (column-stochasticity).
    This guarantees:
    \[
    \boldsymbol\xi \preceq_{CCX} \boldsymbol X, ~ 
    \mathbb{E}[\xi_i]=\mathbb{E}[X_i], ~
    \mathrm{Var}(\xi_i)\le \mathrm{Var}(X_i),
    \]
    ensuring both individual fairness and risk reduction.

    \item \textbf{Column-stochasticity alone preserves budget, but not fairness.}  
    Pure CS matrices balance the global budget but can distort allocations.
    Some agents may bear more than their fair share, as in the naive star
    example where one hub absorbs all risk.

    \item \textbf{Row-stochasticity preserves individual means.}  
    Even without CS, RS ensures each $\xi_i$ has the same expected value
    as $X_i$ (Lemma~\ref{lem:RS-mean}). This condition enforces fairness
    in expectation and allows Jensen’s inequality to yield convex order
    comparisons when risks are i.i.d.

    \item \textbf{Degree heterogeneity drives inequity.}  
    In networks like scale-free graphs, high-degree hubs achieve strong
    diversification ($\mathrm{Var}(\xi_i)\approx \sigma^2/(d_i+1)$), while
    peripheral nodes with small $d_i$ remain nearly as risky as in autarky.
    Thus, variance inequality mirrors degree inequality.

    \item \textbf{Local vs.\ global sharing.}  
    Structures with only local sharing (ring/chain) reduce variance moderately,
    but cannot achieve the $1/n$ scaling of complete pooling.
    The efficiency of risk sharing scales with the connectivity of the graph.

    \item \textbf{Symmetry and regularity promote equity.}  
    Regular graphs (all degrees equal) combined with DS weighting
    yield uniform variance reduction and symmetric outcomes.
    Such designs are preferable when equitable treatment of participants
    is a primary objective.
\end{enumerate}

\medskip
\noindent
An heuristic message is that budget balance (CS) is a minimal requirement; row-stochasticity (RS) guarantees
per-agent fairness; doubly-stochasticity (DS) combines both and delivers the
strongest risk-order improvements. Network topology then modulates how much
diversification is achievable: dense, regular graphs approach complete pooling,
whereas sparse or highly skewed graphs yield unequal outcomes.

\section{Random Graph Extensions and Robustness of Results}\label{sec:random}

The classical insurance literature often assumes an \emph{exchangeability} (or
\emph{anonymity} property): $X_i$ denotes the generic risk of an agent in a
homogeneous pool, and risk-sharing mechanisms are symmetric in $i$.
In such settings, convex order and law-of-large-numbers arguments apply cleanly,
and “representative-agent” interpretations are natural.

\subsection{Losses on a Fixed Network}

In our network framework, the vector $\boldsymbol{X}=(X_1,\ldots,X_n)$ is still random,
but the risk-sharing operator $\mathrm{M}$ reflects a fixed network.
Agents are not anonymous: each node has a distinct set of neighbors,
and allocations $\boldsymbol{\xi}=\mathrm{M}\boldsymbol{X}$ reflect this heterogeneity.
Properties such as convex order dominance or per-agent variance reduction
must then be proved \emph{row by row}, rather than by appealing to exchangeability.

This asymmetry is not a flaw but a feature: it encodes heterogeneity of exposure
and diversification possibilities inherent in the graph.
Mathematically, the LRS representation remains sound: for any nonnegative $\mathrm{M}$,
$\boldsymbol{\xi}=\mathrm{M}\boldsymbol{X}$ is well-defined, and budget balance and convexity are recovered if
$\mathrm{M}$ is (column-/row-/doubly-) stochastic.

\subsection{Adding a Second Layer of Randomness: Random Graphs}

A richer framework arises if $\mathrm{M}$ itself is random, drawn from a distribution of networks.
This yields a \emph{two-layer model}:
\[
\boldsymbol \xi \;=\; \mathrm{M}\boldsymbol{X}, ~ \text{where $\boldsymbol{X}$ is random and $\mathrm{M}$ is random, independent of $\boldsymbol{X}$}.
\]
Some measure-related mathematical details are provided in Section \ref{sec:prob:second:layer}.

\begin{definition}[Random-graph LRS]
Let $\mathbb{M}$ be a distribution on the set $\mathcal{M}_n^+$ of matrices
(e.g., induced by a random graph model and weighting rule).
An allocation $\boldsymbol{\xi}$ is said to follow a random-graph LRS if
$\boldsymbol{\xi}= \mathrm{M}\boldsymbol{X}$ with $\mathrm M\sim\mathbb{M}$ independent of $X$.
\end{definition}

\begin{proposition}[Law of iterated expectation]
\label{prop:iterated-expectation}
For any random-graph LRS, conditional on $\mathrm{M}$,
\[
\mathbb{E}[\boldsymbol\xi\mid \mathrm M] = \mathrm M\,\mathbb{E}[\boldsymbol{X}].
\]
In particular, if $\mathbb{E}[X_i]=\mu$ for all $i$ and $\mathrm{M}$ is row-stochastic a.s.,
then $\mathbb{E}[\xi_i\mid \mathrm M]=\mu$, and hence $\mathbb{E}[\xi_i]=\mu$ unconditionally.
\end{proposition}

\begin{proof}
Condition on $\mathrm{M}$ and apply linearity of expectation.
If $\mathrm{M}$ is RS, then $\displaystyle\sum_j \mathrm{M}_{ij}=1$ almost surely, so the conditional expectation of
$\xi_i$ is $\mu$ for each $i$. Taking unconditional expectation preserves equality.
\end{proof}

Random networks such as Erdős–Rényi graphs and preferential attachment models provide canonical test cases for network‐based risk sharing \citep{ErdosRenyi1959,BarabasiAlbert1999}.

\subsection{The Case of Erd\H{o}s–R\'enyi Graphs}

Let $G\sim G(n,p)$ be an Erd\H{o}s–R\'enyi graph, where each edge $\{i,j\}$ is present
independently with probability $p$.
Define $d_i=\deg(i)$ and
\[
\mathrm{M}_{ij} = \begin{cases}
\displaystyle\frac{1}{d_i+1}, & \text{if $j=i$ or $\{i,j\}\in E(G)$},\\[6pt]
0, & \text{otherwise}.
\end{cases}
\]
Then $\mathrm{M}$ is row-stochastic for every realization of $G$.

\begin{proposition}[Variance under random degree]
\label{prop:ER-variance}
If $X_i$ are i.i.d.\ with variance $\sigma^2$, then conditional on $d_i$,
\[
\mathrm{Var}(\xi_i\mid d_i) = \frac{\sigma^2}{d_i+1}.
\]
Unconditionally,
\[
\mathbb{E}[\mathrm{Var}(\xi_i\mid d_i)] = \sigma^2\,\mathbb{E}\!\left[\frac{1}{d_i+1}\right],
\]
where $d_i\sim \mathrm{Bin}(n-1,p)$.
\end{proposition}

\begin{proof}
Conditional on $d_i$, $\xi_i$ is the average of $d_i+1$ independent losses,
so its variance is $\sigma^2/(d_i+1)$. The unconditional expectation follows
by taking expectation over $d_i$.
\end{proof}

\subsection{Interpretation}

\begin{itemize}
    \item Exchangeability at the $\boldsymbol X$-level is broken once $\mathrm{M}$ is fixed:
    two agents with different degrees face different allocations.
    \item Randomizing $\mathrm{M}$ re-introduces a form of anonymity \emph{in distribution}:
    ex ante, each agent’s variance depends only on the distribution of degrees,
    not on identity.
    \item In homogeneous random graphs (e.g.\ $G(n,p)$ with $p$ constant),
    degrees concentrate around $np$, so $\mathrm{Var}(\xi_i)\approx \sigma^2/(np)$
    for most agents. This recovers approximate exchangeability.
    \item In heterogeneous random graphs (e.g.\ scale-free), degree dispersion
    persists, and inequity remains even after averaging over $\mathrm{M}$.
\end{itemize}

\subsection{The Case of Scale-Free (Preferential-Attachment) Random Graphs}

We now contrast Erd\H{o}s--R\'enyi with a scale-free family, e.g.\ Barab\'asi--Albert (BA)
preferential-attachment graphs. Let $G$ be generated by a BA process with parameter
$m\in\mathbb{N}$ (each arriving node connects to $\mathrm{M}$ existing nodes with probability
proportional to their degree). It is well known that the degree distribution exhibits a
power-law tail: $\mathbb{P}(d_i \ge k) \sim k^{-(\gamma-1)}$ with $\gamma\approx 3$ for BA.

We adopt the same network-induced LRS rule as before:
\[
\mathrm{M}_{ij} \;=\; \begin{cases}
\dfrac{1}{d_i+1}, & \text{if } j=i \text{ or } \{i,j\}\in E(G),\\[6pt]
0, & \text{otherwise},
\end{cases}
~ \text{ ~ ~ so that} ~
\xi_i \;=\; \frac{1}{d_i+1}\Big(X_i+\!\!\sum_{j\in N(i)} X_j\Big).
\]
Thus $\mathrm{M}$ is row-stochastic a.s. (each row sums to $1$) and respects the graph support.

\begin{proposition}[Conditional and unconditional variance in scale-free graphs]
\label{prop:BA-variance}
Assume $X_1,\dots,X_n$ are i.i.d.\ with variance $\sigma^2<\infty$, independent of $G$.
Then, conditional on the degree $d_i$,
\[
\mathrm{Var}(\xi_i \mid d_i) \;=\; \frac{\sigma^2}{d_i+1}.
\]
Unconditionally,
\[
\mathbb{E}\!\left[ \mathrm{Var}(\xi_i \mid d_i) \right]
\;=\; \sigma^2\, \mathbb{E}\!\left[ \frac{1}{d_i+1} \right]
~ \text{and}~ 
\mathrm{Var}(\xi_i) \;=\; \sigma^2\, \mathbb{E}\!\left[ \frac{1}{d_i+1} \right]
\;+\; \mathrm{Var}\!\left( \mathbb{E}[\xi_i \mid d_i] \right).
\]
If $\mathbb{E}[X_i]=\mu$ and $X_i$ are i.i.d., then $\mathbb{E}[\xi_i\mid d_i]=\mu$,
so $\mathrm{Var}(\xi_i) = \sigma^2\, \mathbb{E}[1/(d_i+1)]$.
\end{proposition}

\begin{proof}
Identical to the Erd\H{o}s--R\'enyi case: conditional on $d_i$, $\xi_i$ averages
$d_i+1$ independent losses, hence the conditional variance. The unconditional formulas
follow from the law of total variance. If $\mathbb{E}[X_i]=\mu$, row-stochasticity
ensures $\mathbb{E}[\xi_i\mid d_i]=\mu$ (Lemma~\ref{lem:RS-mean}), making the second term vanish.
\end{proof}

\begin{proposition}[Bounds under power-law degrees]
\label{prop:powerlaw-inverse-degree}
Let $d$ be a nonnegative integer-valued random variable with tail $\mathbb{P}(d\ge k)\sim C k^{-(\gamma-1)}$ as $k\to\infty$ for some
$\gamma>2$ and $C>0$. Then:
\begin{enumerate}
\item $\mathbb{E}\!\left[\frac{1}{d+1}\right]\in(0,1]$ and is finite.
\item (Jensen lower bound) Since $x\mapsto 1/x$ is convex on $(0,\infty)$,
\[
\frac{1}{\mathbb{E}[d]+1}\ \le\ \mathbb{E}\!\left[\frac{1}{d+1}\right].
\]
\item (Tunable upper bound) For any $K\in\mathbb{N}$,
\[
\mathbb{E}\!\left[\frac{1}{d+1}\right]\ \le\ \mathbb{P}(d\le K)\ +\ \frac{1}{K+1}.
\]
In particular, if the small-degree mass $\mathbb{P}(d\le K)$ is bounded away from $1$ and $K$ is chosen moderately large,
the expectation is controlled by the small-degree probabilities.
\end{enumerate}
\end{proposition}

\begin{proof}
Aee Appendix~\ref{prop:powerlaw-inverse-degree}.

\end{proof}

Note that if the degree sequence contains a few hubs 
with degrees $d_{\max}$ growing as $n^{1/2}$, then tight asymptotic 
approximations can be derived for the corresponding variances, since the contribution 
of such hubs dominates the normalization terms in large $n$.

\begin{corollary}[Typical variance and dispersion in BA graphs]
\label{cor:BA-typical}
In a BA graph with parameter $m\ge 1$, most nodes have degree on the order of $\mathrm{M}$
(with significant probability mass also at small degrees), while a few hubs satisfy
$d_{\max}=\Theta(n^{1/2})$. Under the LRS rule above and i.i.d.\ losses with variance $\sigma^2$,
\[
\mathrm{Var}(\xi_i) \approx \frac{\sigma^2}{d_i+1}.
\]
Hence, with high probability as $n\to\infty$,
\[
\min_i \lbrace\mathrm{Var}(\xi_i)\rbrace \;\sum\; \frac{\sigma^2}{d_{\max}} \;=\; \Theta\!\big(\sigma^2 n^{-1/2}\big),
~ 
\max_i \lbrace\mathrm{Var}(\xi_i)\rbrace \;\sim\; \sigma^2,
\]
so the spread between well-diversified hubs and poorly connected peripheral nodes persists.
\end{corollary}

\begin{proof}[Proof sketch]
Use $d_{\max}=\Theta(n^{1/2})$ for BA and the variance formula
$\mathrm{Var}(\xi_i)=\sigma^2/(d_i+1)$. Typical nodes have $d_i=\Theta(m)$,
yielding $\Theta(\sigma^2/m)$ variance; hubs have $d_i=\Theta(n^{1/2})$,
yielding $\Theta(\sigma^2 n^{-1/2})$ variance.
\end{proof}

\begin{proposition}[Variance gap between hubs and non-hubs in power-law networks]\label{prop:ineq}
For each $n$, let $d_1^{(n)},\dots,d_n^{(n)}$ be degrees with empirical distribution $F_n$ such that:
\begin{enumerate}
\item (\emph{Power-law tail}) There exists $\gamma\in(2,3]$ and $C>0$ with
$1-F_n(k)\sim C k^{-(\gamma-1)}$ uniformly in $n$ as $k\to\infty$.
\item (\emph{Top-order growth}) The $\lfloor \alpha n\rfloor$-th largest degree (for any fixed $\alpha\in(0,1)$)
satisfies $d^{(n)}_{(\lfloor \alpha n\rfloor)}\to\infty$ at a polynomial rate in $n$
(e.g., $d^{(n)}_{(1)}=\Theta(n^{1/(\gamma-1)})$ and similarly for the top $\alpha n$ quantile).
\item (\emph{Small-degree mass}) There exist $k_0$ and $c_0>0$ such that $\displaystyle\liminf_{n\to\infty} F_n(k_0)\ge c_0$
(i.e., a non-vanishing fraction of nodes have degree at most $k_0$).
\end{enumerate}
Let $\xi_i$ be the LRS allocations with $\operatorname{Var}(\xi_i)=\sigma^2/(d_i^{(n)}+1)$ (independent risks, row-stochastic local averaging).
Then for any fixed $\alpha\in(0,1)$,
\[
\frac{1}{\lfloor \alpha n\rfloor}\sum_{i\in\text{Top-}\alpha}\operatorname{Var}(\xi_i)\ \xrightarrow[n\to\infty]{}\ 0
~\text{while}~
\liminf_{n\to\infty}\ \frac{1}{n-\lfloor \alpha n\rfloor}\sum_{i\notin\text{Top-}\alpha}\operatorname{Var}(\xi_i)\ \ge\ \frac{\sigma^2}{k_0+1}\,c_0\ >0,
\]
hence the average variance among hubs is asymptotically much smaller than among non-hubs.
\end{proposition}

\begin{proof}
See Appendix~\ref{app:prop:ineq}
\end{proof}

Heuristically, unlike Erd\H{o}s--R\'enyi, where degrees concentrate and ex ante anonymity is approximately
restored, scale-free graphs \emph{retain} substantial heterogeneity even after averaging
over the graph draw. The LRS variance $\sigma^2/(d_i\!+\!1)$ maps degree inequality into
risk inequality: hubs nearly ``self-insure'' by spreading risk across many neighbors,
while low-degree nodes remain poorly diversified. Thus, unless the weighting rule $\mathrm{M}$
is adjusted (e.g., making rows DS with wider support beyond immediate neighbors, or adding
global mixing), ex ante fairness remains limited.

\paragraph{Design implications.}
In heavy-tailed networks, to mitigate post-sharing inequality one may:
(i) add a light global mixing layer $\mathrm D$ that is DS and independent of $G$, using
$\tilde {\mathrm M} = \mathrm D \mathrm M$ (which lowers the covariance trace and improves convex order;
see Proposition~\ref{prop:trace-DS});
(ii) reweight edges to reduce variance dispersion, e.g., assign larger self-retention to hubs
and higher neighbor weight to peripheral nodes while keeping rows stochastic;
(iii) introduce occasional long-range links (``rewiring'') to lift small degrees,
shifting the effective degree distribution’s lower tail and improving worst-case variance.

\medskip
\noindent
With random $\mathrm{M}$ from a scale-free ensemble, the two-layer model remains mathematically sound:
$\mathrm{M}$ is row-stochastic a.s., budget balance can be enforced in expectation or a.s.\ (with CS),
and the variance structure follows directly from the degree distribution. The key qualitative
difference from Erd\H{o}s--R\'enyi is persistent ex-ante heterogeneity of $\mathrm{Var}(\xi_i)$,
driven by power-law degrees.

\subsection{Numerical Simulations}

\begin{figure}[!ht]
    \centering
\includegraphics[width=0.95\linewidth]{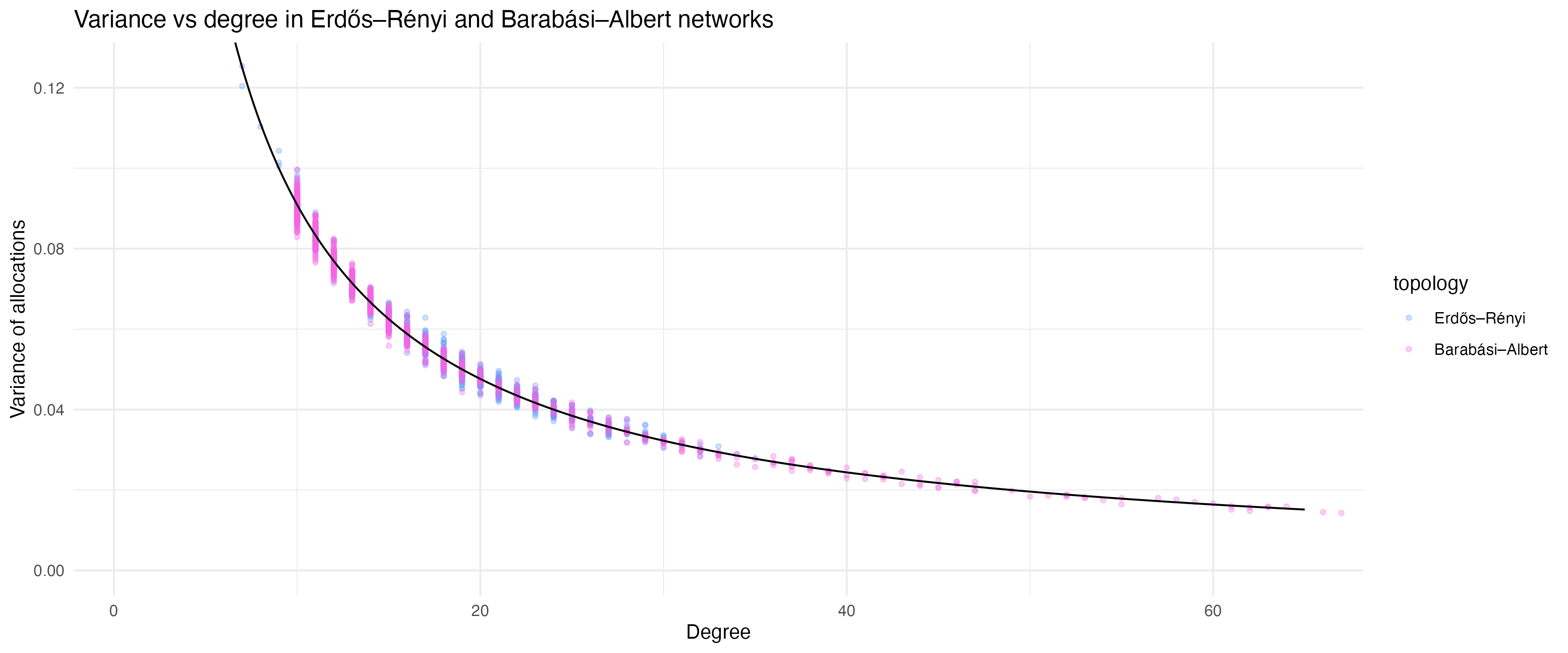}
   \caption{Variance of allocations as a function of node degree in Erd\H{o}s--R\'enyi 
($G(1000,0.02)$, blue) and Barab\'asi--Albert ($n=1000$, $m=10$, pink) graphs. 
Both models have comparable expected average degree ($\approx 20$). Each point corresponds 
to one node, with variance estimated from $B=2000$ independent exponential loss simulations 
under equal-weight reciprocal sharing. The solid black curve shows the theoretical benchmark 
$\mathrm{Var}(\xi_i) = \sigma^2/(d_i+1)$ with $\sigma^2=1$, confirming the inverse 
degree--variance relationship. While Erd\H{o}s--R\'enyi nodes cluster tightly around the curve, 
the Barab\'asi--Albert graph displays greater dispersion, reflecting the heterogeneity of 
its degree distribution and the persistent inequality between hubs and peripheral nodes.}
    \label{fig:er_ba}
\end{figure}

Figure \ref{fig:er_ba} consolidates the comparison between Erd\H{o}s--R\'enyi and
Barab\'asi--Albert networks by plotting degree against post-sharing variance in a single panel.
In both cases, the theoretical relationship $\mathrm{Var}(\xi_i) \approx \sigma^2/(d_i+1)$ is
clearly visible (plain black thin line): higher degree nodes diversify more effectively and face lower residual variance.
For the Erd\H{o}s--R\'enyi model with $p=0.02$, degrees are narrowly concentrated around $20$,
so most nodes fall along a tight band near $\sigma^2/21$, reflecting approximate homogeneity.
In contrast, the Barab\'asi--Albert network with $m=10$ exhibits a heavy-tailed degree
distribution: while hubs with large degree achieve extremely low variances, many peripheral
nodes remain close to autarky levels. The joint plot underscores how scale-free topologies
translate degree heterogeneity into persistent inequality in risk outcomes, whereas random
graphs with concentrated degrees approximate the fairness of homogeneous pools.

\subsection{A probabilistic setup for the second layer of randomness}\label{sec:prob:second:layer}

The use of adjacency and stochastic matrices derived from random graphs is classical in probabilistic graph theory \citep{ChungLu2006,Lovasz2012}, and provides natural tools to analyze the double randomness inherent in network-based risk sharing. But it is necessary to, formally, distinguish two probability spaces:
\begin{itemize}
  \item A \emph{loss space} $(\Omega_X,\mathcal{F}_X,\mathbb{P}_X)$ carrying 
  the random loss vector $\boldsymbol{X}=(X_1,\dots,X_n):\Omega_X\to\mathbb{R}^n$.
  \item A \emph{graph space} $(\Omega_G,\mathcal{F}_G,\mathbb{P}_G)$ carrying 
  the random network (or sharing) matrix $\mathrm M:\Omega_G\to\mathbb{R}_+^{n\times n}$.
\end{itemize}
We assume $\boldsymbol X$ and $\mathrm{M}$ are independent (two-layer model). The combined experiment lives
on the product space
\[
(\Omega,\mathcal{F},\mathbb{P}) 
:= \big(\Omega_X\times \Omega_G,\ \mathcal{F}_X\otimes\mathcal{F}_G,\ \mathbb{P}_X\otimes\mathbb{P}_G\big),
\]
and the post-sharing allocation is the measurable map
\[
\xi:\Omega\to\mathbb{R}^n,~  
\xi(\omega_X,\omega_G):= \mathrm M(\omega_G)\,X(\omega_X).
\]
When needed, we use the standard CS/RS/DS terminology for $\mathrm M(\omega_G)$: 
column-stochastic (CS) $\mathrm M^\top\mathbf{1}=\mathbf{1}$, row-stochastic (RS) $\mathrm M\mathbf{1}=\mathbf{1}$,
or doubly-stochastic (DS) (both). These properties are evaluated pointwise in $\omega_G$
and then (optionally) in expectation. \emph{Budget balance} corresponds to CS; RS ensures
each $\xi_i$ is a convex combination of the $X_j$'s (Jensen). 
%
%
\medskip

\paragraph{Iterated expectations (three viewpoints).}
Let $\mathbb{E}_X[\cdot]$ denote expectation on $(\Omega_X,\mathcal{F}_X,\mathbb{P}_X)$
and $\mathbb{E}_G[\cdot]$ on $(\Omega_G,\mathcal{F}_G,\mathbb{P}_G)$.
By independence and linearity:
\begin{itemize}
\item[(i)] Over the loss space, fixing the network:
$\mathbb{E}_X[\boldsymbol\xi\mid\mathrm  M] 
= \mathbb{E}_X[\mathrm{M}\boldsymbol{X}\mid \mathrm{M}] 
= \mathrm M\,\mathbb{E}_X[\boldsymbol{X}]. $
\item[(ii)] Over the graph space, fixing the losses:
$\mathbb{E}_G[\boldsymbol{}\xi\mid \boldsymbol X] 
= \mathbb{E}_G[\mathrm M\mid \boldsymbol X]\; X 
= \big(\mathbb{E}_G[\mathrm M]\big)\,\boldsymbol X.$
\item[(iii)]  Over the product space:  
$\mathbb{E}[\boldsymbol\xi] 
= \mathbb{E}_G\big[\mathrm M\big]\;\mathbb{E}_X[\boldsymbol{X}].$
\end{itemize}
Item (i) is the \emph{law of iterated expectation}: conditional on the realized network, 
we simply apply the linear operator $\mathrm{M}$ to the mean of $\boldsymbol{X}$.

\paragraph{Consequences for means (row-/column-stochasticity).}
Write $\mu_j:=\mathbb{E}_X[X_j]$ and $\mu:=(\mu_1,\dots,\mu_n)^\top$.
\begin{itemize}
  \item \emph{Per-agent means under RS a.s.} If $\mathrm{M}$ is RS almost surely and $\mu_j=\mu$ for all $j$,
  then, for every $i$,
  \[
  \mathbb{E}[\xi_i\mid \mathrm M] 
  = \sum_{j} \mathrm{M}_{ij}\mu 
  = \mu,
  ~ \text{hence}~  
  \mathbb{E}[\xi_i]=\mu.
  \]
  (Row sums $=1$ give per-agent mean preservation.) 
  \item \emph{Budget balance under CS (pointwise or in expectation).}
  If $\mathrm{M}$ is CS almost surely, then $\displaystyle\sum_i \xi_i=\sum_i X_i$ for all realizations; in expectation,
  $\displaystyle\mathbb{E}\big[\sum_i\xi_i\big]=\mathbb{E}\big[\sum_i X_i\big]$. 
  If only $\mathbb{E}_G[\mathrm M]$ is CS, then budget balance holds in expectation:
  \[
  \mathbf{1}^\top \mathbb{E}[\xi]
  = \mathbf{1}^\top \mathbb{E}_G[\mathrm M]\;\mathbb{E}_X[\boldsymbol{X}]
  = \mathbf{1}^\top \mathbb{E}_X[\boldsymbol{X}].
  \]
\end{itemize}

\paragraph{Mini-example (two-stage view).}
Let $G\sim G(n,p)$ be an Erd\H{o}s–R\'enyi graph drawn on $(\Omega_G,\mathcal{F}_G,\mathbb{P}_G)$.
Given $G$, set 
\[
\mathrm{M}_{ij} \;=\;
\begin{cases}
\displaystyle\frac{1}{d_i(G)+1}, & j=i \text{ or } \{i,j\}\in E(G),\\
0, & \text{otherwise},
\end{cases}
~ \Rightarrow~  \mathrm M\ \text{is RS a.s.}
\]
Draw i.i.d.\ $X_1,\dots,X_n$ on $(\Omega_X,\mathcal{F}_X,\mathbb{P}_X)$ with $\mathbb{E}_X[X_j]=\mu$.
Then, for each agent $i$:
\[
\mathbb{E}_X[\xi_i\mid \mathrm M] = \mu
~ \Rightarrow~  
\mathbb{E}[\xi_i]=\mu
~ \text{(per-agent mean preserved).}
\]
If, in addition, $\mathrm{M}$ (or $\mathbb{E}_G[\mathrm M]$) is CS, then aggregate budget is preserved 
(pointwise or in expectation, respectively). 

The ``second layer'' is a clean product-measure construction: condition on $\mathrm{M}$ to
apply all fixed-matrix arguments, then average over $\Omega_G$. No new notion of stochastic order is
needed; the usual convex/CCX results follow from conditioning plus RS/CS hypotheses, just as in the
fixed-network case.

\paragraph{Variance on the two-layer space.}
Write $\Sigma := \mathrm{Var}_X[\boldsymbol{X}]$ (on $(\Omega_X,\mathcal{F}_X,\mathbb{P}_X)$) and $\mu := \mathbb{E}_X[\boldsymbol{X}]\in\mathbb{R}^n$.
With $\boldsymbol{\xi}= \mathrm{M}\boldsymbol{X}$ on the product space $(\Omega_X\times\Omega_G,\mathcal{F}_X\otimes\mathcal{F}_G,\mathbb{P}_X\otimes\mathbb{P}_G)$:

\begin{itemize}
\item[(a)] Conditional on the network ($\mathrm{M}$ fixed):
\[
\mathrm{Var}_X(\boldsymbol\xi \mid \mathrm M) \;=\; \mathrm M\,\Sigma\,\mathrm M^\top, 
\]
\item[(b)] Conditional on the losses ($X$ fixed):
\[
\begin{cases}
    \mathbb{E}_G[\boldsymbol\xi\mid \boldsymbol X] \;=\; \mathbb{E}_G[\mathrm M]\,\boldsymbol X, \\
\mathrm{Var}_G(\boldsymbol\xi\mid \boldsymbol X) \;=\; 
\mathbb{E}_G\!\big[(\mathrm M-\mathbb{E}_G[\mathrm M])\,\boldsymbol X \boldsymbol X^\top\,(\mathrm M-\mathbb{E}_G[\mathrm M])^\top \,\bigm|\, \boldsymbol X\big],
\end{cases}
\]
\item[(c)] Unconditional (law of total variance):
\[
\mathrm{Var}(\boldsymbol\xi) 
= \mathbb{E}_G\!\big[\mathrm{Var}_X(\boldsymbol\xi\mid \mathrm M)\big] 
\;+\; \mathrm{Var}_G\!\big(\mathbb{E}_X[\boldsymbol\xi\mid \mathrm M]\big)
= \mathbb{E}_G\!\big[\mathrm M\,\Sigma\,\mathrm M^\top\big] \;+\; \mathrm{Var}_G(\mathrm M\mu).
\]
\end{itemize}

In particular, if $\mathrm{M}$ is row-stochastic a.s.\ and the components of $\mu$ are equal, then $\mathrm{M}\mu=\mu$ 
and the second term vanishes, so $\mathrm{Var}(\boldsymbol\xi)=\mathbb{E}_G[\mathrm{M}\Sigma \mathrm{M}^\top]$ (mean preserved agentwise).

\medskip
\noindent
\textbf{I.i.d.\ specialization.} If $X_i$ are i.i.d.\ with variance $\sigma^2$, so $\Sigma=\sigma^2 I$, then
\[
\mathrm{Var}_X(\boldsymbol\xi\mid \mathrm{M}) \;=\; \sigma^2\,\mathrm{M} \mathrm{M}^\top,
\text{ and }
\mathrm{tr}\big( \mathrm{Var}_X(\boldsymbol\xi\mid \mathrm{M})\big) \;=\; \sigma^2\,\mathrm{tr}(\mathrm{M}\mathrm{M}^\top).
\]
Under a (possibly random) doubly-stochastic post-mixer $\mathrm D$ independent of $(\mathrm{M},\boldsymbol X)$,
$\mathrm{tr}\big(\mathrm{Var}(\mathrm{D}\boldsymbol \xi)\big)\le \mathrm{tr}\big(\mathrm{Var}(\boldsymbol \xi)\big)$ (trace contraction). 

\medskip
\noindent
\textbf{Equicorrelation.} If $\Sigma=\sigma^2\big[(1-\rho)\mathbb{I}+\rho\,\mathbf{1}\mathbf{1}^\top\big]$ with $-1/(n-1)\le\rho<1$
and $\mathrm{M}$ is row-stochastic (so each row $\mathrm{M}_i$ sums to $1$), then
\[
\mathrm{Var}_X(\xi_i\mid \mathrm{M}) \;=\; \sigma^2\Big((1-\rho)\,\|\mathrm{M}_i\|_2^2+\rho\Big)
~ \Rightarrow~ 
\text{mixing shrinks only the idiosyncratic part via $\|\mathrm{M}_i\|_2^2$.}
\]

\medskip
\noindent
\textbf{Erd\H{o}s–R\'enyi $G(n,p)$.}
Let $G\sim G(n,p)$ and define, given $G$, the RS weights
$\mathrm{M}_{ij}=\frac{1}{d_i(G)+1}$ for $j=i$ or $\{i,j\}\in E(G)$, and $0$ otherwise.
With i.i.d.\ $X_i$ and $\mathrm{Var}(X_i)=\sigma^2$,
\[
\mathrm{Var}_X(\xi_i\mid \mathrm{M}) \;=\; \frac{\sigma^2}{d_i(G)+1},
\text{ and }
\mathbb{E}\!\big[\mathrm{Var}_X(\xi_i\mid \mathrm M)\big] \;=\; \sigma^2\,\mathbb{E}\!\Big[\frac{1}{d_i(G)+1}\Big],
\]
and if $\mathbb{E}_X[X_i]=\mu$ (constant) then $\mathrm{Var}(\xi_i)=\sigma^2\,\mathbb{E}[1/(d_i+1)]$ since $\mathbb{E}_X[\xi_i\mid \mathrm M]=\mu$ by RS.

\subsection{Numerical Illustration}

\begin{figure}[!ht]
    \centering
    \includegraphics[width=0.25\linewidth]{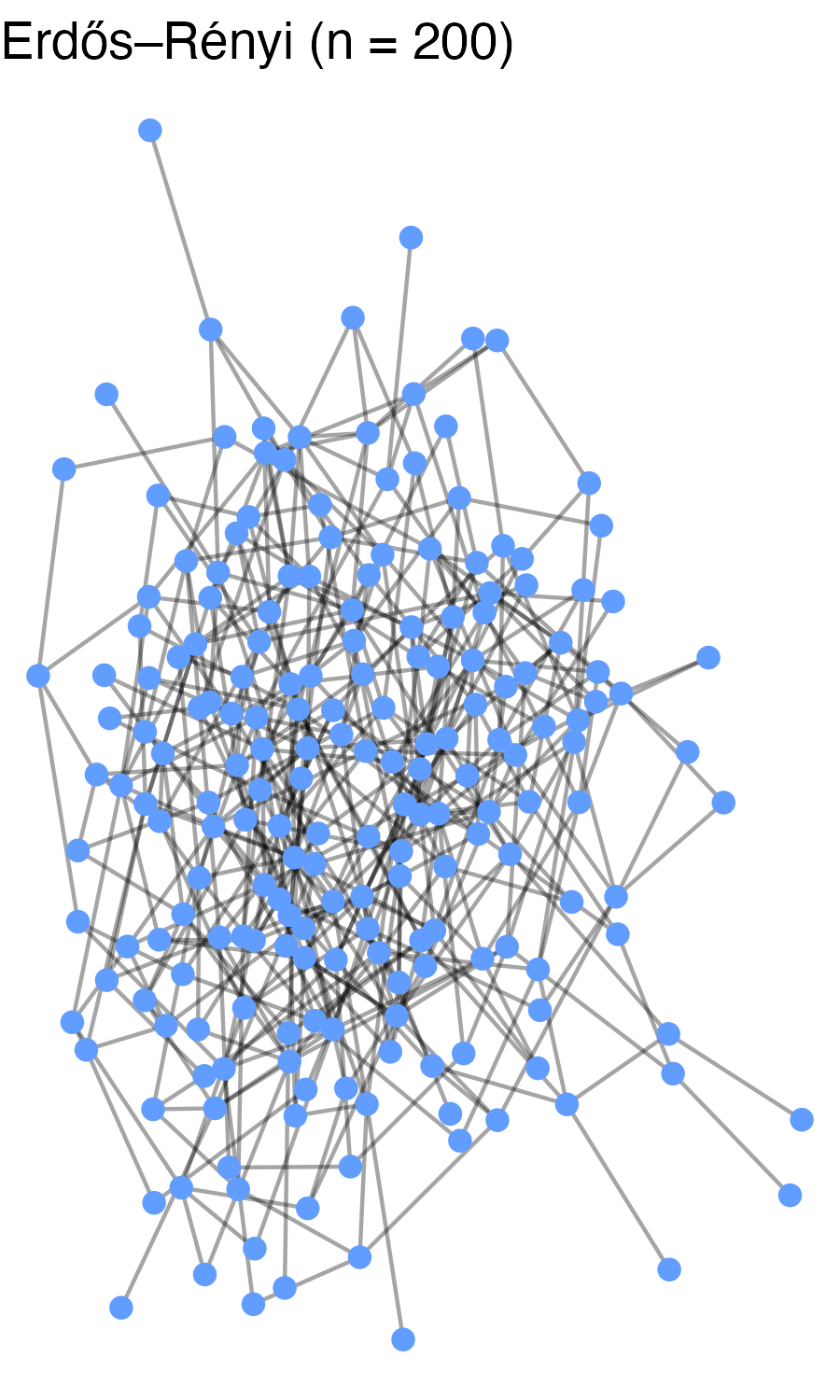}
    \includegraphics[width=0.25\linewidth]{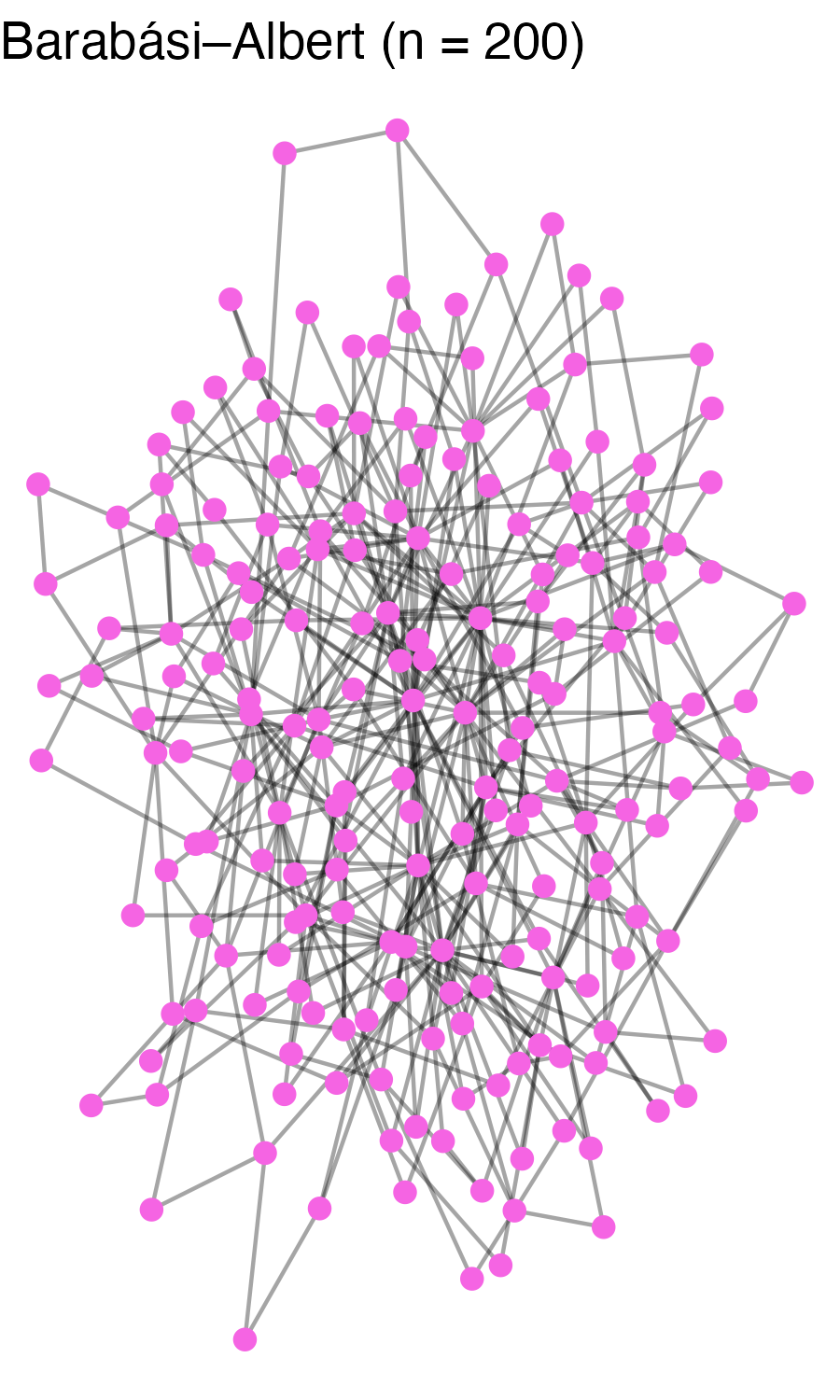}
    \includegraphics[width=0.475\linewidth]{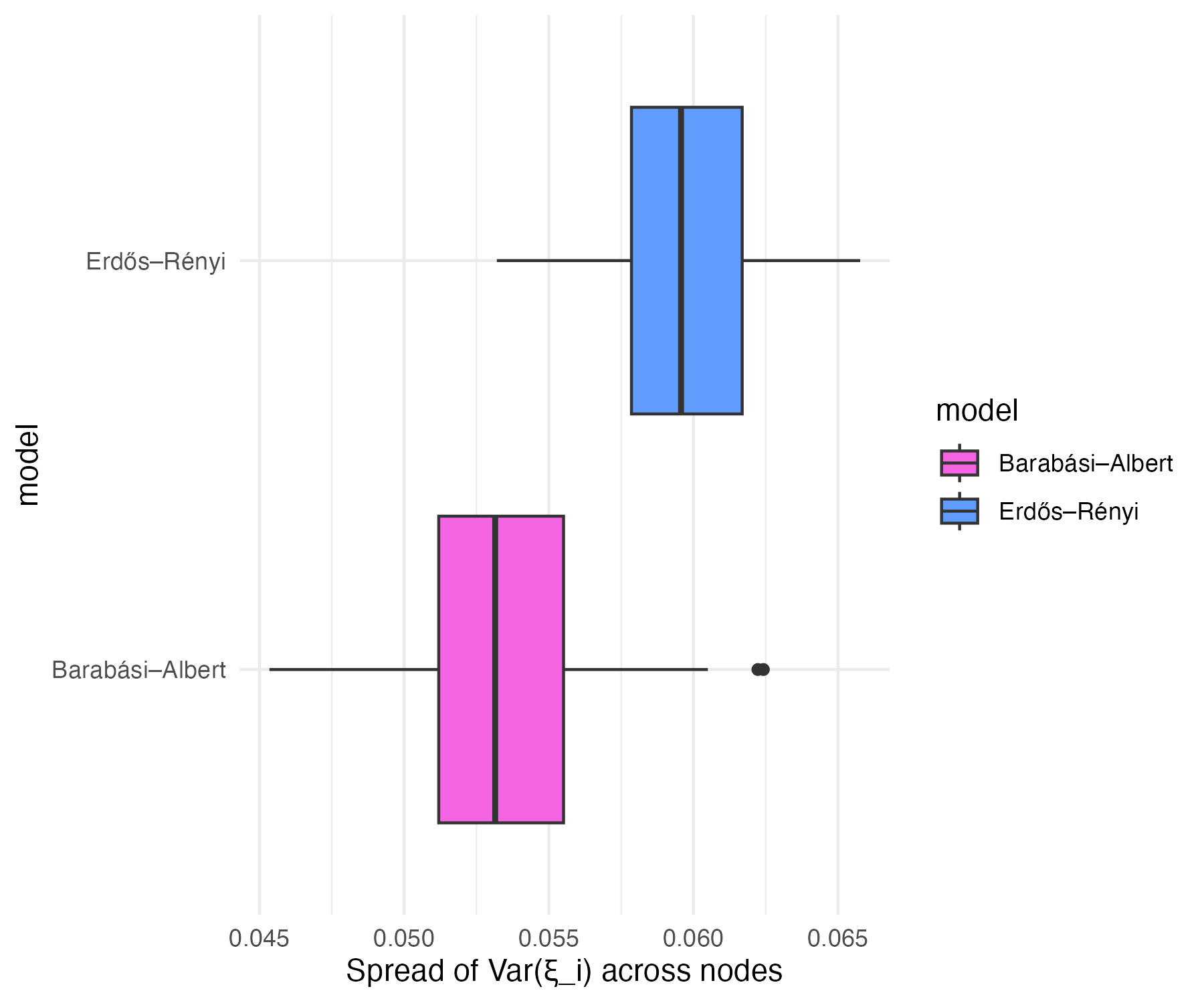}
    \caption{Comparison of network topologies and risk-sharing heterogeneity.
Left: examples of Erdős–Rényi and Barabási–Albert networks with $n=200$ nodes.
Right: distribution of the spread (90\%-10\% quantile range) of $\mathrm{Var}(\xi_i)$
across nodes, computed from $R=50$ replications with $n=600$ and $B=1000$ samples.
Barabási–Albert networks display a more heterogeneous distribution of variances across nodes
compared to Erdős–Rényi networks.}

    \label{fig:double:random}
\end{figure}

Figure~\ref{fig:double:random} illustrates the impact of network topology on the heterogeneity of individual variances under linear risk sharing. 
The left panels show typical realizations of an Erdős–Rényi graph and a Barabási–Albert graph, both with $n=200$ nodes. 
The right panel reports the spread of $\mathrm{Var}(\xi_i)$ across nodes, measured by the 90\%-10\% quantile range, for $n=600$ and $B=1000$. 
While Erdős–Rényi networks yield relatively homogeneous risk exposures across nodes, Barabási–Albert networks generate a wider dispersion, 
reflecting the influence of hubs on the redistribution of risk.

\subsection{Convex Ordering under Random Graph Matrices}

We now extend the convex order arguments to the case where the sharing
matrix $\mathrm{M}$ itself is random, independent of the loss vector $X$.

\begin{definition}[Random-matrix convex order]
Let $\boldsymbol{X}=(X_1,\dots,X_n)$ be i.i.d.\ integrable random variables and 
let $\mathrm{M}$ be a random nonnegative $n\times n$ matrix, independent of $X$.
Define $\boldsymbol{\xi}= \mathrm{M}\boldsymbol{X}$. 
We say that $\boldsymbol{\xi}$ is \emph{convexly dominated by $X$ under random networks}
if, almost surely in $\mathrm{M}$, each row of $\mathrm{M}$ is a probability vector 
(i.e.\ $\mathrm{M}$ is row-stochastic a.s.). 
\end{definition}

\begin{lemma}[Conditional convex order]\label{lem:condCO}
Suppose $\mathrm{M}$ is row-stochastic almost surely. Then, conditional on $\mathrm{M}$,
each component $\xi_i = (\mathrm{M}\boldsymbol{X})_i$ satisfies
\[
  \xi_i \;\preceq_{\mathrm{CX}}\; X_1.
\]
\end{lemma}

\begin{proof}
Fix $\mathrm{M}$ and consider row $i$, which is a probability vector $p^{(i)}$.
Then $\xi_i = p^{(i)\top} X$. By Jensen's inequality conditional on $X$,
for any convex $f$,
\[
  f(\xi_i) = f\!\big(p^{(i)\top}\boldsymbol{X}\big) 
  \;\leq\; \sum_j p^{(i)}_j f(X_j).
\]
Taking expectations and using i.i.d.\ of the $X_j$, we obtain
$\mathbb{E}[f(\xi_i)\mid \mathrm{M}] \leq \mathbb{E}[f(X_1)]$, hence $\xi_i \preceq_{\mathrm{CX}} X_1$.
\end{proof}

\begin{proposition}[Random-network convex dominance]\label{prop:randCO}
Let $\mathrm{M}$ be a random matrix, independent of $X$, row-stochastic almost surely.
Then
\[
  \boldsymbol\xi = \mathrm{M}\boldsymbol{X} \;\preceq_{\mathrm{CCX}}\; \boldsymbol{X} .
\]
If, in addition, $\mathrm{M}$ is column-stochastic a.s.\ (budget balance),
then $\mathbb{E}[\sum_i \xi_i] = \mathbb{E}[\sum_i X_i]$ and the dominance holds with
mean preservation both individually and in aggregate.
\end{proposition}

\begin{proof}
Condition on $\mathrm{M}$ and apply Lemma~\ref{lem:condCO}. Since $X_1 \stackrel{d}{=} X_i$
for each $i$, we conclude $\xi_i \preceq_{\mathrm{CX}} X_i$ given $\mathrm{M}$. 
Unconditioning preserves the convex order inequalities.
Column-stochasticity ensures budget balance as in Proposition~1.
\end{proof}

\begin{corollary}[Doubly-stochastic post-mixing]\label{DSpostmix}
If $\mathrm D$ is any (possibly random) doubly-stochastic matrix independent of
$(\mathrm M,\boldsymbol{X})$, and if $ \boldsymbol\xi = \mathrm{M}\boldsymbol{X}$, then
\[
   \mathrm{D}\boldsymbol\xi \;\preceq_{\mathrm{CCX}}\; \boldsymbol\xi
   \text{ and }
   \mathrm{tr}\big(\mathrm{Var}(\mathrm{D}\boldsymbol\xi) \big)\leq\; \mathrm{tr}\big(\mathrm{Var}(\boldsymbol\xi)\big),
\]
with expectation over both $\mathrm{M}$ and $\mathrm{D}$.
\end{corollary}

\begin{proof}
Condition on $(\mathrm{M},\boldsymbol{X})$ and apply the existing ``more mixing'' result, from Proposition~\ref{prop:postmix}
(of Section~\ref{sec:sec:dsm}). Taking expectations preserves the inequalities.
\end{proof}

No new notion of ``random convex order'' is required: the usual convex order
applies row by row, conditional on $\mathrm{M}$. The randomness of $\mathrm{M}$ is handled by
iterated expectation. Row-stochasticity ensures per-agent convex combinations;
column-stochasticity ensures budget balance. Doubly-stochasticity combines both
and unlocks the strongest dominance and variance-reduction results, just as in
the fixed-matrix case.

\subsection{Numerical Illustration}

\begin{figure}[!ht]
    \centering
    \includegraphics[width=0.475\linewidth]{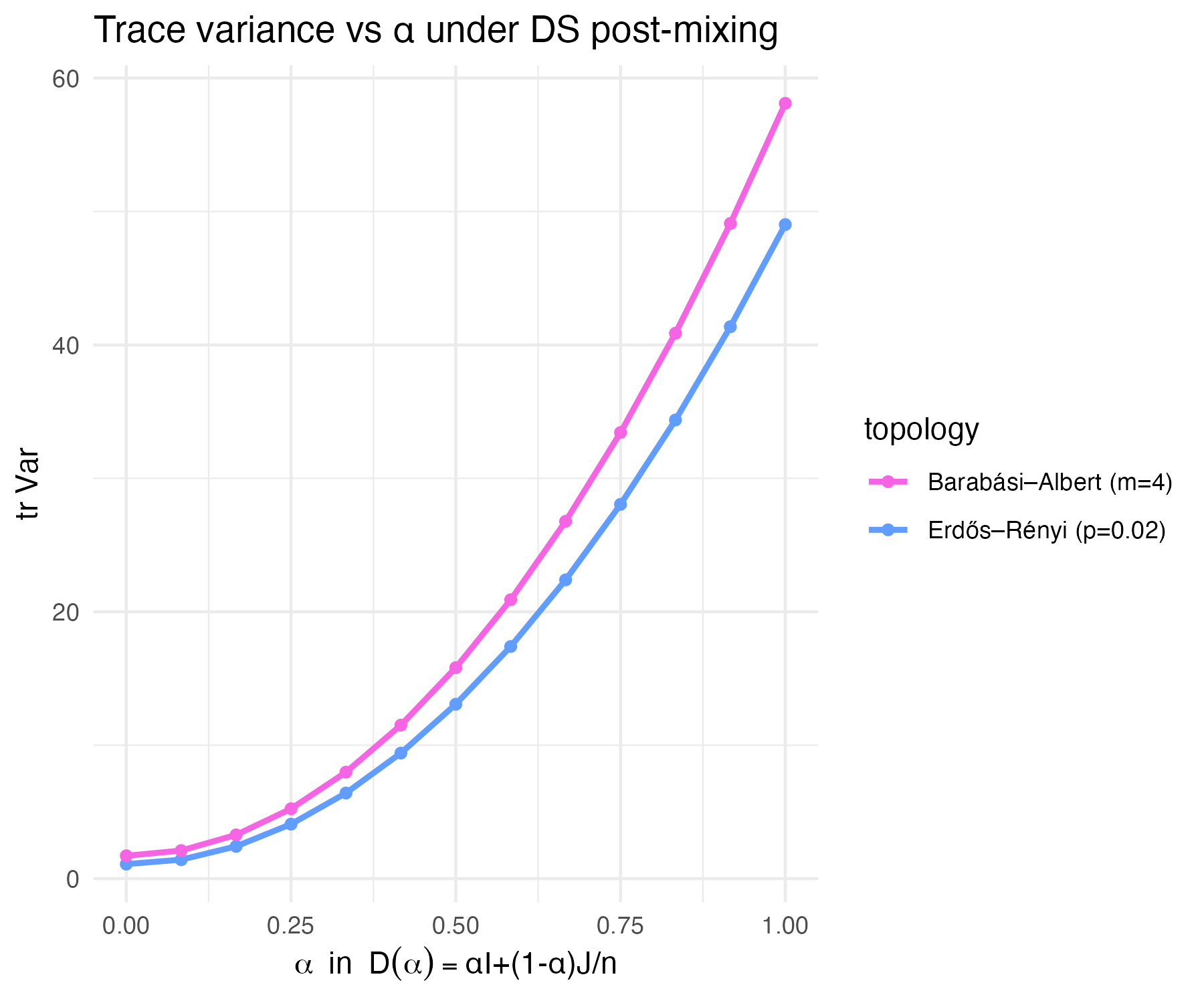}
    \includegraphics[width=0.475\linewidth]{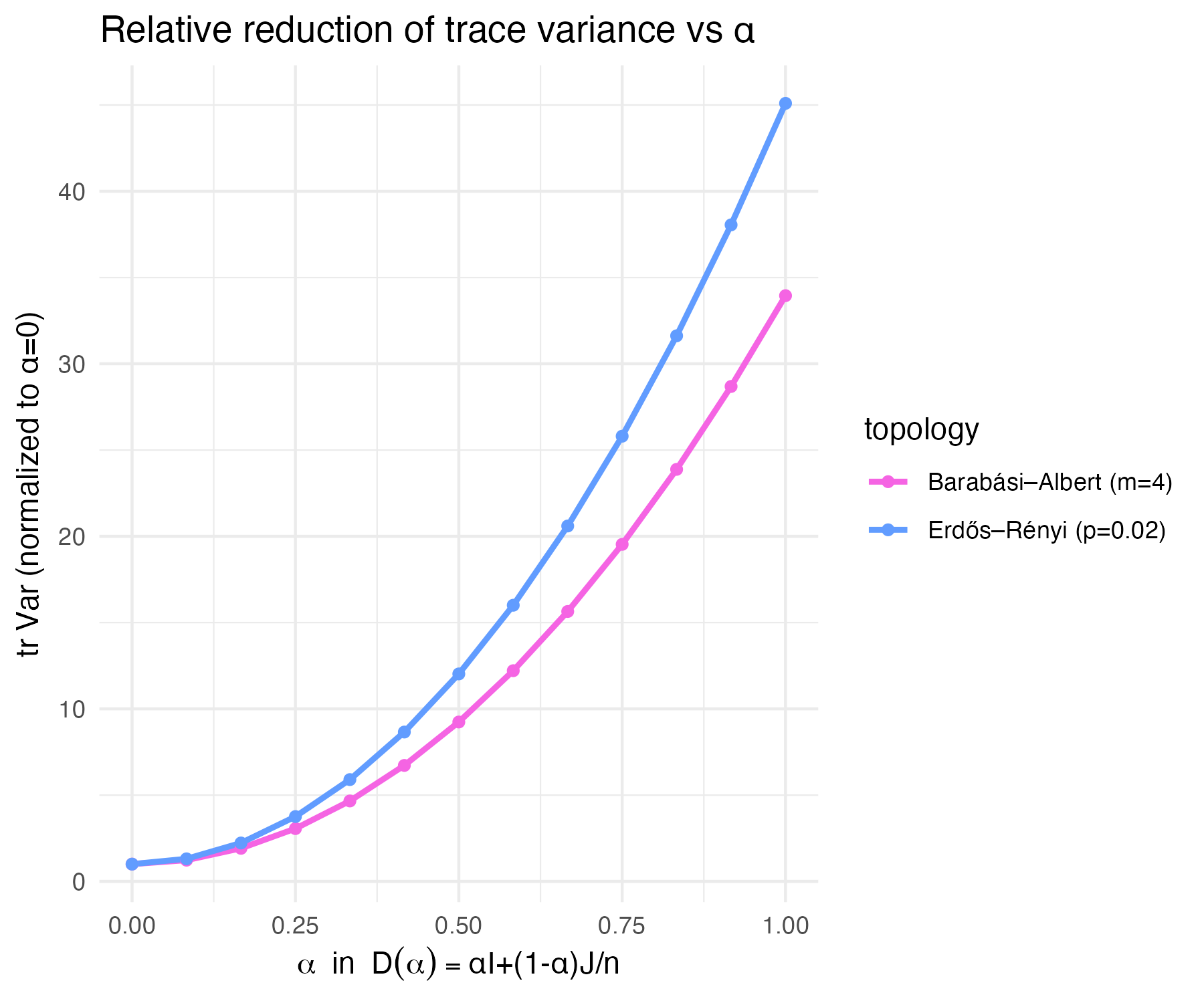}
   \caption{Effect of doubly–stochastic post–mixing on total variance. 
Left: absolute trace variance $\operatorname{tr}\,\mathrm{Var}\big(D(\alpha)\,\xi\big)$ as a function of $\alpha$; 
right: the same curves normalized by their value at $\alpha=0$. 
Allocations are $\boldsymbol\xi =\mathrm M \boldsymbol X$ where $\mathrm M$ is the row–stochastic (equal–neighbor) matrix built from the network and $X$ has i.i.d.\ $\mathcal{E}\mathrm{xp}(1)$ components. 
The post–mixer is $D(\alpha)=\alpha I+(1-\alpha)\mathbf{J}/n$ (doubly stochastic). 
We compare Erd\H{o}s–R\'enyi $G(n=400,p=0.02)$ (blue) and Barab\'asi–Albert $(n=400,m=4)$ (pink), which have comparable expected average degree ($\approx 8$). 
Curves are computed from $B=1500$ Monte Carlo draws and $\alpha\in\{0,0.083,\dots,1\}$. 
Both panels show that $\operatorname{tr}\,\mathrm{Var}$ is monotone in $\alpha$ and minimized at full averaging ($\alpha=0$), in line with the theory that doubly–stochastic post–mixing reduces total risk dispersion.}
    \label{fig:er_ba_trace}
\end{figure}

Figure \ref{fig:er_ba_trace} quantifies how a global, doubly–stochastic post–mixing step (from Corollary~\ref{DSpostmix}) $D(\alpha)=\alpha \mathbb{I}+(1-\alpha)\mathbf{J}/n$ affects the dispersion of allocations that were first produced by local, network–based sharing. 
Starting from $\boldsymbol \xi = \mathrm M \boldsymbol X$ with $\mathrm M$ the equal–neighbor (row–stochastic) matrix derived from the underlying graph and $X$ i.i.d.\ $\mathrm{Exp}(1)$, we form the post–mixed allocations $\zeta(\alpha)=D(\alpha)\,\xi$ and report $\operatorname{tr}\,\mathrm{Var}\big(\zeta(\alpha)\big)$. 
Two generating mechanisms are compared at matched density: Erd\H{o}s–R\'enyi $G(400,0.02)$ and Barab\'asi–Albert with $m=4$ (average degree $\approx 8$ in both cases). 
The left panel shows the absolute level of $\operatorname{tr}\,\mathrm{Var}$, which increases smoothly with $\alpha$ and is minimized at $\alpha=0$ (complete averaging). 
The right panel normalizes by the $\alpha=0$ value, highlighting that although both topologies benefit from the same DS mixer, the relative gain differs slightly with network structure: ER (more homogeneous degrees) shrinks faster for small $\alpha$, while BA (heavy–tailed degrees) exhibits a modestly slower reduction. 
Overall, the monotone decrease toward $\alpha=0$ empirically illustrates the proposition that DS post–mixing cannot increase total variance and typically yields substantial risk–sharing improvements over purely local schemes.

\section{Balancing Self-Retention and Collective Diversification}
\label{sec:selfretention}

We now study linear risk sharing where the sharing matrix is a convex combination of the
identity (self-retention) and a network-induced mixer:
\[
\mathrm{M}(\lambda) \;=\; (1-\lambda)\,\mathbb{I} \;+\; \lambda\,\mathrm{P}, \text{ where } \lambda\in[0,1],
\]
where $\mathrm P$ captures the network mixing rule. The parameter $\lambda$ is the \emph{mixing
intensity}: $\lambda=0$ is autarky (no sharing), $\lambda=1$ is pure network sharing.
We analyze stochasticity (budget balance vs.\ convex-combination property), variance/convex order
comparisons, and incentives.

Recent studies have proposed Pareto‐optimal allocation rules for decentralized insurance systems \citep{Feng2022Decentralized}.

\subsection{Which $\mathrm P$ are admissible? (RS/CS/DS and network support)}
\label{subsec:admissibleP}

Let $G=(V,E)$ be the (possibly directed) network on $n$ agents. We require $\mathrm P_{ij}\ge0$ and
$\mathrm P_{ij}=0$ whenever $\{i,j\}\notin E$ (or $(i,j)\notin E$ for directed graphs). Three canonical
normalizations:

\begin{enumerate}
\item \textbf{Row-stochastic (RS) random-walk mixer.}
Given the adjacency $\mathrm A=(\mathrm A_{ij})$ and out-degrees $d_i=\displaystyle\sum_j \mathrm A_{ij}$, define
$\mathrm P^{\mathrm{rw}} = \mathrm D^{-1}\mathrm A$ (with $\mathrm D=\mathrm{diag}(d_1,\dots,d_n)$), and optionally
\emph{lazy} version $\mathrm P^{\mathrm{lrw}} =\displaystyle \frac12 \mathbb I + \frac12 \mathrm D^{-1}\mathrm A$. Then $\mathrm P$ is RS
(each row sums to $1$), and each row is a convex combination over neighbours.

\item \textbf{Doubly-stochastic (DS) mixing on the graph support.}
When DS is desired (budget balance and convex-combo rows), one can construct
$\mathrm P^{\mathrm{ds}}$ supported on the set of edges $E$ by Sinkhorn-Knopp scaling on the (nonnegative) support
matrix (see Section~\ref{sec:sec:scaling:sin}), or by taking a convex combination of permutation matrices whose edges lie in $E$
(Birkhoff–von Neumann on the bipartite expansion of $G$, see Section~\ref{sec:sec:BvN}). In all cases, $\mathrm P^{\mathrm{ds}}$
satisfies $\mathrm P^{\mathrm{ds}}\mathbf{1}=\mathbf{1}$ and $(\mathrm P^{\mathrm{ds}})^\top\mathbf{1}=\mathbf{1}$.

\item \textbf{Column-stochastic (CS) mixer.}
Less common in practice; CS alone preserves budget balance but does \emph{not}
guarantee per-agent convexity (Jensen) or individual mean preservation.
\end{enumerate}

\begin{proposition}[Stochasticity of $M(\lambda)$]
\label{prop:Mlambda-stoch}
If $\mathrm P$ is RS (resp.\ CS, DS), then $\mathrm M(\lambda)=(1-\lambda)\mathbb{I}+\lambda \mathrm P$ is also RS
(resp.\ CS, DS) for every $\lambda\in[0,1]$.
\end{proposition}

\begin{proof}
Immediate: row (resp.\ column) sums are convex combinations of those of $\mathbb{I}$ and $\mathrm P$.
If both row and column sums are $1$ for $\mathrm P$, they remain $1$ for $\mathrm M(\lambda)$.
\end{proof}

\begin{remark}[Why this matters]
RS ensures each $\xi_i(\lambda)$ is a convex combination of neighbors and self, enabling
Jensen’s inequality and per-agent mean preservation (under identical means). CS ensures
budget balance. DS guarantees both and unlocks the strongest dominance results
(componentwise convex order, trace contraction).
\end{remark}

\subsection{Numerical Illustration}

\begin{figure}[!ht]
    \centering
    \includegraphics[width=0.475\linewidth]{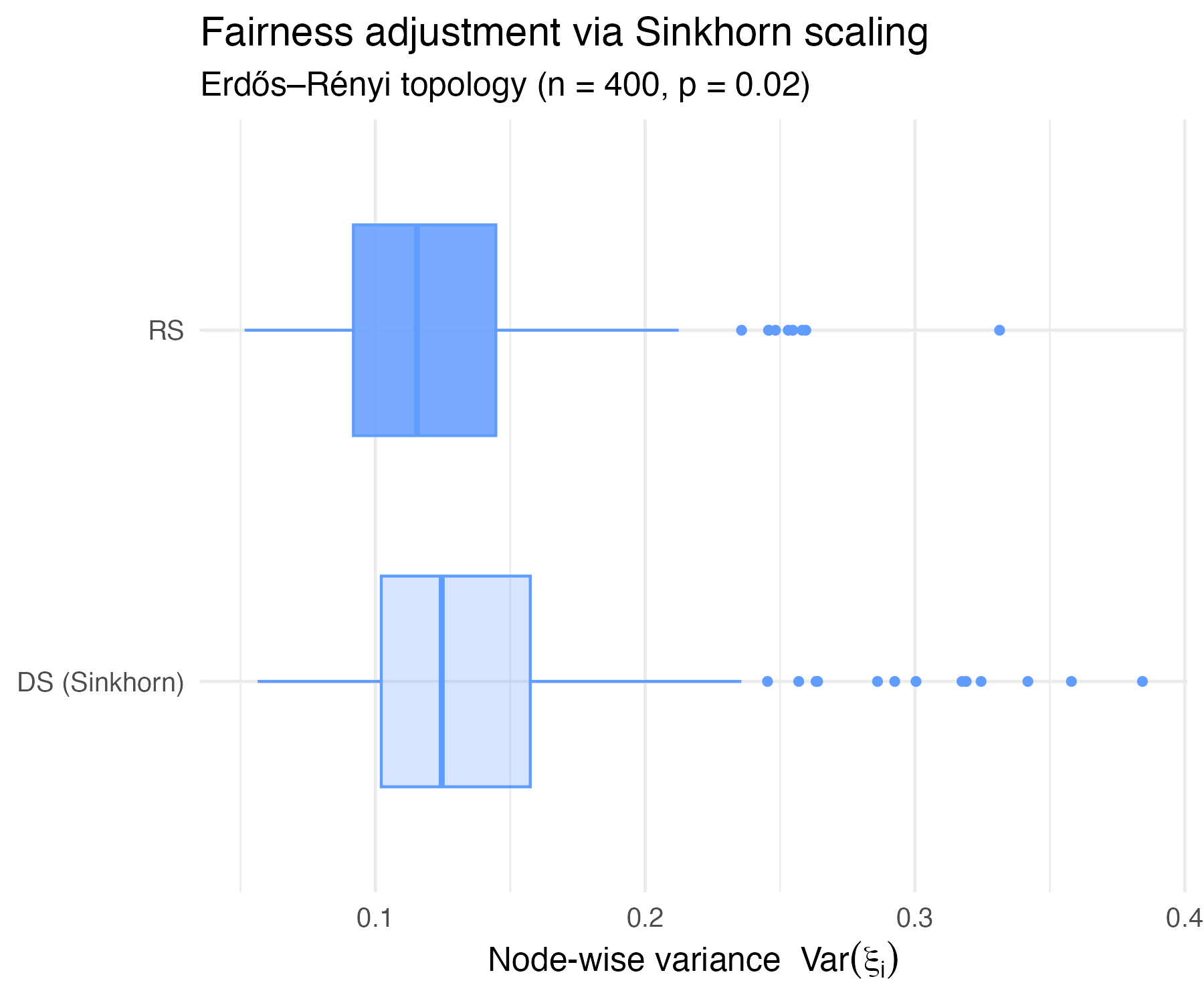}
    \includegraphics[width=0.475\linewidth]{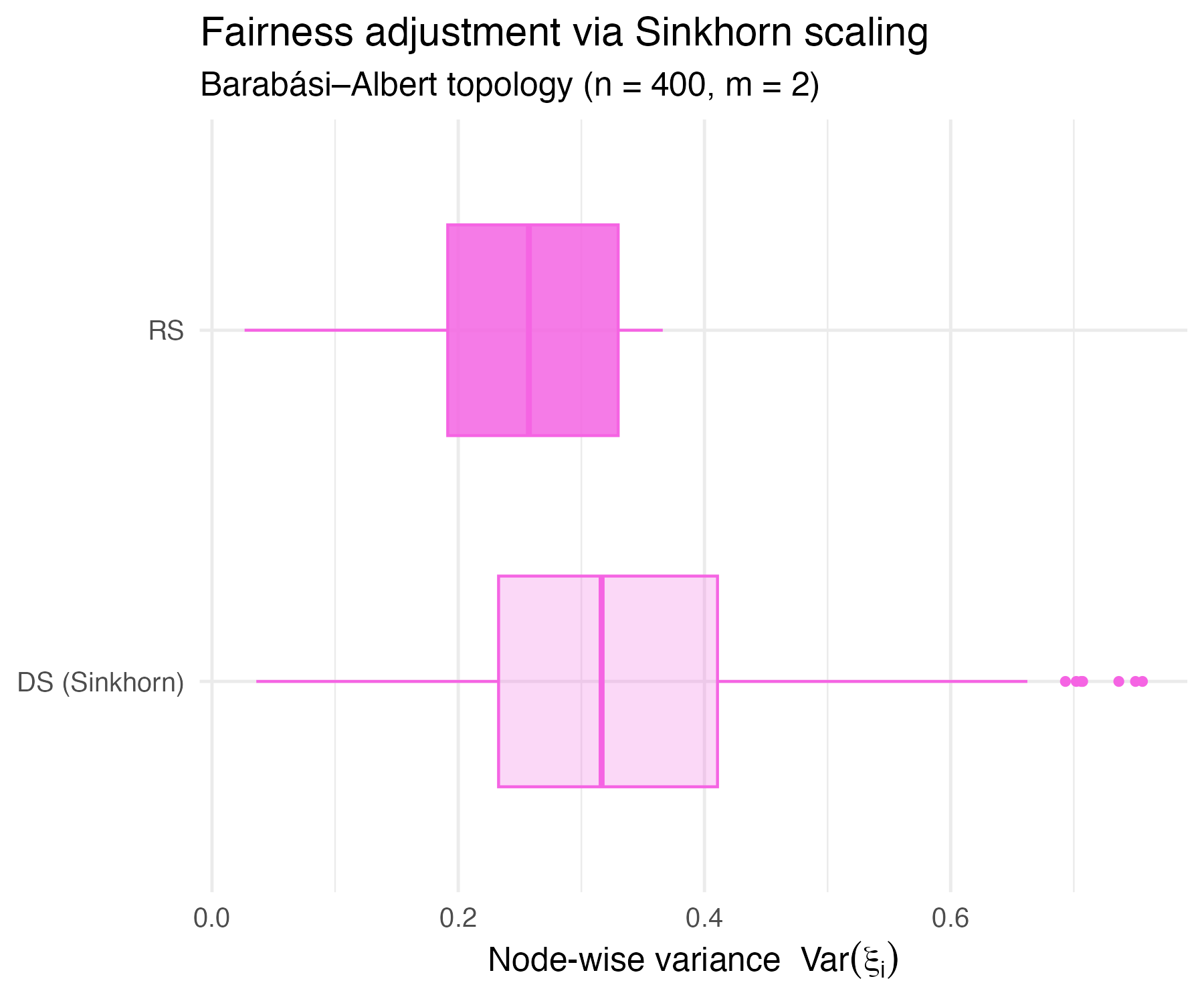}
\caption{Effect of Sinkhorn scaling on the distribution of node-wise variances.
Each panel compares row-stochastic (RS) and doubly-stochastic (DS) allocations obtained via Sinkhorn scaling.
Left: Erdős–Rényi graph with $n=400$ and edge probability $p=0.02$.
Right: Barabási–Albert graph with $n=400$ and $m=2$.
For both topologies, Sinkhorn scaling reduces the dispersion of $\mathrm{Var}(\xi_i)$ across nodes,
leading to a fairer allocation of risk.}

    \label{fig:RS:DS}
\end{figure}

Figure~\ref{fig:RS:DS} illustrates how fairness adjustments based on Sinkhorn scaling
affect the heterogeneity of node-wise variances. 
In each case, the baseline (RS) corresponds to a row-stochastic sharing rule, 
while the adjusted scheme (DS) enforces both row and column stochasticity.
For Erdős–Rényi graphs, the adjustment slightly narrows the distribution of $\mathrm{Var}(\xi_i)$,
whereas in Barabási–Albert graphs, where heterogeneity is more pronounced due to hubs,
the reduction in variance dispersion is more visible.
This highlights how Sinkhorn scaling acts as a fairness correction mechanism,
mitigating structural imbalances induced by the underlying network topology.

\subsection{Variance and its monotone response to $\lambda$}
\label{subsec:variance-lambda}

Let $\boldsymbol{X}=(X_1,\dots,X_n)^\top$ be i.i.d.\ with mean $\mu$ and variance $\sigma^2<\infty$,
independent across agents. Set $\boldsymbol \xi(\lambda)=\mathrm M(\lambda)\boldsymbol X$.

\begin{proposition}[Variance formula under i.i.d.\ risks]
\label{prop:var-formula}
With $\Sigma=\mathrm{Var}[\boldsymbol{X}]=\sigma^2 I$, we have
\[
\mathrm{Var}[\boldsymbol \xi(\lambda)] = \sigma^2\,\mathrm M(\lambda)\mathrm M(\lambda)^\top,
~ 
\mathrm{Var}\big(\xi_i(\lambda)\big) = \sigma^2\,\big\| \mathrm{M}_i(\lambda)\big\|_2^2,
\]
where $\mathrm{M}_i(\lambda)$ is the $i$-th row of $M(\lambda)$.
\end{proposition}

\begin{proof}
$\mathrm{Var}[\mathrm{M}\boldsymbol{X}]=\mathrm M\,\Sigma\,\mathrm M^\top=\sigma^2 \mathrm M\mathrm M^\top$; the $i$-th diagonal entry is the squared
$\ell_2$-norm of the $i$-th row.
\end{proof}

Write $\boldsymbol p_i$ for the $i$-th row of $\mathrm P$. Then
\[
\mathrm{M}_i(\lambda) = (1-\lambda)\boldsymbol e_i^\top + \lambda \boldsymbol p_i^\top,
~ 
\|\mathrm{M}_i(\lambda)\|_2^2
= (1-\lambda)^2 + 2\lambda(1-\lambda)\,p_{ii} + \lambda^2 \|\boldsymbol p_i\|_2^2.
\]

\begin{theorem}[Per-agent variance is nonincreasing in $\lambda$ for DS mixers]
\label{thm:monotone-variance-DS}
If $\mathrm P$ is DS, then for each $i$,
$\mathrm{Var}(\xi_i(\lambda))$ is nonincreasing in $\lambda\in[0,1]$ and
\[
\mathrm{Var}(\xi_i(1)) \;=\; \sigma^2\|p_i\|_2^2 \;\le\; \sigma^2,
\]
with strict inequality for any row $\boldsymbol p_i\neq \boldsymbol e_i^\top$.
\end{theorem}

\begin{proof}
Differentiate $g_i(\lambda)=\|\mathrm{M}_i(\lambda)\|_2^2$:
\[
g_i'(\lambda)=2(1-\lambda)(-1+p_{ii})+2\lambda(p_{ii}-\|p_i\|_2^2).
\]
For a DS row $\boldsymbol p_i$ (a probability vector), $0\le p_{ii}\le \|p_i\|_2^2 \le 1$.
Hence the first term is $\le0$ for $\lambda<1$, the second is $\le0$ for $\lambda>0$.
Thus $g_i'(\lambda)\le0$ on $(0,1)$, with strict negativity unless $\boldsymbol p_i=\boldsymbol e_i^\top$.
Multiply by $\sigma^2$ to obtain the claim for variances.
\end{proof}

\begin{corollary}[Best and worst cases]
\label{cor:best-worst}
For $\mathrm P=\mathrm J/n$ (complete equal-sharing), $\|p_i\|_2^2=1/n$ and
$\mathrm{Var}(\xi_i(1))=\sigma^2/n$. For $P=I$ (no mixing), $\mathrm{Var}(\xi_i(\lambda))=\sigma^2$.
\end{corollary}

\begin{remark}[RS-only mixers]
If $\mathrm P$ is RS only (e.g.\ $\mathrm P=\mathrm D^{-1}\mathrm A$), then $\mathrm{M}_i(\lambda)$ remains a convex combination,
so Jensen-based convex order statements can still be derived \emph{relative to a symmetric
benchmark}. However, per-agent variance need not be monotone for every $i$ without DS; what
does hold is a \emph{representative-agent} decrease for independent risks:
\[
\frac{1}{n}\mathrm{tr}\big(\mathrm{Var}[\xi(\lambda)]\big)
= \frac{\sigma^2}{n}\,\mathrm{tr}\big(\mathrm M(\lambda)\mathrm M(\lambda)^\top\big)
~ \text{is nonincreasing in $\lambda$ when $\mathrm P$ is column-stochastic as well.}
\]
If $\mathrm P$ is RS but not CS, budget balance can fail (Section~\ref{sec:convex order}).
\end{remark}

\subsection{Numerical Illustration}

\begin{figure}[!ht]
    \centering
    \includegraphics[width=0.975\linewidth]{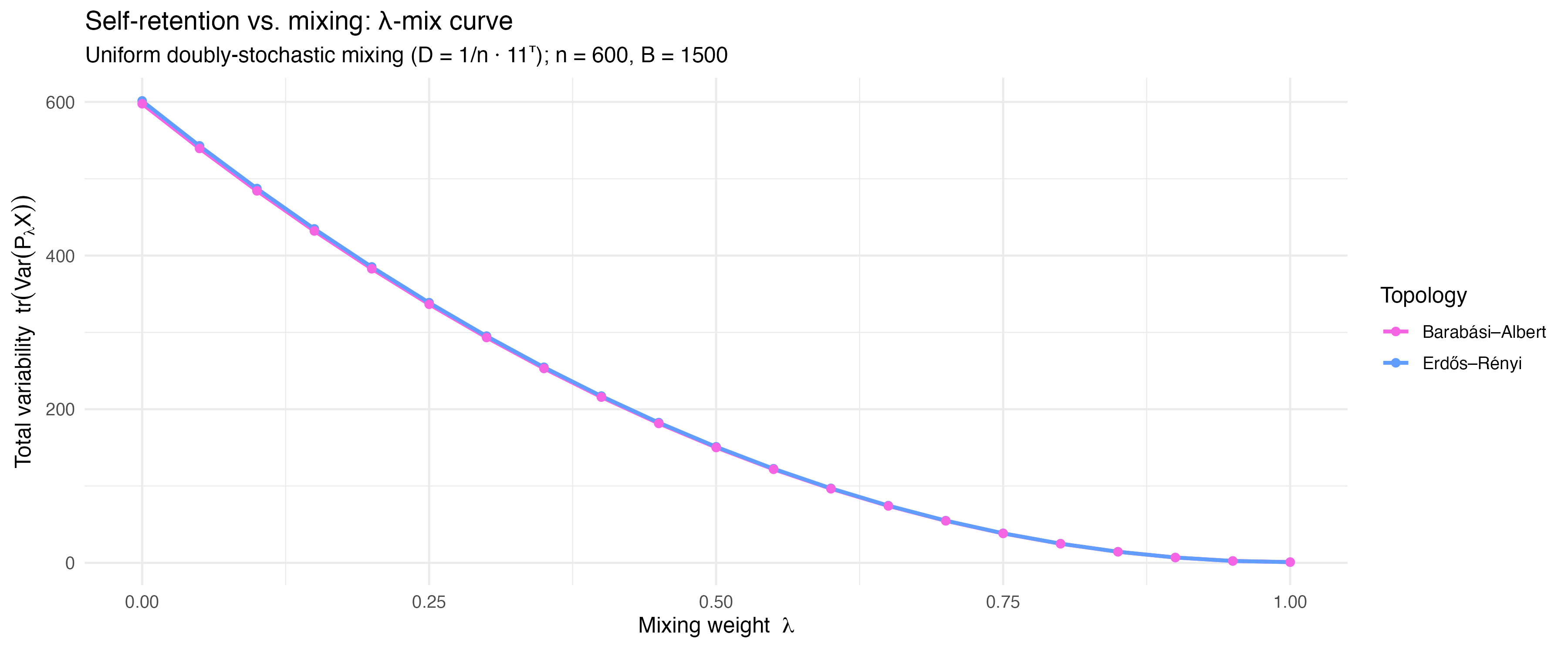}
\caption{Self-retention versus mixing: $\lambda$-mix curve.
The plot shows the total variability $\mathrm{tr}(\mathrm{Var}(\mathrm P_\lambda X))$
as a function of the mixing weight $\lambda \in [0,1]$, with
$\mathrm P_\lambda = (1-\lambda)\mathbb I + \lambda \mathrm D$, where $\mathrm D$ is the uniform
doubly-stochastic matrix.
Simulations are based on $n=600$ nodes and $B=1500$ replications,
for Erdős–Rényi and Barabási–Albert topologies.}

    \label{fig:SRvsmix}
\end{figure}

Figure~\ref{fig:SRvsmix}illustrates the trade-off between self-retention
and risk mixing through the $\lambda$-mix operator.
For $\lambda=0$, each agent fully retains its own risk, leading to maximal
aggregate variability.
As $\lambda$ increases, risk is progressively redistributed via the
uniform doubly-stochastic matrix, and the total variability
$\mathrm{tr}(\mathrm{Var}(\mathrm P_\lambda \boldsymbol X))$ decreases monotonically.
Remarkably, the curves for Erdős–Rényi and Barabási–Albert networks are almost
indistinguishable, showing that under uniform mixing the specific topology
plays no role in aggregate risk reduction: only the mixing weight $\lambda$ matters.

\subsection{Convex order and “more mixing”}
\label{subsec:cx-more-mixing}

\begin{theorem}[convex order improvement for DS mixers]
\label{thm:cx-DS}
Assume i.i.d.\ $X_i$ (finite mean). If $\mathrm P$ is DS, then for each $i$ and every $\lambda>0$,
\[
\xi_i(\lambda) \;\preceq_{CX}\; X_i,
\]
with strict improvement unless $\boldsymbol p_i=\boldsymbol e_i^\top$ (no off-diagonal weight).
Moreover, if $0\le \lambda_1<\lambda_2\le1$, then
$\boldsymbol \xi(\lambda_2)\preceq_{CCX}\boldsymbol \xi(\lambda_1)$.
\end{theorem}

\begin{proof}
Since rows of $\mathrm M(\lambda)$ are probability vectors (RS) and $\boldsymbol X$ are i.i.d.,
Lemma~\ref{lem:prob-weight-cx} (in Section~\ref{sec:convex order}) gives $\mathrm{M}_i(\lambda)\boldsymbol X \preceq_{CX} X_1 \stackrel{d}{=} X_i$.
Monotonicity in $\lambda$ follows from writing
$\mathrm M(\lambda_2)=\mathrm D\,\mathrm M(\lambda_1)$ with $\mathrm D\displaystyle=(1-\frac{\lambda_2-\lambda_1}{1-\lambda_1})\mathbb I
+ \frac{\lambda_2-\lambda_1}{1-\lambda_1}\mathrm P$, which is DS when $\mathrm P$ is DS; then apply the
``more mixing'' result of Proposition~\ref{prop:postmix} (from Section~\ref{sec:convex order}).
\end{proof}

\subsection{Incentives: why the $\lambda$-mix helps everyone (DS case)}
\label{subsec:incentives}

The tokenization of risk and reward in distributed insurance introduces new forms of sharing matrices in which investors act via stake and governance mechanisms \citep{FengLi2023Distributed}.

Suppose each agent has a concave, increasing Bernoulli utility $u$ with
$\mathbb{E}|u(X_i)|<\infty$. Under DS and i.i.d.\ $X$, Theorem~\ref{thm:cx-DS}
implies $\xi_i(\lambda)\preceq_{CX} X_i$, hence
$\mathbb{E}[u(\xi_i(\lambda))] \ge \mathbb{E}[u(X_i)]$ for every $i$:
\emph{any} $\lambda>0$ is a (weak) Pareto improvement. In particular,
the certainty equivalent $CE_i(\lambda)$ increases in $\lambda$.

For small $\lambda$, a first-order condition makes this explicit.

\begin{proposition}[Local incentive to increase $\lambda$]
\label{prop:local-incentive}
Assume $u$ is twice continuously differentiable with $u''<0$ and
$\mathbb{E}[X_i]=\mu$, $\mathrm{Var}(X_i)=\sigma^2$. Then for each $i$,
\[
\left.\frac{d}{d\lambda}\,\mathrm{Var}\big(\xi_i(\lambda)\big)\right|_{\lambda=0}
= 2\sigma^2\,(p_{ii}-1) \;<\;0
~ \text{whenever }p_{ii}<1,
\]
so infinitesimal mixing strictly reduces variance and raises the certainty equivalent.
\end{proposition}

\begin{proof}
Differentiate $\|\mathrm{M}_i(\lambda)\|_2^2$ at $\lambda=0$:
$g_i'(0)=2(-1+p_{ii})$. Multiply by $\sigma^2$. Since $p_{ii}\le1$
with equality if and only if $\boldsymbol p_i=\boldsymbol e_i^\top$, the derivative is strictly negative otherwise.
\end{proof}

Heuristically, the identity component preserves individual incentives: at $\lambda=0$, no one is worse off,
and any small move towards the DS mixer strictly lowers \emph{each} agent’s variance (unless
the mixer leaves them untouched). The $\lambda$-knob thus aligns the “collective good”
(diversification) with individual incentives.

\subsection{What if the mixer is adjacency-based?}
\label{subsec:adjacency}

Raw adjacency $\mathrm A$ is neither RS nor CS. But two practical normalizations can be considered:

\begin{enumerate}
\item \textbf{Random-walk (RS) normalization:} $\mathrm P=\mathrm D^{-1}\mathrm A$ (optionally lazy).
Then $\mathrm M(\lambda)$ is RS for all $\lambda$, hence each $\xi_i(\lambda)$ is a convex
combination of own loss and neighbors. Budget balance may fail (CS not guaranteed).

\item \textbf{Doubly-stochastic (DS) projection on support:}
Use Sinkhorn scaling or a convex combination of support-respecting permutation matrices
to obtain a DS $\mathrm P$ on the graph. Then $\mathrm M(\lambda)$ is DS for all $\lambda$, and
Theorems~\ref{thm:monotone-variance-DS}–\ref{thm:cx-DS} apply.
\end{enumerate}

\begin{proposition}[Representative-agent guarantee under independent risks]
\label{prop:rep-agent-lambda}
Let $\boldsymbol X$ have independent coordinates with variance $\sigma^2$.
If $\mathrm P$ is CS (hence $\mathrm M(\lambda)$ is CS by Proposition~\ref{prop:Mlambda-stoch}),
then the representative-agent variance
\[
\mathrm{Var}\big(\boldsymbol \xi'(\lambda)\big)\;=\;\frac{1}{n}\,\mathrm{tr}\big(\mathrm{Var}[\boldsymbol \xi(\lambda)]\big)
\]
is nonincreasing in $\lambda$.
\end{proposition}

\begin{proof}
Write $\boldsymbol \xi(\lambda_2)=\mathrm D \boldsymbol \xi(\lambda_1)$ for $\lambda_2>\lambda_1$ with
$\mathrm D=(1-\theta)\mathbb{I}+\theta \mathrm P$ and $\theta=(\lambda_2-\lambda_1)/(1-\lambda_1)$.
If $\mathrm P$ is CS, then so is $\mathrm D$. For independent risks,
trace arguments as in Section~3 (representative-agent variance) give the decrease.
\end{proof}

\subsection{Summary and design guidance}

\begin{itemize}
\item \textbf{Take $\mathrm M(\lambda)=(1-\lambda)\mathbb I+\lambda\mathrm  P$ with $\mathrm P$ DS on the graph support.}
Then $\mathrm M(\lambda)$ is DS for all $\lambda$, every agent enjoys convex order improvement
and variance strictly decreases with $\lambda$ (Theorems~\ref{thm:monotone-variance-DS},
\ref{thm:cx-DS}), with bounds interpolating from $\sigma^2$ to $\sigma^2\|p_i\|_2^2$.

\item \textbf{If only RS is feasible (random-walk normalization),} $\mathrm M(\lambda)$ is RS. One retain
per-agent convex-combination structure and individual mean preservation, but budget balance is
not automatic; representative-agent risk still falls with $\lambda$ under independence.

\item \textbf{Identity mixing creates incentives.} At $\lambda=0$, no one is harmed; for any agent
actually mixed by $\mathrm P$ ($p_{ii}<1$), variance falls to first order in $\lambda$ (Proposition~\ref{prop:local-incentive}).

\item \textbf{Complete pooling as the limit case.} For $\mathrm P=\mathrm J/n$ one recovers the classical $1/n$
variance law at $\lambda=1$. Intermediate $\lambda$ yields a smooth path between autarky and full pool.
\end{itemize}



\subsection{Quadratic structure, optimal $\lambda$, and spectral formulas}
\label{subsec:quadratic-optimal-lambda}

Let $\mathrm M(\lambda)=(1-\lambda)\mathbb I+\lambda \mathrm P$ with $0\le \lambda \le 1$, where $P\ge 0$
respects the network support and is RS/CS/DS as specified. Assume first that
$X_1,\dots,X_n$ are independent with $\mathbb{E}[X_i]=\mu$ and
$\mathrm{Var}(X_i)=\sigma^2<\infty$, so $\Sigma=\mathrm{Var}[\boldsymbol{X}]=\sigma^2 I$.

\begin{lemma}[Exact quadratic form for individual variance]
\label{lem:indiv-quadratic}
Let $\boldsymbol p_i$ be the $i$-th row of $\mathrm P$, $a_i:=\|\boldsymbol p_i\|_2^2$ and $d_i:=p_{ii}$.
Then
\[
\mathrm{Var}\big(\xi_i(\lambda)\big)
= \sigma^2 \Big( (1-\lambda)^2 + 2\lambda(1-\lambda)d_i + \lambda^2 a_i \Big)
= \sigma^2\Big( 1 + 2\lambda(d_i-1) + \lambda^2(1+a_i-2d_i) \Big).
\]
Moreover,
\[
\frac{d^2}{d\lambda^2}\mathrm{Var}\big(\xi_i(\lambda)\big)
= 2\sigma^2 \big(1+a_i-2d_i\big)
= 2\sigma^2\big[(1-d_i)^2 + (a_i-d_i^2)\big]\;\ge\;0,
\]
so $\mathrm{Var}(\xi_i(\lambda))$ is convex in $\lambda$ for every $i$.
\end{lemma}

\begin{proof}
See Appendix~\ref{app:lem:indiv-quadratic}
\end{proof}

\begin{lemma}[Aggregate (= representative-agent) variance is a convex quadratic]
\label{lem:trace-quadratic}
Let $\Vert\cdot\Vert_F$ be the Frobenius norm. Then
\[
\frac{1}{n}\mathrm{tr}\big(\mathrm{Var}\big[\boldsymbol\xi(\lambda)\big]\big)
=\frac{\sigma^2}{n}\,\mathrm{tr}\Big( \mathrm M(\lambda)\mathrm M(\lambda)^\top\Big)
= \frac{\sigma^2}{n}\Big\{ n + 2\lambda\big(\mathrm{tr}(\mathrm P)-n\big)
+ \lambda^2 \Vert \mathrm P-\mathbb I\Vert_F^2\Big\}.
\]
In particular, the coefficient of $\lambda^2$ is $\displaystyle\frac{\sigma^2}{n}\,\Vert \mathrm P-\mathbb{I}\Vert_F^2\ge 0$,
so the representative-agent variance is convex in $\lambda$.
\end{lemma}

\begin{proof}
Expand $\mathrm M(\lambda)\mathrm M(\lambda)^\top=(1-\lambda)^2 \mathbb I+\lambda(1-\lambda)(\mathrm P+\mathrm P^\top)+\lambda^2 \mathrm P\mathrm P^\top$,
take traces, use $\mathrm{tr}(\mathrm P\mathrm P^\top)=\Vert \mathrm P\Vert_F^2$ and
$\Vert \mathrm P-\mathbb I\Vert_F^2=\Vert\mathrm  P\Vert_F^2+\Vert \mathbb I\Vert_F^2-2\,\mathrm{tr}(\mathrm P)
=\Vert \mathrm P\Vert_F^2+n-2\,\mathrm{tr}(\mathrm P)$.
\end{proof}

\begin{theorem}[Social-planner optimal $\lambda^\star$ for the representative agent]
\label{thm:lambda-star-trace}
Assume independence and equal variances $\sigma^2$. The value of $\lambda$
that minimizes the representative-agent variance (equivalently,
$\displaystyle\frac{1}{n}\mathrm{tr}\big(\mathrm{Var}[\boldsymbol\xi(\lambda)]\big)$) is
\[
\lambda^\star_{\mathrm{RA}}
\;=\; \Pi_{[0,1]}\!\left(\frac{n-\mathrm{tr}(\mathrm P)}{\Vert \mathrm P-\mathbb I\Vert_F^2}\right),
\]
where $\Pi_{[0,1]}$ denotes Euclidean projection onto $[0,1]$.
\end{theorem}

\begin{proof}
Differentiate the quadratic in Lemma~\ref{lem:trace-quadratic} and set to zero:
$2(\mathrm{tr}(\mathrm P)-n)+2\lambda \Vert \mathrm P-\mathbb I\Vert_F^2=0$, whence the ratio. Project to $[0,1]$.
\end{proof}

\begin{remark}
If $\mathrm P=\mathbb{I}$ then $\Vert \mathrm P-\mathbb I\Vert_F^2=0$ and every $\lambda$ is optimal (variance is constant).
If $\mathrm P=\mathrm J/n$ (complete equal-share), then $\mathrm{tr}(\mathrm P)=1$,
$\Vert \mathrm P-\mathbb I\Vert_F^2=n-1$, so $\lambda^\star_{\mathrm{RA}}=1$.
\end{remark}

\paragraph{Weights/heterogeneous risk aversion.}
Let a planner minimize the weighted sum $\displaystyle\sum_{i=1}^n w_i\,\mathrm{Var}(\xi_i(\lambda))$
with $w_i>0$. Using Lemma~\ref{lem:indiv-quadratic}:

\begin{proposition}[Closed form for weighted-variance optimum]
\label{prop:weighted-opt}
Let $a_i=\|p_i\|_2^2$ and $d_i=p_{ii}$. Then
\[
\lambda^\star_{w}
\;=\; \Pi_{[0,1]}\!\left(
\frac{\sum_{i=1}^n w_i (1-d_i)}{\sum_{i=1}^n w_i \big(1+a_i-2d_i\big)}
\right).
\]
\end{proposition}

\begin{proof}
The objective is a quadratic in $\lambda$ with linear coefficient
$2\sigma^2\sum_i w_i(d_i-1)$ and quadratic coefficient
$\sigma^2\sum_i w_i(1+a_i-2d_i)\ge 0$. Take derivative, set to zero, project to $[0,1]$.
\end{proof}

\paragraph{Spectral form under symmetric DS.}
Suppose $\mathrm P$ is \emph{symmetric} and DS. Then $\mathrm P$ is diagonalizable with
real eigenvalues $1=\mu_1\ge \mu_2\ge \cdots \ge \mu_n\ge -1$ and an orthonormal basis (by the spectral theorem for real symmetric matrices).
We get a clean eigenvalue formula.

\begin{theorem}[Eigenvalue representation and optimal $\lambda$ (symmetric DS)]
\label{thm:spectral}
If $\mathrm P$ is symmetric DS, then
\[
\frac{1}{n}\mathrm{tr}\big(\mathrm{Var}\big[\boldsymbol\xi(\lambda)\big]\big)
=\frac{\sigma^2}{n}\sum_{k=1}^n \big(1-\lambda+\lambda \mu_k\big)^2.
\]
The minimizer is
\[
\lambda^\star_{\mathrm{RA}}
=\Pi_{[0,1]}\!\left(
\frac{\sum_{k=2}^n (1-\mu_k)}{\sum_{k=2}^n (1-\mu_k)^2}
\right),
\]
(where the $k=1$ term cancels since $1-\mu_1=0$). Moreover
$0\le \lambda^\star_{\mathrm{RA}}\le 1$, with $\lambda^\star_{\mathrm{RA}}=1$
when $\mu_2,\ldots,\mu_n=0$ (complete pooling), and $\lambda^\star_{\mathrm{RA}}\to 0$
as $\mathrm P\to \mathbb I$.
\end{theorem}

\begin{proof}
See Appendix~\ref{app:thm:spectral}.
\end{proof}

\subsection{Correlated risks: equicorrelation and general covariance}
\label{subsec:correlated}

We now allow dependence. Let $\Sigma=\mathrm{Var}[\boldsymbol{X}]$ be any PSD matrix.

\begin{lemma}[Trace identity for general covariance]
\label{lem:trace-general}
For any $\Sigma\succeq \mathrm 0$,
\[
\mathrm{tr}\big(\mathrm{Var}\big[\boldsymbol\xi(\lambda)\big]\big)
=\mathrm{tr}\Big( \mathrm M(\lambda)\Sigma \mathrm M(\lambda)^\top\Big)
= (1-\lambda)^2 \mathrm{tr}(\Sigma)
+ 2\lambda(1-\lambda)\,\mathrm{tr}(\mathrm P\Sigma)
+ \lambda^2\,\mathrm{tr}(\mathrm P\Sigma \mathrm P^\top).
\]
\end{lemma}

\begin{proof}
Expand $\mathrm M(\lambda)$, use bilinearity of the trace and $\mathrm{tr}(AB)=\mathrm{tr}(BA)$.
\end{proof}

A particularly transparent case is \emph{equicorrelation}:
$\Sigma=\sigma^2\big[(1-\rho)\mathbb I+\rho \mathbf{1}\mathbf{1}^\top\big]$ with $-1/(n-1)\le \rho<1$.

\begin{proposition}[Individual variance under equicorrelation]
\label{prop:equicorr}
Assume $\mathrm M(\lambda)$ is RS. Then the $i$-th variance is
\[
\mathrm{Var}\big(\xi_i(\lambda)\big)
= \sigma^2\Big( (1-\rho)\,\|\mathrm{M}_i(\lambda)\|_2^2 + \rho \Big).
\]
In particular, mixing only reduces the idiosyncratic component by shrinking
$\|\mathrm{M}_i(\lambda)\|_2^2$; the common component $\rho\sigma^2$ is irreducible by LRS.
\end{proposition}

\begin{proof}
For any weight vector $\boldsymbol w$ summing to $1$,
$\boldsymbol w^\top \Sigma \boldsymbol w = \sigma^2\big[(1-\rho)\|\boldsymbol w\|_2^2 + \rho (\boldsymbol w^\top \mathbf{1})^2\big]
= \sigma^2\big[(1-\rho)\|\boldsymbol w\|_2^2 + \rho\big]$.
Take $\boldsymbol w=\mathrm{M}_i(\lambda)$; RS ensures $\boldsymbol w^\top\mathbf{1}=1$.
\end{proof}

\begin{corollary}[Planner’s optimal $\lambda$ with equicorrelation and symmetric DS]
\label{cor:equicorr-opt}
Let $\mathrm P$ be symmetric DS with eigenvalues $1=\mu_1\ge\mu_2\ge\cdots\ge\mu_n$ and
$\Sigma$ equicorrelated as above. Then
\[
\frac{1}{n}\mathrm{tr}\big(\mathrm{Var}\big[\boldsymbol \xi(\lambda)\big]\big)
= \sigma^2\left\{ \rho + (1-\rho)\,\frac{1}{n}\sum_{k=1}^n \big(1-\lambda+\lambda \mu_k\big)^2 \right\}.
\]
The minimizer is the same $\lambda^\star_{\mathrm{RA}}$ as in Theorem~\ref{thm:spectral}
(the $\rho$ term factors out), showing that correlated risk does not change the
optimal $\lambda$—it only creates an irreducible floor in variance.
\end{corollary}

\begin{proof}
Plug the spectral form of $\mathrm M(\lambda)$ into $\boldsymbol w^\top \Sigma\boldsymbol  w$ aggregated over $i$,
or sum Proposition~\ref{prop:equicorr} over $i$ and use orthogonality of the eigenbasis.
\end{proof}

\subsection{“More mixing” as a contraction: matrix norms and bounds}
\label{subsec:norm-bounds}

\begin{lemma}[Frobenius norm bound and monotonicity of the trace under DS post-mixing]
\label{lem:frobenius}
Let $\mathrm D$ be DS. For any PSD $\Sigma$ and any $\mathrm{M}$,
$\mathrm{tr}\big(\mathrm D\mathrm M\,\Sigma\,(\mathrm D\mathrm M)^\top\big) \le \mathrm{tr}\big(\mathrm M\Sigma \mathrm M^\top\big)$.
In particular, taking $\mathrm M(\lambda_2)=\mathrm D \mathrm M(\lambda_1)$ with
$\mathrm D=(1-\theta)\mathbb I+\theta \mathrm P$ DS (any $\theta\in[0,1]$) yields
$\mathrm{tr}\big(\mathrm{Var}[\boldsymbol \xi(\lambda_2)]\big)\le \mathrm{tr}\big(\mathrm{Var}[\boldsymbol \xi(\lambda_1)]\big)$.
\end{lemma}

\begin{proof}
Consider $h(\boldsymbol Z)=\mathrm{tr}(\boldsymbol Z\Sigma \boldsymbol Z^\top)$; $h$ is convex in $\boldsymbol Z$ for $\Sigma\succeq \mathrm 0$.
By Birkhoff–von Neumann, $\mathrm D=\displaystyle\sum_r \omega_r \mathrm P_r$ (permutations). Hence
$h(\mathrm D\mathrm M)\le \displaystyle\sum_r \omega_r h(\mathrm P_r \mathrm M)=\sum_r \omega_r h(\mathrm M)=h(\mathrm M)$
since $h$ is invariant under left multiplication by permutations.
\end{proof}

\paragraph{Rowwise bounds (DS).}
When $\mathrm P$ is DS, each $\boldsymbol p_i$ is a probability vector, so
$1/n \le \|\boldsymbol p_i\|_2^2 \le 1$ and $0\le d_i\le 1$.
Thus, by Lemma~\ref{lem:indiv-quadratic},
\[
\sigma^2\Big( (1-\lambda)^2 + \frac{\lambda^2}{n} \Big)
\;\le\; \mathrm{Var}(\xi_i(\lambda))
\;\le\; \sigma^2\Big( (1-\lambda)^2 + 2\lambda(1-\lambda) + \lambda^2 \Big) = \sigma^2.
\]
The lower bound is attained by $\mathrm P=\mathrm J/n$ (equal mixing), the upper by $\mathrm P=\mathbb I$.

\subsection{Calibration and interpretation}
\label{subsec:calibration}

The quadratic formulas above make calibration straightforward.

\begin{itemize}
\item \textbf{Planner’s $\lambda^\star$.} Use Theorem~\ref{thm:lambda-star-trace}
or its spectral form (Theorem~\ref{thm:spectral}). If one can estimate
$\mathrm{tr}(\mathrm P)$ and $\Vert \mathrm P-\mathbb I\Vert_F$ (or the spectrum of a symmetric DS $\mathrm P$),
one obtain $\lambda^\star$ in closed form.

\item \textbf{Agent-weighted $\lambda^\star$.} If a regulator wants to penalize high-variance
agents more, pick weights $w_i$ and use Proposition~\ref{prop:weighted-opt}.

\item \textbf{Correlation-robustness.} With equicorrelation, the \emph{optimal} $\lambda$
is unchanged (Corollary~\ref{cor:equicorr-opt}); only the irreducible variance floor rises
with $\rho$. In fully general $\Sigma$, Lemma~\ref{lem:trace-general} keeps the objective quadratic.

\item \textbf{Incentives.} Proposition~\ref{prop:local-incentive} already shows
each agent strictly benefits for small $\lambda$ whenever $p_{ii}<1$.
The convexity of individual variances (Lemma~\ref{lem:indiv-quadratic})
implies diminishing marginal reductions as $\lambda$ approaches $1$.
\end{itemize}

\subsection{Worked Examples and Calibration}
\label{subsec:worked-examples}

We illustrate the formulas with three canonical mixers $\mathrm P$:
(i) complete equal-sharing, (ii) ring (self+two neighbors), and
(iii) adjacency-based random-walk on an Erd\H{o}s--R\'enyi graph.
Unless noted, $X_i$ are i.i.d.\ with mean $\mu$ and variance $\sigma^2$.

\subsubsection{Complete equal-sharing ($\mathrm P=\mathrm J/n$)}

Let $\mathrm P=\displaystyle\frac{1}{n}\mathbf{1}\mathbf{1}^\top$. Then $\mathrm P$ is symmetric DS with spectrum
\(\{\,\mu_1=1,\;\mu_2=\cdots=\mu_n=0\,\}\).
From Theorem~\ref{thm:spectral},
\[
\frac{1}{n}\mathrm{tr}\big(\mathrm{Var}[\boldsymbol\xi(\lambda)]\big)
= \frac{\sigma^2}{n}\Big((1-\lambda+\lambda)^2 + \sum_{k=2}^n (1-\lambda)^2\Big)
= \sigma^2\Big(\frac{1}{n} + (1-\lambda)^2 \frac{n-1}{n}\Big).
\]
The minimizer is \(\lambda^\star_{\mathrm{RA}}=1\) and
\(\mathrm{Var}(\xi_i(1))=\sigma^2/n\) for all \(i\) (Corollary~\ref{cor:best-worst}).

\medskip
\noindent\emph{Interpretation.} As mixing strength \(\lambda\uparrow 1\),
variance shrinks to the classical \(1/n\) law; incentives are fully aligned
(Proposition~\ref{prop:local-incentive}) and fairness is perfect.

\subsubsection{Ring (cycle) with self+two neighbours ($\mathrm  P$ circulant DS)}

Take the $n$-cycle and set
\[
\mathrm P_{ii}=\frac{1}{3},~  \mathrm P_{i,i\pm1}=\frac{1}{3}~ (\text{mod }n),~  \mathrm P_{ij}=0\text{ otherwise}.
\]
Then $\mathrm P$ is symmetric DS and circulant with eigenvalues (Eq. (3.11) in \cite{gray2006toeplitz})
\[
\mu_k \;=\; \frac{1+2\cos\big(2\pi(k-1)/n\big)}{3},~  k=1,\dots,n.
\]
Using Theorem~\ref{thm:spectral},
\[
\lambda^\star_{\mathrm{RA}}
=\Pi_{[0,1]}\!\left(
\frac{\sum_{k=2}^n (1-\mu_k)}{\sum_{k=2}^n (1-\mu_k)^2}
\right).
\]
We can evaluate these sums in closed form (by trace/Frobenius identities):
\(\displaystyle\sum_{k=1}^n \mu_k = \mathrm{tr}(\mathrm P) = n/3\) and
\(\displaystyle\sum_{k=1}^n \mu_k^2 = \mathrm{tr}(\mathrm P^2)=\Vert \mathrm P\Vert_F^2 = n\cdot 3\cdot(1/3)^2 = n/3\).
Since \(\mu_1=1\), we get
\[
\sum_{k=2}^n (1-\mu_k) = (n-1) - \Big(\frac{n}{3}-1\Big) = \frac{2n}{3},
~ 
\sum_{k=2}^n (1-\mu_k)^2
= (n-1) - 2\Big(\frac{n}{3}-1\Big) + \Big(\frac{n}{3}-1\Big) = \frac{2n}{3}.
\]
Hence \(\lambda^\star_{\mathrm{RA}}=1\). Individual variances at \(\lambda=1\) equal
\(\sigma^2\|p_i\|_2^2=\sigma^2/3\) (each row has three entries of \(1/3\)).

\medskip
\noindent\emph{Interpretation.} Even with local sharing only, the planner’s variance
objective prefers the maximal feasible mixing \(\lambda=1\), but the floor is
\(\sigma^2/3\) (locality limits diversification relative to complete pooling).

\subsubsection{Erd\H{o}s--R\'enyi random-walk mixer (RS, typically not DS)}

Let \(G\sim G(n,p)\) with degrees \(d_i\sim \mathrm{Bin}(n-1,p)\).
Consider the (non-lazy) random-walk normalization \(\mathrm P=\mathrm D^{-1}\mathrm A\) (set \(\mathrm P_{ii}=0\)).
Then \(\mathrm P\) is RS, generally not CS/DS, and (see Sections 2.2 and 2.3 in \cite{ChungLu2006})
\[
\mathrm{tr}(\mathrm P)=0, ~  \Vert \mathrm P\Vert_F^2 = \sum_{i=1}^n \sum_{j} \mathrm P_{ij}^2
= \sum_{i=1}^n d_i \cdot \frac{1}{d_i^2} = \sum_{i=1}^n \frac{1}{d_i}.
\]
From Lemma~\ref{lem:trace-quadratic}, for independent risks
\[
\frac{1}{n}\mathrm{tr}\big(\mathrm{Var}[\boldsymbol\xi(\lambda)]\big)
= \frac{\sigma^2}{n}\Big\{ n - 2\lambda n + \lambda^2\big(\Vert \mathrm P\Vert_F^2 + n\big)\Big\}.
\]
The unconstrained minimizer is
\[
\lambda^\star_{\mathrm{RA}}
= \frac{n}{\Vert \mathrm P\Vert_F^2 + n}
= \frac{1}{1 + \displaystyle\frac{1}{n}\Vert \mathrm P\Vert_F^2}
= \frac{1}{1 + \displaystyle\frac{1}{n}\sum_{i=1}^n \frac{1}{d_i}}.
\]
By Jensen’s inequality for the convex map \(x\mapsto 1/x\) on \((0,\infty)\),
\[
\frac{1}{n}\sum_{i=1}^n \frac{1}{d_i}
\;\ge\; \frac{1}{\displaystyle\frac{1}{n}\sum_i d_i}
\;=\; \frac{1}{(n-1)p}\,,
\]
whence the upper bound
\[
\lambda^\star_{\mathrm{RA}}
\;\le\; \frac{1}{1 + \displaystyle\frac{1}{(n-1)p}}
= \frac{(n-1)p}{1+(n-1)p}
= 1 - \frac{1}{1+(n-1)p}.
\]
In particular, for dense enough graphs with \((n-1)p\gg 1\),
\[
\lambda^\star_{\mathrm{RA}} \;=\; 1 - \Theta\!\Big(\frac{1}{(n-1)p}\Big),
\]
so the planner chooses a mixing level very close to \(1\).
A matching lower bound follows from concentration of \(d_i\) around \((n-1)p\) (e.g.,
Chernoff): with high probability, \(d_i\in[(1-\epsilon)(n-1)p,(1+\epsilon)(n-1)p]\),
which implies
\[
\frac{1}{n}\sum_i \frac{1}{d_i}
\;\le\; \frac{1}{(1-\epsilon)(n-1)p}\,
~ \Rightarrow~ 
\lambda^\star_{\mathrm{RA}}
\;\ge\; \frac{1}{1 + \displaystyle\frac{1}{(1-\epsilon)(n-1)p}}
= 1 - \Theta\!\Big(\frac{1}{(n-1)p}\Big).
\]

\medskip
\noindent\emph{Interpretation.}
Random-walk sharing on $G(n,p)$ already drives the planner’s optimal mixing
\(\lambda^\star_{\mathrm{RA}}\) to be near $1$ as soon as expected degree grows.
However, since $\mathrm P$ is not DS, per-agent monotonicity and universal convex order
improvements need not hold; the guarantees remain at the \emph{representative-agent}
and aggregate (trace) level (Lemmas~\ref{lem:trace-quadratic}, \ref{lem:frobenius}).
To recover the stronger DS guarantees, replace $\mathrm P$ by a DS scaling on the support (e.g.,
Sinkhorn) or apply a small global DS post-mixer $\mathrm D$ to $\mathrm M(\lambda)$ (Lemma~\ref{lem:frobenius}).

\subsubsection{Lazy symmetric normalization (approximate DS)}

If one prefers a symmetric operator, consider the lazy normalized adjacency
\[
\widehat {\mathrm P} \;=\; \frac{1}{2}\mathbb I + \frac{1}{2}\,\mathrm D^{-1/2} \mathrm A \mathrm D^{-1/2}.
\]
Then $\widehat{\mathrm P}$ is symmetric with spectrum in \([0,1]\) and
\(\widehat{\mathrm P}\mathbf{1}=\mathbf{1}\) if and only if the graph is regular (see Sections 2.2 and 2.3 in \cite{ChungLu2006}).
While $\widehat{\mathrm P}$ is not DS in general, Theorem~\ref{thm:spectral} still applies
formally to compute \(\lambda^\star_{\mathrm{RA}}\) once the eigenvalues are known
(or estimated); the optimizer remains near \(1\) when the second-largest eigenvalue
is bounded away from \(1\).

\section{Discussion and Conclusion}

This paper develops a unified theory of linear risk sharing on networks. By encoding sharing mechanisms as nonnegative matrices, we showed how row- and column-stochasticity govern fundamental properties: budget balance, mean preservation, and convex order dominance. Doubly stochastic operators guarantee the strongest results, ensuring fairness and variance reduction at both the individual and collective levels.

Our network-based analysis revealed that topology crucially shapes outcomes: complete graphs maximize diversification, stars highlight inequity, rings provide only local sharing, while random and scale-free networks display degree-driven heterogeneity. Introducing a
second layer of randomness, with the sharing matrix itself drawn from a distribution of graphs, restored approximate symmetry in homogeneous random graphs but preserved inequity in scale-free ones. The study of convex combinations of self-retention and network mixing provided both theoretical guarantees and incentive compatibility, yielding design guidance for peer-to-peer insurance and decentralized risk pooling.

Our framework is intentionally stylized: it focuses on variance and convex order dominance, rather than tail risk or systemic contagion. It assumes i.i.d.\ losses in many results, and does not model strategic behavior or endogenous network formation. Despite these simplifications, the analysis suggests clear policy implications. Regulators and designers of P2P platforms can
use doubly-stochastic corrections (e.g.\ Sinkhorn scaling or global post-mixing) to ensure fairness,
and can tune the $\lambda$-mix parameter to balance self-retention with collective diversification.
These tools help bridge mathematical tractability with the design of sustainable, equitable,
and transparent insurance mechanisms.

Future work may explore dynamic extensions, where both risks and networks evolve over
time, as well as applications to correlated losses, systemic risk, and regulatory design.
Another promising direction is the calibration of network-based LRS models to empirical
insurance or financial data, testing the predictive power of variance reductions and convex order
dominance in practice. Multi-period risk sharing, including intertemporal transfers and annuitization,
has been recently formalized in the spatio-temporal risk sharing literature, offering insight into how
allocations evolve under diverse fairness criteria \citep{FengLiu2025SpatioTemporal}.

\pagebreak

\section*{Funding}

This research was supported by the Natural Sciences and Engineering Research Council of Canada (NSERC) and the SCOR Foundation for Science. Grant numbers and additional details will be provided after the review process, to preserve authors anonymity.

\section*{CRediT authorship contribution statement}
Philipp Ratz: Conceptualization, Methodology, Formal analysis, Writing – original draft. 

\noindent Arthur Charpentier: Conceptualization, Methodology, Supervision, Writing – review \& editing.

\section*{Declaration of competing interest}
The authors declare that they have no known competing financial interests or personal
relationships that could have appeared to influence the work reported in this paper.

\section*{Ethics approval}
This study did not involve human participants or animals.

\section*{Data Availability Statement}
Data sharing not applicable to this article as no datasets were generated or analyzed. 

\section*{Declaration of generative AI and AI-assisted technologies in the manuscript preparation process}
During the preparation of this work, the authors employed ChatGPT-5 to refine the language and improve the readability of the manuscript, as none of the authors are native speakers of English. Following the use of this tool, the authors carefully reviewed and edited the content and take full responsibility for the final published version.

\

\pagebreak

\bibliographystyle{plainnat}
\bibliography{biblio}

\pagebreak

\appendix

\section{Appendices}

\subsection{Proof of Lemma \ref{lem:indiv-quadratic}}\label{app:lem:indiv-quadratic} 

\begin{proof}
    Write $\mathrm{M}_i(\lambda)=(1-\lambda)e_i^\top+\lambda p_i^\top$.
Since $\Sigma=\sigma^2I$, the $i$-th variance is $\sigma^2\|\mathrm{M}_i(\lambda)\|_2^2$,
and the identities follow by expanding $\|\mathrm{M}_i(\lambda)\|_2^2$ and
collecting terms. For convexity, note
$1+a_i-2d_i=(1-d_i)^2+(a_i-d_i^2)\ge 0$ by $a_i\ge d_i^2$.
\end{proof}

\subsection{Proof of Theorem \ref{thm:spectral}}\label{app:thm:spectral} 

\begin{proof}
Since $\mathrm P=\mathrm P^\top$, $\mathrm M(\lambda)$ diagonalizes in the same basis with eigenvalues
$1-\lambda+\lambda \mu_k$. The trace is the sum of squared eigenvalues, giving the formula.
Differentiate the sum and set to zero:
$\displaystyle\sum_{k=2}^n (1-\mu_k)(1-\lambda+\lambda\mu_k)=0$, which yields the ratio.
For the bounds, note $0\le 1-\mu_k\le 2$ and $(1-\mu_k)^2\ge (1-\mu_k)$,
so the ratio lies in $[0,1]$ unless the denominator is zero (which happens only for $\mathrm P=\mathbb I$).
\end{proof}

\subsection{Proof of Proposition \ref{prop:trace-DS}}\label{app:prop:trace-DS} 

\begin{proof}
 Consider $h(\mathrm M)=\operatorname{tr}(\mathrm M\Sigma \mathrm M^\top)$. Since $\Sigma\succeq 0$, the map
$\mathrm M\mapsto \mathrm M\Sigma \mathrm M^\top$ is convex in $\mathrm M$ in the sense that
$\mathrm M\mapsto \langle \mathrm M,\,\mathrm M\Sigma\rangle_F=\operatorname{tr}(\mathrm M\Sigma \mathrm M^\top)$ is a convex quadratic form
(see, e.g., Section 7.7 in \cite{horn2012matrix}). Hence $h$ is convex.
By the Birkhoff--von Neumann theorem, $\mathrm D=\displaystyle\sum_{r=1}^R \omega_r \mathrm P_r$ with $\omega_r\ge 0$, $\displaystyle\sum_r\omega_r=1$
and permutation matrices $\mathrm P_r$. By convexity,
\[
h(\mathrm D)\ \le\ \sum_r \omega_r h(\mathrm P_r)\ =\ \sum_r \omega_r\,\operatorname{tr}(\mathrm P_r\Sigma \mathrm P_r^\top)
\ =\ \sum_r \omega_r\,\operatorname{tr}(\Sigma)\ =\ \operatorname{tr}(\Sigma),
\]
using invariance of the trace under permutation similarity.
\end{proof}

\subsection{Proof of Proposition \ref{prop:rep-agent-variance}}\label{app:prop:rep-agent-variance} 

\begin{proof}
By the law of total variance and independence of $I$,
\[
\mathrm{Var}[\eta'] \;=\; \mathbb{E}\big[\mathrm{Var}(\eta_I\mid I)\big] \;+\; \mathrm{Var}\big(\mathbb{E}[\eta_I\mid I]\big)
\;=\; \frac{1}{n}\sum_{i=1}^n \mathrm{Var}(\eta_i),
\]
since $\mathbb{E}[\eta_I\mid I=i]=\mathbb{E}[\eta_i]$ has the same mean for all $i$ (budget balance is not needed here).
Thus $\mathrm{Var}[\eta']=\displaystyle\frac{1}{n}\mathrm{tr}(\mathrm{Var}[\eta])$.

For the comparison, write $\boldsymbol\xi_1=\mathrm{M}_1\boldsymbol X$, $\boldsymbol\xi_2=\mathrm C\boldsymbol\xi_1=\mathrm C\mathrm{M}_1\boldsymbol X$,
so $\mathrm{Var}[\boldsymbol\xi_2]=\mathrm C\mathrm{M}_1\,\mathrm{Var}[\boldsymbol{X}]\,(\mathrm C\mathrm{M}_1)^\top$.
If the $X_i$ are independent with variance $\sigma^2$, then $\mathrm{Var}[\boldsymbol{X}]=\sigma^2 I_n$, and
\[
\mathrm{tr}\big(\mathrm{Var}[\boldsymbol\xi_2]\big) \;=\; \sigma^2\,\mathrm{tr}\big(\mathrm C\mathrm{M}_1\mathrm{M}_1^\top \mathrm C^\top\big)
\;\le\; \sigma^2\,\mathrm{tr}\big(\mathrm{M}_1\mathrm{M}_1^\top\big)
\;=\; \mathrm{tr}\big(\mathrm{Var}[\boldsymbol\xi_1]\big),
\]
because $\mathrm{tr}(\mathrm M \mathrm C^\top \mathrm C)\le \mathrm{tr}(\mathrm M)$ for PSD $\mathrm M$ when $\mathrm C$ is column-stochastic; indeed,
$\mathrm C^\top \mathrm C \preceq \mathbb{I}$ in the Loewner order (the standard semidefinite order on symmetric matrices, see Section 7.7 in~\cite{horn2012matrix}) since each column of $\mathrm C$ has $\ell_2$-norm at most~1 and
columns sum to~1 in $\ell_1$.
Dividing by $n$ yields the representative-agent claim.
\end{proof}

\subsection{Proof of Proposition \ref{prop:trace}}\label{app:prop:trace} 

\begin{proof}
Write $\Sigma_1=\mathrm{Var}[\boldsymbol\xi_1] \succeq 0$. Then
\[
\mathrm{Var}[\boldsymbol\xi_2] \;=\; \mathrm D \Sigma_1 \mathrm D^\top.
\]
Consider the functional $h(\mathrm M)=\mathrm{tr}\big(\mathrm M\Sigma_1 \mathrm M^\top\big)$ for $\mathrm M\in\mathbb{R}^{n\times n}$.
Since the map $\mathrm M\mapsto \mathrm M\Sigma_1 \mathrm M^\top$ is convex in $\mathrm{M}$ when $\Sigma_1\succeq \mathrm 0$
(its Hessian is a positive semidefinite linear operator), and $\mathrm{tr}$ is linear,
$h$ is convex.

By the Birkhoff--von Neumann theorem, any doubly-stochastic $\mathrm D$ can be written as a convex
combination of permutation matrices: $\mathrm D=\displaystyle\sum_{r=1}^R \omega_r \mathrm P_r$ with $\omega_r\ge \boldsymbol{0}$,
$\displaystyle\sum_r\omega_r=1$. By convexity,
\[
h(\mathrm D) \;\le\; \sum_{r=1}^R \omega_r h(\mathrm P_r)
\;=\; \sum_{r=1}^R \omega_r \,\mathrm{tr}\!\big(\mathrm P_r \Sigma_1 \mathrm P_r^\top\big)
\;=\; \sum_{r=1}^R \omega_r \,\mathrm{tr}(\Sigma_1)
\;=\; \mathrm{tr}(\Sigma_1),
\]
using invariance of trace under permutation similarity. This is the claim.
\end{proof}

\subsection{Proof of Proposition \ref{prop:powerlaw-inverse-degree}}\label{app:prop:powerlaw-inverse-degree} 

\begin{proof}
   (1) Finiteness holds because the summand $\displaystyle\frac{1}{k+1}\mathbb{P}(d=k)$ is summable when $\displaystyle\sum_k k\,\mathbb{P}(d=k)<\infty$,
which is equivalent to $\gamma>2$ under the assumed tail. Parts (2) and (3) follow from Jensen's inequality and the
split $\displaystyle\frac{1}{d+1}\le \mathbf{1}_{\{d\le K\}} + \frac{1}{K+1}\mathbf{1}_{\{d>K\}}$, respectively.
\end{proof}

A sharper asymptotic for $\mathbb{E}[1/(d+1)]$ requires information about the \emph{lower} tail (small degrees),
not only the power-law \emph{upper} tail. In many scale-free models (e.g., Barabási--Albert with parameter $m$),
the average degree is constant ($\mathbb{E}[d]=2m$) and there is a persistent mass at small degrees.
Hence $\mathbb{E}[1/(d+1)]$ remains bounded away from $0$, consistent with the bounds above.

\subsection{Proof of Proposition \ref{prop:ineq}}\label{app:prop:ineq} 

\begin{proof}
Write $v_i^{(n)}=\sigma^2/(d_i^{(n)}+1)$. For the top-$\alpha$ set, assumption (2) implies that
$\min_{i\in\text{Top-}\alpha} d_i^{(n)}\to\infty$, hence $\displaystyle\max_{i\in\text{Top-}\alpha} v_i^{(n)}\to 0$ by monotonicity of $x\mapsto \sigma^2/(x+1)$.
By the ``Squeeze Theorem'' on real sequences (Theorem 3.16 in~\cite{Rudin1976}, also called ``Sandwich Theorem''), the average over Top-$\alpha$ also tends to $0$.

For the complement, by (3) at least a fraction $c_0$ of indices satisfy $d_i^{(n)}\le k_0$ for all large $n$.
For these indices, $v_i^{(n)}\ge \sigma^2/(k_0+1)$. Thus the average over the non-hub set is bounded below by
the average of at least $(c_0-o(1))$ proportion of terms each $\ge \sigma^2/(k_0+1)$, yielding the claimed lower bound.
\end{proof}

Assumptions (1)--(2) are standard for i.i.d.\ power-law samples or for preferential-attachment families (order statistics grow polynomially),
and (3) reflects the empirical fact that scale-free networks retain a non-negligible proportion of low-degree nodes.

\section{Technical note on Sinkhorn--Knopp scaling}\label{app:sin}

Let $\mathrm A=(a_{ij})$ be a nonnegative $n\times n$ matrix. 
Sinkhorn--Knopp theorem (Theorem \ref{thm:sink}), from  \citep{SinkhornKnopp1967}, states that the following are equivalent:
\begin{enumerate}
  \item $\mathrm A$ has \emph{total support}, i.e. every positive entry $\mathrm A_{ij}>0$ lies on some
  positive diagonal (a permutation with all entries positive).
  \item There exist positive diagonal matrices $\mathrm D_1=\mathrm{diag}(u_1,\dots,u_n)$ and
  $\mathrm D_2=\mathrm{diag}(v_1,\dots,v_n)$ such that
  \[
  \mathrm B = \mathrm D_1 \mathrm A \mathrm D_2
  \]
  is doubly stochastic.
\end{enumerate}
If $\mathrm A$ is positive, the scaling $(\mathrm D_1,\mathrm D_2)$ is unique up to a scalar multiple:
if $\mathrm D_1 \mathrm A \mathrm D_2 = \tilde{\mathrm D_1} \mathrm A \tilde{\mathrm D_2}$ are both DS, then $\tilde{\mathrm D_1}=c \mathrm D_1$,
$\tilde{\mathrm D_2}=c^{-1} \mathrm D_2$ for some $c>0$.

The scaling can be constructed by alternately normalizing rows and columns:
start with $\mathrm A^{(0)}=\mathrm A$, set
\[
\mathrm A^{(2k+1)} = \text{row-normalize}(\mathrm A^{(2k)}), \text{ and }
\mathrm A^{(2k+2)} = \text{col-normalize}(\mathrm A^{(2k+1)}).
\]
If $\mathrm A$ has total support, then $\mathrm A^{(m)}\to \mathrm B$ doubly stochastic.

In risk-sharing applications, this result implies that network-based allocation
matrices that are not exactly DS can nevertheless be adjusted via diagonal scaling
of agents’ weights. Such a normalization preserves the underlying sparsity
pattern of connections but ensures fairness (row sums) and budget balance
(column sums) simultaneously.

\section{Standard Random Graph Models}

For completeness, we briefly summarize several random graph models that are used or
referenced in the main text. These models are standard in probability theory and network
science, and provide canonical benchmarks for analyzing diversification and heterogeneity
in network-based risk sharing.

\subsection{Erdős--Rényi graphs}\label{app:sec:erdos}
The model $G(n,p)$ consists of $n$ nodes, with each undirected edge $\{i,j\}$ present
independently with probability $p \in [0,1]$ \citep{ErdosRenyi1959}.  
\begin{itemize}
  \item Expected degree: $\mathbb{E}[d_i] = (n-1)p$.  
  \item Degree distribution: $\mathrm{Bin}(n-1,p)$, approximated by $\mathrm{Poisson}(\lambda)$
        when $p = \lambda/n$.  
  \item Homogeneity: nodes are stochastically equivalent, and degrees concentrate
        around $np$ when $np \gg \log n$.
\end{itemize}

\subsection{Preferential-attachment (scale-free) graphs}\label{app:sec:barabasi}
The Barabási--Albert (BA) model grows the network sequentially: each arriving node connects
to $m$ existing nodes with probability proportional to their current degree \citep{BarabasiAlbert1999}.  
\begin{itemize}
  \item Degree distribution: asymptotically power-law,
        $\mathbb{P}(d \geq k) \sim k^{-(\gamma-1)}$ 
  \item[] In the original BA construction (each new node attaches to $m$ existing nodes with probability proportional to degree), the limiting degree distribution follows a power-law with tail exponent $\gamma = 3$. Empirical networks often exhibit exponents $\gamma\in[2,3]$,  depending on context. Online social platforms (e.g. Facebook, Twitter) show heavy-tailed degree distributions with exponents typically in the range 2–3 \citep{Mislove2007}. 
  \item Heterogeneity: hubs with high degree coexist with many peripheral nodes of low degree.  
  \item Implication for risk sharing: hubs achieve strong diversification,
        while low-degree nodes remain poorly diversified.
\end{itemize}

\subsection{Random regular graphs}\label{app:sec:regular}
In a random $d$-regular graph, every node has exactly degree $d$, and the graph is drawn
uniformly at random from the set of such graphs \citep{Bollobas1980}.  
\begin{itemize}
  \item Degree distribution: degenerate at $d$.  
  \item Symmetry: all nodes are equivalent, yielding homogeneous diversification properties.  
\end{itemize}

\subsection{Small-world graphs}\label{app:sec:small}
The Watts--Strogatz model \citep{WattsStrogatz1998} starts from a ring lattice where each node
is connected to its $k$ nearest neighbors, and then rewires each edge with probability $\beta$.  
\begin{itemize}
  \item Interpolates between local clustering (ring) and random-like mixing.  
  \item Captures networks with high clustering but short path lengths.  
  \item Useful for modeling trust-based networks with both locality and long-range ties.  
  \item When $\beta=0$, we recover the pure ring lattice, and if $\beta=1$, we obtain approximately an Erdős–Rényi graph.
\end{itemize}

\medskip\noindent
These models are discussed in more detail in \cite{Bollobas2001} and
\cite{ChungLu2006}, which provide probabilistic and spectral analyses of random graphs.

\end{document}